\documentclass[11pt]{article}
\usepackage{jheppub}
\usepackage{amsmath,amssymb,amsfonts,graphicx,slashed,amsthm,mathtools,upgreek, enumerate, tensor, subfigure, bbm}
\usepackage[dvipsnames]{xcolor}
\usepackage{arydshln}
\usepackage{slashed}

\usepackage{cancel}
\usepackage{comment}
\usepackage{hyperref}
\usepackage{float}
\usepackage[utf8]{inputenc}
\usepackage[titletoc]{appendix}
\usepackage{caption}

\numberwithin{equation}{section} 
\allowdisplaybreaks

\newcommand{\beq}{\begin{equation}}
\newcommand{\eeq}{\end{equation}}
\newcommand{\beqn}{\begin{eqnarray}}
\newcommand{\eeqn}{\end{eqnarray}}
\newcommand{\bea}{\begin{equation}\begin{aligned}}
\newcommand{\eea}{\end{aligned}\end{equation}}
\newcommand{\bra}[1]{\langle #1 |}
\newcommand{\ket}[1]{| #1 \rangle}

\newcommand{\avg}[1]{\langle #1 \rangle}

\DeclareMathOperator{\tr}{tr}

\newcommand{\SY}{{\bf{Y}}_{\mu}}

\title{
Complexity Growth in Integrable and Chaotic Models} 
\author[a,b]{Vijay Balasubramanian}
\author[a]{\!, Matthew DeCross}
\author[c]{\!, Arjun Kar}
\author[a]{\!, Yue (Cathy) Li}
\author[d]{\!, Onkar Parrikar} 

\affiliation[\,a]{David Rittenhouse Laboratory, University of Pennsylvania,\\
209 S. 33rd Street, Philadelphia PA 19104, USA.}
\affiliation[\,b]{Theoretische Natuurkunde, Vrije Universiteit Brussel (VUB), and \\ International Solvay Institutes, Pleinlaan 2, B-1050 Brussels, Belgium.}
\affiliation[\,c]{Department of Physics and Astronomy, University of British Columbia, \\
6224 Agricultural Road, Vancouver, BC V6T 1Z1, Canada.}
\affiliation[\,d]{Varian Physics Lab, Stanford University, \\
382 Via Pueblo, Stanford, CA 94305, USA.}

\emailAdd{vijay@physics.upenn.edu}
\emailAdd{mdecross@sas.upenn.edu}
\emailAdd{arjunkar@phas.ubc.ca}
\emailAdd{yl244@sas.upenn.edu}
\emailAdd{parrikar@stanford.edu}

\abstract{
We use the SYK family of models with $N$ Majorana fermions to study the complexity of time evolution, formulated as the shortest geodesic length on the unitary group manifold between the identity and the time evolution operator, in free, integrable, and chaotic systems. Initially, the shortest geodesic follows the time evolution trajectory, and hence complexity grows linearly in time. We study how this linear growth is eventually truncated by the appearance and accumulation of conjugate points, which signal the presence of shorter geodesics intersecting the time evolution trajectory. By explicitly locating such ``shortcuts'' through analytical and numerical methods, we demonstrate that:
(a) in the free theory, time evolution encounters conjugate points at a polynomial time; consequently complexity growth truncates at $O(\sqrt{N})$, and we find an explicit operator which ``fast-forwards'' the free $N$-fermion time evolution with this complexity, (b) in a class of interacting integrable theories, the complexity is upper bounded by $O({\rm poly}(N))$, and (c) in chaotic theories, we argue that conjugate points do not occur until exponential times $O(e^N)$, after which it becomes possible to find infinitesimally nearby geodesics which approximate the time evolution operator. Finally, we explore the notion of eigenstate complexity in free, integrable, and chaotic models.

}


\begin{document}

\maketitle

\parskip=10pt


\section{Introduction}

Quantum complexity has been proposed as a quantity relevant for understanding non-perturbative phenomena in quantum gravity, such as the growth of wormholes behind horizons \cite{Susskind:2014rva, Stanford:2014jda, Brown:2015bva,Belin:2018bpg,Saad:2019pqd}, the structure of spacetime singularities \cite{Barbon:2015ria}, and the possible appearance of firewalls \cite{Susskind:2015toa} at late times in an evaporating black hole. 
The challenge in understanding these conjectures is to have a well-defined measure of complexity in the underlying quantum gravity theory, or, equivalently, in its holographic field theory dual, if the latter exists \cite{Maldacena:1997re,Witten:1998qj}.  If the conjectures relating complexity to black hole physics are correct, then we expect that maximally chaotic theories with a holographic dual \cite{Maldacena:2016hyu} feature linear growth of complexity for a time exponential in the entropy of the system.

One possibility is that the relevant notion we seek is quantum state complexity. Some progress has been made in computing the circuit complexity of constructing states in some simple free field theories on a lattice \cite{Jefferson:2017sdb, Chapman:2017rqy, Khan:2018rzm, Hackl:2018ptj}, but defining state complexity in infinite-dimensional Hilbert spaces that appear in the continuum limit is in general difficult. Free systems have also been used by complexity theorists to build intuition about criteria for complexity growth \cite{Atia:2016sax}. An alternative notion that might be relevant is the quantum circuit complexity of the time evolution operator.  There has been some progress in computing this quantity in the context of quantum chaotic systems like black holes in holography, and there is evidence that it grows linearly for a long time as expected, given appropriate assumptions \cite{Susskind:2018fmx,Balasubramanian:2019wgd,Caginalp:2020tzw}.  It is also interesting to consider integrable theories, as some of these do admit quantum gravitational descriptions \cite{Klebanov:2002ja}, as well as to discern by comparison what aspects of chaos lead to an exponential time scale for complexity growth.  In addition, it may be potentially possible to use complexity as an order parameter in families of theories that interpolate between free, integrable and chaotic limits to distinguish between each regime. The purpose of this paper is to further develop  methods for computing the complexity of the time evolution operator in the context of a concrete family of models (the SYK$_q$ models) which can be parametrically tuned between free, integrable, and chaotic regimes.
On general grounds, the complexity of time evolution is expected to grow linearly with time and then plateau at some fixed value, and subsequently undergo Poincar\'{e} recurrences back to small values.
The questions of how long this linear growth persists and what height the plateau reaches depend sensitively on the theory under consideration, and will be central issues in this work.

A major drawback of complexity, from a physicist's viewpoint, is its high degree of non-uniqueness.
Measuring complexity generally requires many choices, such as a choice of gate set, reference state/operator, or tolerance in preparing the final state/operator. 
Determining a natural set of these choices for computing complexity in quantum gravity is beyond the scope of this work.
Furthermore, computer scientists generally think of complexity in terms of small, discrete operations which are composed to create a complex quantum circuit.
As physics generally happens in the continuum, it is advantageous to work with a naturally continuous notion of complexity for operators in physical quantum systems.
Such a notion was formulated in terms of minimal geodesic lengths on high-dimensional manifolds of operators \cite{nielsen2005geometric,Nielsen_2006,Nielsen2007}, and many recent results on complexity make use of this formalism \cite{Jefferson:2017sdb, Chapman:2017rqy, Khan:2018rzm, Hackl:2018ptj, Bhattacharyya:2018bbv,  Balasubramanian:2018hsu,Magan:2018nmu,Caputa:2018kdj, Ali:2018fcz, Balasubramanian:2019wgd, Ali:2019zcj, Bhattacharyya:2019kvj,  Bernamonti:2019zyy,Bernamonti:2020bcf,Erdmenger:2020sup, Flory:2020eot, Flory:2020dja}.\footnote{An alternative approach to defining complexity draws intuition from path integrals in quantum field theory, and interprets quantum circuits as optimized procedures for performing such path integrals \cite{Takayanagi:2018pml,Caputa:2017urj,Camargo:2019isp}.  This approach builds on the tensor network formulation of holography \cite{Swingle:2009bg, Pastawski:2015qua, Hayden:2016cfa, Milsted:2018san, Milsted:2018yur, Bao:2018pvs, Caputa:2020fbc}.  For yet another approach to the analysis of complexity growth, this time making use of unitary $k$-designs and random circuits, see \cite{Brandao:2016ghi,Roberts:2016hpo,Brandao:2019sgy}. }
In this setting, there is a relatively natural choice which leads to a unique definition of quantum complexity that is equivalent to the quantum circuit definition: the degree of locality of the Hamiltonian defines a set of ``easy" operators (operators which are at most as local as the Hamiltonian).
Operators which are more non-local than the Hamiltonian are considered ``hard".
This choice of splitting into easy and hard operators corresponds to a choice of metric (the ``complexity metric") on the group manifold of unitary operators, where directions corresponding to easy operators have low weight and directions corresponding to hard operators have weight of order the Hilbert space dimension.\footnote{There are proposals for complexity which utilize instead the bi-invariant geometry, which treats easy and hard operators on an equal footing \cite{Yang:2019iav,Yang:2019udi,Yang:2020tna}.  A proposal which defines the ``infinite cost factor" limit has also been explored \cite{Bueno:2019ajd,Erdmenger:2020sup}. }

In this geometric formalism for complexity, studying complexity growth is related to studying the growth of the distance function from the identity operator in the complexity metric.
As globally length-minimizing geodesics are often difficult to find on generic Riemannian manifolds, the strategy employed by \cite{Balasubramanian:2019wgd} was to look for geodesics that were at least initially globally minimizing, and then to search along those geodesics for possible obstructions to global minimality.
On a general Riemannian manifold, such obstructions are either local or global: local obstructions, also known as ``conjugate points", imply that the geodesic is not a local minimum of the distance function (i.e., it is a saddle point), while global obstructions, or ``geodesic loops", imply that the geodesic is not globally minimal.
Any complete picture of complexity growth must include an accounting of both local and global obstructions. Locating global geodesic loops (which are not signaled by conjugate points) in a systematic way is computationally intractable, but (as shown in \cite{Balasubramanian:2019wgd}) conjugate points can be more tractable under certain assumptions, and have a significant effect on complexity growth.

Once global minimality of a given geodesic is obstructed, either by a conjugate point or a geodesic loop, we are guaranteed that the growth rate of the distance function will no longer be exactly linear along this geodesic.
However, it may still be approximately linear (with a smaller growth rate), if we encounter an isolated conjugate point or geodesic loop, since the new geodesics involved in computing the distance may have growing lengths.
We expect, however, that the first conjugate point or geodesic loop along a fixed geodesic associated with time evolution will quickly be followed by the end of complexity growth in general, rather than just a reduction in growth rate, possibly due to a rapid accumulation of subsequent conjugate points/loops. 
This intuition comes partially from the expected behavior for chaotic Hamiltonians, where after meeting the first obstruction to complexity growth, the complexity is expected to quickly plateau \cite{Brown:2017jil}.
We will see that free and integrable models also reproduce this expectation, with the first conjugate point signaling the end of complexity growth entirely and a transition to a plateau regime in the distance function within an $O(1)$ time afterward.

Since this paper explores a variety of topics using both analytic and numerical techniques, we now provide a road map by summarizing our main results by section.
In Sec.~\ref{sec:conjugate-review}, we begin with a review of the geometric formalism developed in \cite{Nielsen2007,Balasubramanian:2019wgd} to keep the discussion self-contained. Since conjugate points play an important role in this work, we explain their significance to complexity growth in detail. We also give new sufficient-but-not-necessary criteria for locating conjugate points in terms of more familiar quantities from thermalization and quantum chaos such as adjoint eigen-operators of the Hamiltonian and infinite-temperature thermal two-point functions.

In Sec.~\ref{sec:free}, we apply these criteria to the free ($q=2$) SYK model. Since free models have relatively simple Hamiltonians, their time evolution operators are simple enough that we can locate all conjugate points and even the geodesic loops which take over after some of these conjugate points.
In fact, we find a large number of conjugate points (associated to easy operators) which occur at early (i.e., polynomial) times and signal a rapid end to the linear growth of the complexity of time evolution in the free models, followed by a long plateau. 
The geodesic loops we study are in one-to-one correspondence with these early conjugate points, which demonstrates that these global obstructions to complexity growth, which are otherwise very difficult to locate, can sometimes be found by leveraging the study of the local obstructions, i.e., conjugate points.
These effects place a sharp upper bound on the complexity growth of free systems of $N$ fermions, which is $O(\sqrt{N})$ in the plateau regime.

In Sec.~\ref{sec:perturbation} we consider a class of interacting-but-integrable deformations of the free SYK model. We first study a subset of conjugate points in perturbation theory in the coupling constant (which controls the deformation), and find that the deformation causes these conjugate points to move to later times.
Going beyond perturbation theory, we also identify certain geodesic loops using the structure of the integrable interaction, which bound the complexity of time evolution in this interacting model. 
These geodesic loops may not be signaled by conjugate points; if so, this feature of complexity growth distinguishes integrable interacting theories from free theories.
The bound on complexity in this interacting integrable model predicts a plateau of height order $O(N)$ which begins at a time significantly later than in the free model (but still at polynomial time) .
Some straightforward generalizations of this simple model show plateaus of height $O(\text{poly}(N))$ for any polynomial in $N$.

In Sec.~\ref{sec:local-chaotic}, we study the possibility of finding conjugate points at sub-exponential times in chaotic theories. In \cite{Balasubramanian:2019wgd}, it was argued that in chaotic models, ``almost all'' of the conjugate points occur at exponential times. One might worry that there are a small number of conjugate points which can nevertheless appear at an earlier time; in particular, prime suspects for this are conjugate points for which the Jacobi field involves only local operators. Indeed, these are precisely the type of conjugate points which obstruct complexity growth in the free SYK model at an early time. Using  ideas from random matrix theory and the Eigenstate Thermalization Hypothesis (ETH), we show that in chaotic models, such conjugate points cannot occur before exponential time. This strengthens the arguments of \cite{Balasubramanian:2019wgd} that local obstructions to complexity growth in chaotic models do not occur at sub-exponential times.

In Sec.~\ref{sec:syk-numerical}, we numerically study conjugate points for various integrable and chaotic SYK Hamiltonians up to $N=8$ (i.e., four qubits). We emphasize that this gives us a concrete (albeit numerical) way to locate obstructions to complexity growth for SYK models, which can in principle be extended to larger $N$. The numerical results show that a class of conjugate points associated to simple operators (i.e., where the Jacobi field mostly involves simple operators) stay at a fixed time scale as we crank up the weighting of the hard directions, while those associated to hard operators (i.e., where the Jacobi field mostly involves hard operators) rapidly shift to late times proportional to the weighting factor (which is taken to be exponential in $N$).
Together with the results of Sec.~\ref{sec:local-chaotic}, this provides further evidence that the complexity in chaotic models does not plateau until exponential times, modulo global obstructions.
Our results on the behavior of conjugate points and geodesic loops in the complexity geometry illustrate how rich geometric structure underlies the growth of the complexity of time evolution in free, integrable, and chaotic theories.

In Sec.~\ref{sec:ECH} we revisit and explore the Eigenstate Complexity Hypothesis (ECH) of \cite{Balasubramanian:2019wgd}. We present the ECH matrices calculated from the eigenstates of SYK models with varying degrees of integrability. The off-diagonal matrix entries for free SYK show fluctuations that scale with the size of the system $N$, while those of the chaotic models are suppressed uniformly for all $N$, modulo discrete symmetries of the system. The distribution of the off-diagonal elements is discrete for the free system, which echoes the strong reduction of the number of degrees of freedom analyzed in Sec.~\ref{sec:free}. This reduction already does not occur in interacting systems even if they are integrable. As expected, the off-diagonal distributions of the interacting-integrable and the chaotic systems have continuous support.

Finally, we end with a discussion of interesting points and future work (Sec.~\ref{sec:disc}).


\section{Conjugate points and complexity growth}\label{sec:conjugate-review}
In this section, we will discuss conjugate points and their effect on complexity growth. We will see that conjugate points can be studied very concretely in terms of more familiar quantities such as Hamiltonian eigenvectors, thermal two-point functions etc. Of course, global geodesic loops (which are not signaled by conjugate points) should ultimately also play an important role in any complete picture of complexity growth, but a systematic study of these appears to be intractable for now.
\subsection{Conjugate points in the Euler-Arnold formalism}
Let $\mathcal{U}(\mathcal{H})$ be the group of all unitary operators on a finite dimensional Hilbert space $\mathcal{H}$,\footnote{We will usually restrict to the special unitary group.} and let $\{T_i\}$ be an orthogonal basis for its Lie algebra with respect to the Killing norm. Quantum circuit complexity is polynomially equivalent to a distance function on $\mathcal{U}(\mathcal{H})$, with a certain right-invariant metric $G_{ij}$ (the ``complexity metric") which weights tangent space directions corresponding to non-local operators heavily \cite{Nielsen2007}. The choice of which operators are to be considered non-local is not unique; a common choice for spin systems or systems with clear notions of site-based locality is to consider as local all operators which are at most $k$-local (act on at most $k$ sites or $k$ degrees of freedom) for some fixed $k$ that does not scale with $N$, the total number of degrees of freedom. Then, any operators which are $(k+1)$-local or greater are considered nonlocal and are weighted in the complexity metric. The weighting of the ``hard" directions is $O(e^S = \dim \mathcal{H})$ to ensure the polynomial equivalence to circuit complexity. To implement this weighting, we choose a metric that splits the tangent space into easy directions $\{T_{\alpha}\}$ and hard directions $\{T_{\dot{\alpha}}\}$, and weights the hard directions in the length functional by a ``cost factor'' $(1+\mu)$:
\beq \label{metric}
G_{ij} = \left(\begin{matrix}\delta_{\alpha\beta} & 0 \\ 0 & (1+\mu) \delta_{\dot{\alpha}\dot{\beta}}\end{matrix}\right).
\eeq
When the cost factor is $\mu=0$, all operators are equally weighted. In Nielsen's setup, the cost factor is taken to be $\mu \sim e^{\alpha S}$ for some $O(1)$ coefficient $\alpha$, but we will let $\mu$ be arbitrary throughout. 

Quantum circuits in this context are paths on the unitary manifold, and the complexity of a unitary $U$ is measured by the length of a minimal geodesic connecting the identity to $U$. An efficient formulation of the geodesic equation on Lie groups equipped with right-invariant metrics was given by Arnold and is known as the Euler-Arnold equation \cite{Arnold1966,Tao2010}\footnote{The Euler-Arnold equation, while not used in the original formulation of geodesic complexity \cite{Nielsen2007}, has been previously used in the context of geodesic complexity and holography \cite{Balasubramanian:2018hsu,Balasubramanian:2019wgd,Erdmenger:2020sup,Flory:2020eot, Flory:2020dja}. }
\begin{equation}
    G_{ij} \frac{dV^j}{ds} = {f_{ij}}^k V^j G_{k\ell} V^\ell , \label{eq:EA}
\end{equation}
where $G_{ij}$ is the metric on the Lie algebra defined in \eqref{metric}, and ${f_{ij}}^k$ are the structure constants of the Lie algebra.
The Euler-Arnold equation determines a velocity vector $V(s)$, which can then be integrated to give the path followed by the geodesic:
\begin{equation}
    U(s) = \mathcal{P} \exp \left( \int_0^s ds'\, V(s') \right),
\label{eq:path-order}
\end{equation}
where $\mathcal{P}$ stands for path ordering. We will always parametrize our paths with $s \in [0,1]$.

Understanding the growth of complexity for a family of operators $U(t)$ now essentially reduces to the question of when a minimal geodesic becomes non-minimizing, and subsequently finding the new minimal geodesic. While the latter problem is difficult, there is actually a local (in the space of paths) signature that a geodesic is non-minimizing: conjugate points.\footnote{Encountering a conjugate point is sufficient, but not necessary, for a geodesic to become non-minimizing.} Conjugate points, which were the main objects of study in \cite{Balasubramanian:2019wgd}, represent deformations of a geodesic which leave the length and the endpoint locations fixed to first order in the deformation parameter. More precisely:

\textbf{Definition}: Given a geodesic $U(s):[0,1] \to \mathcal{U}(\mathcal{H})$ with $U(0)=P$ and $U(1)=Q$, if there exists a one-parameter family of curves $U(\eta,s):[-\epsilon,\epsilon] \times [0,1] \to \mathcal{U}(\mathcal{H})$ such that $U(\eta,s)$ obeys the geodesic equation at first order in $\eta$ with $U(\eta,0)= P$ and $U(\eta,1)=Q+O(\eta^2)$, then $P$ and $Q$ are said to be \emph{conjugate} along the geodesic $U(s)$. 

\begin{figure}
    \centering
    \includegraphics[height=5cm]{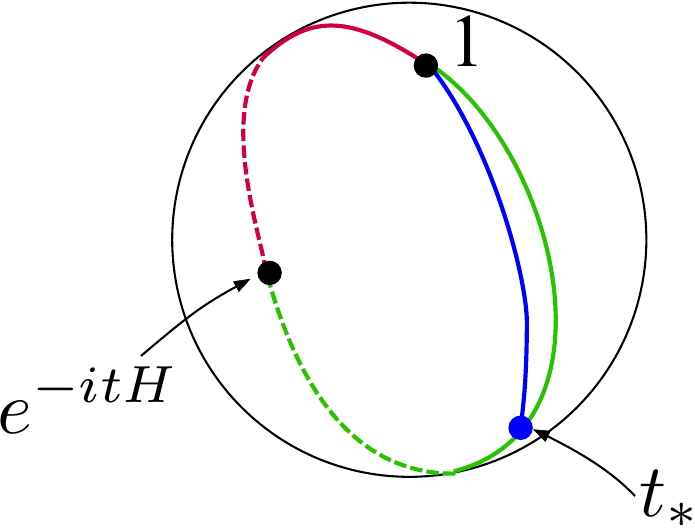}
    \caption{A cartoon of what happens when a geodesic between $1$ and $e^{-itH}$ encounters a conjugate point. The green geodesic is initially the locally minimizing geodesic, before it reaches $t_*$ where it encounters a conjugate point (the blue point). For $t>t_*$, the green geodesic is no longer locally minimizing, and a different geodesic (shown in red) will be the local minimum. Note that even though the conjugate point indicates this transition, the new geodesic which takes over after $t_*$ is not infinitesimally close to the original one (although we can reach it by gradient flow from the original geodesic).}
    \label{fig:conjugate_point}
\end{figure}

Deformations which leave the length (but possibly not the endpoints) fixed to first order in the parameter $\eta$ above can be represented as vector fields $\frac{dU}{d\eta}$ along the geodesic, and are called \emph{Jacobi fields}. If we imagine deforming the geodesic along a Jacobi field, we are not guaranteed that the endpoint of the geodesic will remain fixed at leading order in the deformation parameter. If we do find such a Jacobi field with fixed endpoints along some segment of a geodesic, a shorter path between the initial point and a later point along the path can be found by deforming the geodesic along the Jacobi field between the points which are conjugate, and subsequently smoothing out the resulting kink where the deformed and original paths meet (this relies on the endpoint deviation vanishing). This smoothing reduces the length at a lower order in the deformation parameter than the deformation's leading order effect on the length. Thus, the question of whether the endpoint of a geodesic segment is conjugate to the initial point is equivalent to whether there exists a Jacobi field along the segment that fixes the endpoints at leading order in the deformation parameter. Importantly for us, \emph{the conjugate point is a signature that the original geodesic is no longer locally minimizing, and is in fact a saddle point after that time}. It is worth emphasizing also that the new minimal geodesic which takes over may not be infinitesimally near the original one, and can be highly non-trivial. See Fig.~\ref{fig:conjugate_point} for a depiction of a conjugate point on a compact manifold. 

The  unitary operator studied in both this work and \cite{Balasubramanian:2019wgd} is the time evolution operator $e^{-itH}$, where $H$ is the system Hamiltonian.
At small enough times $t$, the globally minimizing geodesic between the identity and $e^{-itH}$ solving \eqref{eq:EA} is the ``linear geodesic", a specific geodesic with constant velocity $V(s) = Ht$.
Since the linear geodesic is constant in $s$, the path ordering in \eqref{eq:path-order} is trivial, and the path of unitaries is $U(s) = e^{-istH}$.
By perturbing the Euler-Arnold equation with $V \to V+\delta V$ and keeping the $O(\delta V)$ terms, we obtain the Jacobi equation; plugging in $V(s) = Ht$ for the original background geodesic around which we are perturbing, we obtain the Jacobi equation specialized to the linear geodesic:
\begin{align}\label{JE1}
    i \frac{d\delta V_L}{ds} & = \mu t [H,\delta V_{NL}]_L , \\
    i \frac{d\delta V_{NL}}{ds} & = \frac{\mu t}{1+\mu} [H,\delta V_{NL}]_{NL} ,\label{JE2}
\end{align}
where the $L,NL$ subscripts represent projections to the easy and hard subspaces of generators in $\mathfrak{su}(2^{N/2})$.
As this is a first-order ordinary differential equation, any initial condition $\delta V(0)$ can be integrated to a solution $\delta V(s)$.
To find conjugate points, \cite{Balasubramanian:2019wgd} defined a super-operator \textbf{Y}$_\mu$ (where the subscript $\mu$ denotes the cost factor) which takes as input a tangent vector at the identity $\delta V(0)$, produces the corresponding solution of the Jacobi equation $\delta V(s)$, and then computes the first order deviation of the endpoint $e^{-itH}$ under deformation of the linear geodesic by $\delta V(s)$:
\begin{equation}
    \mathcal{P} \exp \left( -i \int_0^1 ds (Ht+\delta V(s)) \right) = e^{-itH} \left( 1 - i \textbf{Y}_\mu (\delta V(0)) + O(\delta V^2) \right) .
\end{equation}
By expanding the path ordered exponential in a Dyson series, the super-operator effectively computes
\begin{equation}
    \textbf{Y}_\mu (\delta V(0)) = \int_0^1 ds\; e^{istH} \delta V(s) e^{-istH} . 
\label{eq:dyson}
\end{equation}
Solving for $\delta V(s)$ in terms of the initial velocity deformation $\delta V(0)$ using equations \eqref{JE1} and \eqref{JE2}, we obtain
\begin{equation}
\begin{split}
    \textbf{Y}_{\mu}(\delta V(0)) = & \int_0^1 ds e^{iHts} \biggl[ \delta V_L(0) - i \mu t \sum_{\dot{\alpha}} \frac{\exp \left( \frac{-i \mu t \lambda_{\dot{\alpha}} s}{1+\mu}  \right)-1}{\frac{-i \mu t \lambda_{\dot{\alpha}} }{1+\mu}} \delta \tilde{V}^{\dot{\alpha}}(0) [H,\tilde{T}_{\dot{\alpha}}]_L \\
    & + \sum_{\dot{\alpha}} \exp \left( \frac{-i \mu t \lambda_{\dot{\alpha}} s}{1+\mu}  \right) \delta \tilde{V}^{\dot{\alpha}}(0) \tilde{T}_{\dot{\alpha}} \biggr] e^{-iHts} ,
\end{split}
\label{eq:super-operator}
\end{equation}
where the $L,NL$ subscripts denote projections to the purely local and purely nonlocal operator subspaces, and $\{\tilde{T}_{\dot{\alpha}}\}$ is a new orthogonal basis of generators for the nonlocal subspace which diagonalizes the super-operator $[H,\,\cdot\: ]_{NL}$ with eigenvalues $\lambda_{\dot{\alpha}}$.
The intuition for this formula, derived in detail in \cite{Balasubramanian:2019wgd}, is essentially to sum up the total deviation along the geodesic by translating the Jacobi field back to the identity and integrating. 
Functionally, it is the first order correction term in a Dyson series expansion of the path ordering \eqref{eq:path-order} in the Jacobi field $\delta V$, as written in \eqref{eq:dyson}. The cost factor $\mu$ should be taken to be $O(e^S)$ in the complexity geometry. 
A conjugate point appears when the first order deviation in the endpoint vanishes for some initial tangent vector $\delta V(0)$. Therefore, \emph{time evolution encounters a conjugate point at time $t$ if the super-operator \textbf{Y}$_\mu$ has a zero mode at time $t$.}
In particular, the zero modes must be Hermitian so that they are valid elements of $\mathfrak{su}(2^{N/2})$. 

\subsection{General criteria for locating conjugate points} \label{sec:general_criteria}

In this section, we give general criteria for locating conjugate points. Our conditions are sufficient for the existence of conjugate points, but not necessary. Their utility lies in the fact that they relate the locations of conjugate points to more familiar properties of quantum systems such as Hamiltonian eigenstates, adjoint eigen-operators, infinite-temperature thermal two-point functions etc.
Further, the hypotheses for these criteria are crucially independent of the cost factor $\mu$ and the precise form of $H$ (so long as it is at most $k$-local). 

\noindent\textbf{Claim 1}: Let $H$ be a $q$-local Hamiltonian where $q\leq k$.

(i) If the Hamiltonian has an adjoint eigen-operator $O$, i.e., $\text{ad}_H O = [H, O] = \lambda O$ for some $\lambda \in \mathbb{R}$, such that $O$ lies entirely within the subspace of $k$-local operators, then time evolution will encounter conjugate points at
\beq
t_{*} = \frac{2\pi}{\lambda}\mathbb{Z}.
\eeq

(ii) If the Hamiltonian has an adjoint eigen-operator $O'$, i.e., $\text{ad}_H O'=[H, O'] = \lambda' O'$ for some $\lambda' \in \mathbb{R}$, such that $O'$ lies entirely within the subspace of non-$k$-local operators, then time evolution will encounter conjugate points at
\beq
t_{*} = \frac{2\pi(1+\mu)}{\lambda'}\mathbb{Z},
\eeq
where $\mu$ is the cost factor. 

\noindent\textbf{Proof}: The proof proceeds by evaluating the super-operator $\textbf{Y}_{\mu}$ on the given adjoint eigen-operators of $H$:

(i) Notice first that evaluation of \textbf{Y}$_\mu$ on a purely local operator $O$ involves only the first term in the square brackets in \eqref{eq:super-operator}.
The second and third terms do not contribute since they depend only on the nonlocal components $\delta \tilde{V}^{\dot{\alpha}}(0)$, which are all zero for local $\delta V(0) = O$ (by assumption).

By evaluating matrix elements of the output in the energy eigenbasis or by expanding out the exponential in the first term of \eqref{eq:super-operator}, we conclude that if $O$ is a $k$-local, adjoint eigen-operator of the Hamiltonian, then $O$ is also an eigen-operator of the super-operator $\mathbf{Y}_\mu$:
\beq
\mathbf{Y}_\mu(O) = \phi(\lambda t)\; O,\;\; \phi(x) = \frac{e^{ix}-1}{ix}.
\eeq
The eigenvalue $\phi(\lambda t)$ becomes zero at the locations $t_* = \frac{2\pi}{\lambda}\mathbb{Z}$, and so we have conjugate points at these locations. 
Of course, to have a conjugate point we must have a Hermitian zero mode of \textbf{Y}$_\mu$, and indeed we do after observing that under these conditions we also have
\begin{equation}
    \textbf{Y}_\mu (O^\dagger) = \phi(-\lambda t) O^\dagger ,
\end{equation}
which means that $O+O^\dagger$ and $i(O-O^\dagger)$ are zero modes at the specified times. 
In this argument, we have not used the form of the Hamiltonian at all except in our assumption that it has a $k$-local adjoint eigen-operator.

(ii) Likewise, the evaluation of \textbf{Y}$_\mu$ on a purely nonlocal operator $O'$ involves only the third term inside the square brackets in \eqref{eq:super-operator}.
The first term inside the square brackets does not contribute because it involves a local projection which will vanish for a purely nonlocal $\delta V(0) = O'$.
To see why the second term does not contribute, observe that it involves the commutator $[H,\tilde{T}_{\dot{\alpha}}]$ followed by a projection to the local subspace.
Since we have assumed $[H,O']=\lambda' O'$ for a purely nonlocal $O'$, we may take a single $\tilde{T}_{\dot{\alpha}}$ to lie along the $O'$ direction, and set the rest of $\delta \tilde{V}^{\dot{\alpha}}(0)$ to zero.
Then, every term of the form $\delta \tilde{V}^{\dot{\alpha}}(0) [H,\tilde{T}_{\dot{\alpha}}]_L$ vanishes; all but one vanish due to $\delta \tilde{V}^{\dot{\alpha}}(0) = 0$, and the final term with $T_{\dot{\alpha}} \propto O'$ vanishes due to the projection after the commutator.
Again evaluating matrix elements in the energy basis or expanding out the exponential in the third term in equation \eqref{eq:super-operator}, we find that if an adjoint eigen-operator $O'$ exists such that $O'$ lies entirely along the hard directions, then $O'$ is also an eigen-operator of the super-operator $\mathbf{Y}_\mu$:
\beq
\mathbf{Y}_\mu(O') = \phi \left( \frac{\lambda' t}{1+\mu} \right)\; O'.
\eeq
In this case, the eigenvalue becomes zero at the locations $t_* = \frac{2\pi(1+\mu)}{\lambda'}\mathbb{Z}$.
Again, we have in mind that the zero modes which lead to conjugate points at these times are really the Hermitian combinations of $O'$ and ${O'}^\dagger$, where we have a minus sign in the argument of $\phi$ for ${O'}^\dagger$.\hfill $\Box$

We will encounter examples of such conjugate points when we discuss the free SYK model in the next section. In fact, all conjugate points at $q=2$ belong to either type (i) or (ii) in Claim 1. As another non-trivial example, consider the $q=4$ SYK model. Let the gate set be chosen such that $2$-local and $4$-local operators are treated as easy, while all other operators are treated as hard.\footnote{Note that this is a different notion of locality than the notion we use in the majority of this work, where instead we pick some constant cutoff $k$ for which all operators that are at most $k$-local are considered easy.}
Since the Hamiltonian has a fermion-number symmetry, we can label eigenstates with the corresponding $\pm 1$ eigenvalue. Any adjoint eigen-operator of $H$ of the form $|m\rangle \langle n|$ where $|m\rangle$ and $|n\rangle$ have opposite fermion number will therefore entirely lie along the hard directions, and will thus give conjugate points at exactly $t_* = \frac{2\pi(1+\mu)}{E_m-E_n}\mathbb{Z}$. 

Note that in case (ii), the conjugate points appear at late times, provided the cost factor is taken to be large. In the geometric setup, this cost factor is often taken to be exponential in $S$, and so we see that these late-time conjugate points appear as an obstruction to complexity growth at exponential times, which is the expected time-scale for complexity saturation in chaotic quantum systems.
So, chaotic theories may have conjugate points of the sort predicted by the hypothesis of Claim 1.(ii), as indeed exemplified by the above example of the $q=4$ SYK model with the gate set protected by fermion number symmetry. On the other hand, in (i), the location of the conjugate points does not depend on $\mu$; in this case, conjugate points could potentially lead to a short-time obstruction to complexity growth, where by ``short-time'' we mean a time of order $\text{poly}(S)$. Indeed this is precisely what happens in the free SYK model (see Sec.~\ref{sec:free}). Since chaotic systems (or, more precisely, systems with geometric, holographic duals) are expected to have complexity growth for exponential time, then we expect such conjugate points which are ``associated to simple operators" do not occur in chaotic systems before exponential times. In order to probe this  further, we re-formulate the existence of such conjugate points as follows:

\noindent\textbf{Claim 2}: Let $M_{\alpha\beta}$ be the positive semi-definite matrix
\beq
M_{\alpha\beta}(t) = \int_0^1 ds\int_0^1ds'\,\mathrm{Tr}\,[e^{i(s-s')tH}T_{\alpha}e^{-i(s-s')tH}T_{\beta}],
\eeq
where $T_{\alpha}$ and $T_{\beta}$ are simple (i.e., at most $k$-local) generators. If $M_{\alpha\beta}(t)$ has a zero mode at time $t_*$, then time evolution encounters a conjugate point at $t_*$. 

\noindent\textbf{Proof}: Let $X^{\alpha}$ be the zero mode of $M_{\alpha\beta}(t)$ at time $t_*$. Now consider 
\beq
\delta V(0) = \sum_{\alpha}X^{\alpha}T^{\alpha}.
\eeq
We evaluate $\textbf{Y}_{\mu}(\delta V(0))$, and compute the Frobenius norm of the resulting operator:
\beqn\label{MY}
||\textbf{Y}_{\mu}(\delta V(0)) ||^2_F &=& \mathrm{Tr}[(\textbf{Y}_{\mu}(\delta V(0)))^{\dagger}\textbf{Y}_{\mu}(\delta V(0))]\nonumber\\
&=& \int_0^1ds\int_0^1ds'\,\mathrm{Tr}[e^{istH}\delta V(0)^{\dagger}e^{-istH}e^{is'tH}\delta V(0)e^{-is'tH}]\nonumber\\
&=&\sum_{\alpha,\beta} (X^{\alpha})^*M_{\alpha\beta}(t)X^{\beta},
\eeqn
where in the second equality we have used the fact that the chosen $\delta V(0)$ lies entirely along the easy directions. Since at time $t_*$ we have $\sum_{\beta} M_{\alpha\beta}(t_*)X^{\beta}=0$, then we conclude that at time $t_*$ we must have
\beq
||\textbf{Y}_{\mu}(\delta V(0)) ||_F = 0,
\eeq
which consequently implies $\textbf{Y}_{\mu}(\delta V(0)) = 0$. Thus, we have a conjugate point at $t_*$. \hfill $\Box$

We will henceforth refer to such conjugate points (which correspond to zero modes of $M_{\alpha\beta}$) as \emph{simple} or \emph{local conjugate points}. Note that $M_{\alpha\beta}(t)$ is the infinite temperature, thermal two-point function between two time-averaged simple operators. Claim 2 above states that the first time $t_*$ at which this matrix develops a zero mode is precisely when the time evolution geodesic $e^{-itH}$ encounters a conjugate point, and thus necessarily stops being a locally minimal geodesic. Conceptually, this relates complexity growth with a more familiar quantity, namely the thermal two-point function. (In Appendix~\ref{sec:twoptfunctions} we write a general expression relating the full super-operator to the infinite-temperature thermal two-point function which may be of interest for future work.) On the practical side, note that $M$ is a much smaller matrix (polynomial in size) as compared to $\textbf{Y}_{\mu}$ (which is exponential in size), and thus gives a useful sufficient-but-not-necessary criterion for locating conjugate points. Such conjugate points, should they exist, will be at a time $t_*$ which is independent of $\mu$.

We can also give a physical interpretation to the smallest eigenvalue of $M_{\alpha\beta}(t)$. Let $\lambda_{\text{min}}(t)$ be the smallest eigenvalue of $M_{\alpha\beta}(t)$. From equation \eqref{MY}, we have
\beq
\lambda_{\text{min}} = \text{min}_{\delta V(0)}\;\frac{||\textbf{Y}_{\mu}(\delta V(0))||_F^2}{e^{-S}||\delta V(0)||^2_F},
\eeq
where we minimize with respect to all (non-zero) local operators $\delta V(0)$. Physically, this means that it is possible to find an infinitesimally nearby curve with a \emph{local} initial velocity $V(0) = Ht + \epsilon \delta V(0)$ (for infinitesimal $\epsilon$) which satisfies the geodesic equation up to $O(\epsilon^2)$, such that the end point displacement from the target unitary $e^{-itH}$ satisfies: 
\beq
|| U(1) - e^{-itH}||^2_F = \epsilon^2\, e^{-S}\lambda_{\text{min}}(t)\,||\delta V(0)||^2_F+ O(\epsilon^3) ,
\eeq
where the subscript $F$ stands for Frobenius norm. Thus, $\lambda_{\text{min}}$ is a measure of the error up to  which we can approximate time evolution by an infinitesimally nearby geodesic. If $\lambda_{\text{min}}$ is exactly zero for some $t_*$, then we have a conjugate point at that location. We will call $\lambda_{\text{min}}$ the \emph{impact parameter} since it measures how close a trajectory with local initial velocity $V(0) = Ht + \epsilon \delta V(0)$ comes to hitting the exact final unitary $e^{-iHt}$.
We will return to this in Sec.~\ref{sec:local-chaotic}, where we will argue that in chaotic models, $\lambda_{\text{min}} \sim O(e^S)$ for $t<e^S$, but becomes small thereafter. Consequently, local conjugate points, should they exist, cannot appear before exponential time in chaotic theories. 

\subsection{Relevance of conjugate points in complexity growth}

In AdS/CFT, several conjectures relate the quantum complexity of the CFT time evolution operator to the growth of a bulk quantity like an extremal volume or action.
While there has been progress in understanding the details of such bulk volume or action calculations, a field-theoretic formulation of circuit complexity in infinite-dimensional Hilbert spaces which reduces to the standard notion of quantum complexity in finite dimensions is still incomplete.
This has led to the development of toy models for the complexity geometry which are designed to reproduce certain coarse-grained features of distances on the full unitary manifold with a right-invariant complexity metric \cite{Brown:2016wib,Lin:2018cbk}.
For example, one such toy model involves a particle moving on a high-genus Riemann surface with metric induced from its universal covering space, the hyperbolic disk $\mathbb{H}^2$ \cite{Brown:2016wib}.

While such toy models have led to interesting insights into the behavior of holographic complexity, they lack a crucial feature of the finite-dimensional complexity geometry: conjugate points.
In the example of the particle moving on a Riemann surface, there are no conjugate points because the sectional curvature of the induced metric is strictly negative.
In an attempt to justify this shortcoming, one might appeal to results of Milnor on sectional curvatures of Lie groups \cite{MILNOR1976293}, which roughly imply that most sectional curvatures on ``complicated enough" Lie groups with right-invariant metrics are negative.
Crucially, however, the results of \cite{MILNOR1976293} do not imply that all sectional curvatures are negative.
In fact, the most important result in \cite{MILNOR1976293} for our purposes is the fact that any right-invariant metric on $SU(n)$ for $n>2$ is required to either have some strictly positive sectional curvature or else be completely flat.
Some of these curvatures were recently computed explicitly for complexity metrics in \cite{Auzzi:2020idm} and were found to be positive.

There is an obvious tension between the lack of conjugate points in the toy models and the fact that in the finite-dimensional complexity geometry (a right-invariant metric on the unitary group), conjugate points are guaranteed to exist and obstruct the complexity growth of time evolution with arbitrary Hamiltonians.\footnote{See \cite{naitoh1981conjugate} for a simpler Lie group geometry where conjugate points are guaranteed to be the first obstruction to complexity growth.}
This fact was emphasized in the original formulation of complexity geometry \cite{Nielsen2007}, and also in its adaptation to the Euler-Arnold formalism \cite{Balasubramanian:2019wgd}.

A possible perspective on this tension is to imagine that, in the context of finite-dimensional holographic systems like the SYK model, conjugate points may move off ``to infinity" or simply disappear from the relevant minimal geodesic as the cost factor $\mu$ is increased, leading to a situation where there are never any conjugate points along the geodesic relevant to complexity.
Unfortunately, as was briefly discussed in \cite{Balasubramanian:2019wgd}, this is impossible due to two facts: 1) the initial linear growth of time evolution's complexity is captured by the linear geodesic, and 2) the right-invariant complexity metric depends continuously on the cost factor $\mu$.
Using these two facts, we will explain in more detail an argument sketched in \cite{Balasubramanian:2019wgd} which demonstrates that conjugate points must exist along the linear geodesic for arbitrary local Hamiltonians at finite distance and cost factor.

We begin by noticing that the case of zero cost factor, $\mu = 0$, corresponds to a bi-invariant metric on the Lie group.
In this case, the exponential maps of the Lie group and Riemannian manifold coincide, which means that all geodesics take the form $e^{-isH}$ for some Hamiltonian $H$.
In the bi-invariant metric, conjugate points are known to exist at finite distance \cite{Nielsen2007}.\footnote{In particular, they appear at $t_* = \frac{2\pi}{(E_m -E_n)}\mathbb{Z}$ for all eigenvalues $E_m,\;E_n$ of $H$.}
Since they begin at finite distance, they cannot move ``to infinity" since they are zero modes of the super-operator \textbf{Y}$_\mu$, and these zero modes depend continuously on $\mu$.
If they were to move to infinity at some finite value of $\mu$, there would be a discontinuity in the super-operator before and after this value.

The only other possibility is that the conjugate points could ``disappear", which would correspond to a zero mode of the super-operator becoming complex.
That is to say, the Jacobi field which gives the conjugate point could pick up a non-Hermitian contribution at some finite value of $\mu$, and in order to have a true conjugate point the Jacobi field must be purely Hermitian.
We do not have a guarantee from simple continuity that this cannot happen, since, for example, the same thing happens for the polynomial equation $x^2+\mu = 0$. 
There is no discontinuity in $\mu$ on the left hand side but the solutions become complex as $\mu$ goes from negative to positive.
So too could the Jacobi fields generating the conjugate points become non-Hermitian at some finite value of $\mu$.
However, it turns out that this also cannot happen.

To understand why conjugate points cannot disappear, we apply Morse theory on the space of paths.
Let $\Omega(U_1,U_2)$ be the space of paths on the Lie group between unitary operators $U_1$ and $U_2$.
The dimensionality of this space is formally infinite, but this subtlety turns out not to affect any conclusions \cite{Milnor1963,PALAIS1963299,Smale1964}.\footnote{The original work of Morse, reviewed by Milnor in section III of \cite{Milnor1963}, relies on finite-dimensional approximations of the full path space, to which Morse's theory is then applied.  By contrast, \cite{PALAIS1963299,Smale1964} prove the same results by working directly in the infinite-dimensional setting.}
For the complexity of time evolution, the relevant path spaces are
\begin{equation}
    \Omega_{t,H} \equiv \Omega(1,e^{-itH}) .
\end{equation}
That is to say, $\Omega_{t,H}$ is the space of all smooth paths $\gamma(s)$ with $\gamma(0) = 1$ and $\gamma(1) = e^{-itH}$.
For convenience, we parametrize all paths with $s \in [0,1]$.
We can consider a real-valued function on $\Omega_{t,H}$ which is often called the energy functional
\begin{equation}
    E_{(\mu)}(\gamma) \equiv \int_0^1 ds \left( \sum_\alpha V_\alpha^2 + (1+\mu) \sum_{\dot{\alpha}} V_{\dot{\alpha}}^2 \right) , 
\end{equation}
where we have made use of the splitting of the Lie algebra into local and nonlocal directions (labeled by $\alpha$ and $\dot{\alpha}$, respectively), the right-invariance of the complexity metric, and also the velocity along the path $V(s) \equiv d\gamma /ds$.

Critical points of the energy functional $E_{(\mu)}$ on $\Omega_{t,H}$ are precisely the paths with velocity $V(s)$ which are geodesics between the identity and $e^{-itH}$.
The most important of these for us is the linear geodesic, which is simply the path $V(s) = Ht$.
Since the linear geodesic is independent of $\mu$, the point in $\Omega_{t,H}$ to which it corresponds is fixed as $\mu$ increases.
Call this point $L \in \Omega_{t,H}$.
The tangent space to $L$, and more generally to any point $\gamma$ in the path space, is the space of vector fields $\delta V(s)$ along $\gamma$ for which $\delta V (0) = \delta V(1) = 0$.\footnote{$\delta V$ must vanish at the endpoints since $\Omega_{t,H}$ is defined as the space of paths with fixed endpoints at $1$ and $e^{-itH}$.}
With this notion of tangent space, one can define the Hessian of the energy functional $E_{(\mu)}$ evaluated at $L$, which we will denote $E''$ (where the derivatives are taken in the space of paths), keeping all dependence on $\mu$, $t$, and $H$ implicit.

One can now apply the Morse index theorem on $\Omega_{t,H}$ using $E_{(\mu)}$ as the Morse function.
The Morse index theorem applied to our situation states that \emph{the number of negative eigenvalues of $E''$ is equal to the number of conjugate points (counted with multiplicity) along the geodesic $L$, and that $E''$ only has a zero eigenvalue if the endpoint $e^{-itH}$ is conjugate to the identity along $L$} \cite{Milnor1963}.
Since $E_{(\mu)}$ depends continuously on $\mu$, and $L$ is independent of $\mu$, the eigenvalues of $E''$ must also depend continuously on $\mu$.
Therefore, the only way we can ``lose" a conjugate point along $L$ is for an eigenvalue of $E''$ to pass continuously through zero.
In other words, the conjugate point must move beyond $e^{-itH}$ along the linear geodesic.
This means that conjugate points cannot simply disappear; the only way to get rid of them is to boost the cost factor $\mu$ high enough to push them past the endpoint of the geodesic $L$.
So, by taking $t$ large enough (but still finite), we can extend the endpoint of $L$ to always find conjugate points along $L$ at finite distance and cost factor, just as we claimed.
This also amounts to a non-perturbative proof that zero modes of \textbf{Y}$_\mu$ are always Hermitian matrices because if a zero mode were non-Hermitian then the corresponding conjugate point would disappear, but the zero modes are in one-to-one correspondence with the conjugate points.

All of this means that conjugate points are relevant for any complexity calculation which employs complexity geometry and involves the linear geodesic $L$, and toy models which ignore them are useful but incomplete representations of the total complexity geometry.
It would be interesting to find a toy model which can include conjugate points.

\section{Free theories}\label{sec:free}

We now study the growth of complexity in free and integrable models, starting with the quadratic free fermion model, with Hamiltonian
\beq
H = i\sum_{i,j} J_{ij}\psi^i\psi^j, \label{eq:freehamil}
\eeq
where $J_{ij}$ is an anti-symmetric matrix and the sums run from $1$ to $N$. We consider this model as a $q=2$ instance of the SYK$_q$ family of models \cite{Maldacena:2016hyu, SYKkitaev},\footnote{See \cite{Sarosi:2017ykf} for a pedagogical review.}
\beq
H = i^{q/2} \sum_{i_1 \ldots i_q} J_{i_1 \ldots i_q} \psi^{i_1} \ldots \psi^{i_q}. \label{eq:SYKHamil}
\eeq
There, $J_{i_1 \ldots i_q}$ is totally antisymmetric and is drawn from a Gaussian distribution with mean zero and variance parameterized by $\mathcal{J}$, 
\beq
\langle J_{i_1 \ldots i_q}^2 \rangle = \frac{2^{q-1} (q-1)!}{q} \frac{\mathcal{J}^2}{N^{q-1}}.
\eeq
In our context, we consider a particular instance of the model where we have sampled the couplings $J_{ij}$ from such a distribution.
The matrix $J_{ij}$ is antisymmetric and therefore can be written as
\beq
J = VDV^T, 
\eeq
where $V$ is an orthogonal matrix, and $D$ is block-diagonal with antisymmetric blocks:
\beq
D = \left(\begin{matrix} 0 & \omega_1/2 & 0 & 0 & \cdots \\ -\omega_1/2 & 0 & 0 & 0 & \cdots \\ 0 & 0 & 0 & \omega_2/2 & \cdots \\ 0 & 0 & -\omega_2/2 & 0 & \cdots\\ \vdots & \vdots & \vdots & \vdots & \ddots \end{matrix}\right).
\eeq
The matrix $V$ is constructed as follows. First, write the usual diagonalization $J = U\Sigma U^{\dagger}$. Since $J$ is antisymmetric, the matrix $U$ is unitary and the eigenvalues of $J$ are $\pm i\omega_p / 2$, $p = 1\ldots N/2$. Next, define the unitary matrix 
\begin{equation} 
M = \frac{1}{\sqrt{2}} \begin{pmatrix} 1 & 1 \\ i & -i \end{pmatrix} .
\end{equation}
Using  $M$, build the matrix $\Omega = \mathbbm{1}_{(\frac{N}{2} \times \frac{N}{2}) } \otimes M$, the $N \times N$ block diagonal matrix formed by $N/2$ copies of $M$. A short computation shows that $\Omega^{\dagger} \Sigma \Omega = D$, so $J = U\Omega D \Omega^{\dagger} U^{\dagger}$. It turns out that $U\Omega$ is always a real matrix, so we can identify $V = U\Omega$ and then $J = VDV^T$.
Now we can define new fermion operators
\beq
\Psi_i = \sum_j\psi_j V_{ji},
\eeq
which also satisfy the same anti-commutation relations 
\beq
\left\{\Psi_i, \Psi_j\right\} = 2\delta_{ij}.
\eeq
The notion of locality is unchanged by this transformation, since the new fermion operators are linear in the old ones and $V$ is orthogonal.
In terms of these new operators, the Hamiltonian becomes
\beq
H = i\sum_{p=1}^{N/2} \omega_p\Psi_{2p-1}\Psi_{2p}.
\eeq
Finally, we define the ladder operators
\beq
A_p = \frac{1}{2}\left(\Psi_{2p-1}+i\Psi_{2p}\right),\;\;A^{\dagger}_p =\frac{1}{2}\left( \Psi_{2p-1}-i\Psi_{2p}\right),
\eeq
which satisfy
\beq
\left\{A_p , A_p^{\dagger}\right\} = 1,
\eeq
with all other anti-commutators vanishing. In terms of these, the Hamiltonian becomes
\beq
H = \sum_{p=1}^{N/2}\omega_p(A_p+A_p^{\dagger})(A_p-A_p^{\dagger})=\sum_{p=1}^{N/2}\omega_p \left(2A_p^{\dagger} A_{p}-1\right).
\eeq
In this Dirac fermion language, there is a new useful basis of the $2^N-1$ operators which span the algebra $\mathfrak{su}(2^{N/2})$.
To define this basis, we begin by writing a vector of 4 operators
\begin{equation}
    \vec{J}^{(p)} \equiv (1,A_p,A^\dagger_p, 2A^\dagger_p A_p - 1) .
\end{equation}
With the entries of this vector labeled by indices in the order $\beta_p \in \{0,-,+,3\}$, the operator basis is then the set of products over all choices of $\{\beta_p\}$,
\begin{equation}
    J^{(1)}_{\beta_1} J^{(2)}_{\beta_2}  \ldots J^{(N/2)}_{\beta_{N/2}} ,
\label{eq:dirac-basis}
\end{equation}
where we discard the identity $\beta_1 = \ldots = \beta_{N/2} = 0$.
The Hamiltonian can be written compactly as
\begin{equation}
    H = \sum_{p=1}^{N/2} \omega_p J^{(p)}_3 ,
\end{equation}
and $J^{(p)}_3$ has eigenvalues $\pm 1$ in the energy eigenbasis.
Thus, the $2^{N/2}$ eigenvalues of $H$ are
\begin{equation}
    \sum_{p=1}^{N/2} \sigma_p \omega_p ,
\end{equation}
for every possible choice of the coefficients $\sigma_p$ from $\{\pm 1\}$.
The natural notion of locality in the Dirac basis, derived by considering an operator with $k$ Majorana operators to be $k$-local, is to consider $J_+$ and $J_-$ as 1-local operators but $J_3$ as a 2-local operator and $J_0$ as a 0-local operator.
Then, the locality of a general product of $J^{(p)}_{\beta_p}$'s is simply the sum of the individual localities. Since the Hamiltonian is 2-local, then we will take $k=2$ in the rest of this section.
Free fermion time evolution was also studied in \cite{Atia:2016sax}; we will see that geodesic complexity techniques both reproduce the results found there and allow us to uncover new features of free theories.

\subsection{Conjugate points}
We are interested in the complexity of the unitary operator 
\beq
U = e^{-itH}.
\eeq
First, we study conjugate points for the linear geodesic. 
Let us look at the super-operator \textbf{Y}$_\mu$ derived in \cite{Balasubramanian:2019wgd}, whose zero modes as a function of $t$ correspond to conjugate point locations. For free theories, it turns out that \textit{every} conjugate point corresponds to a local or non-local eigen-operator of $\text{ad}_H$. 

To understand the free theory, we observe that the adjoint action of the Hamiltonian is already diagonal in the Dirac fermion basis \eqref{eq:dirac-basis} and, recalling that $\beta_p \in \{0,+,-,3\}$, we can write it as
\begin{equation}
    [H,J^{(1)}_{\beta_1}\dots J^{(N/2)}_{\beta_{N/2}}] = 2 \sum_p \left( \omega_p \delta_{\beta_p +} - \omega_p \delta_{\beta_p -} \right) J^{(1)}_{\beta_1}\dots J^{(N/2)}_{\beta_{N/2}} .
\end{equation}
For the $2^{N/2}-1$ operators in the basis that involve only $J_0$ or $J_3$, the adjoint eigenvalue is zero.
We take 3-local and higher operators to be nonlocal, since the Hamiltonian is quadratic in the Majorana fermions, i.e., $k=2$. Since the adjoint eigen-operators of the Hamiltonian split nicely into simple and hard operators, we can obtain all the conjugate points using Claim 1 in Sec.~\ref{sec:general_criteria}.
The locations of conjugate points associated to local operators are given by Claim 1.(i). They are
\begin{equation}
    t_* = \frac{\pi}{\omega_{p_1}+\omega_{p_2}} \mathbb{Z}, \quad \frac{\pi}{\omega_p} \mathbb{Z} , \quad \frac{\pi}{\omega_{p_1}-\omega_{p_2}} \mathbb{Z} ,
\end{equation}
where $p_1 \neq p_2$, $\omega_{p_1} > \omega_{p_2} > 0$, and $\omega_p > 0$.
We may always define all $\omega_p > 0$ for the price of introducing a minus sign in the definition of $\sigma_p^3$, and we order the $\omega_p$ so that $\omega_p > \omega_q$ for $p < q$.
These families of conjugate points are associated with operators of the forms
\begin{equation} 
A_{p_1}^\dagger A_{p_2}^\dagger + A_{p_2} A_{p_1}, \quad A_p^\dagger + A_p, \quad A_{p_1}^\dagger A_{p_2} + A_{p_2}^\dagger A_{p_1} , \label{eq:conj_ops}
\end{equation}
respectively.
These are two-fold degenerate conjugate points; there are corresponding partner operators, such as $i(A^\dagger_p - A_p)$ for the second operator in \eqref{eq:conj_ops}.
Similarly, the locations of conjugate points corresponding to the purely nonlocal operators are given by Claim 1.(ii),
\begin{equation}
    t_* = \frac{\pi (1+\mu)}{\omega_{p_1} + \omega_{p_2} + \omega_{p_3}} \mathbb{Z}, \quad 
    \frac{\pi (1+\mu)}{\omega_{p_1} + \omega_{p_2} - \omega_{p_3}} \mathbb{Z} , \quad
    \dots ,\quad  \quad \frac{\pi (1+\mu)}{\sum_p \omega_p} \mathbb{Z} ,
\end{equation}
where we cannot pick the same $\omega_p$ twice, and all possible combinations of plus and minus signs can occur in the denominators subject to the constraint that the overall result should be positive.
The associated operators are respectively
\begin{equation}
    A_{p_1}^\dagger A_{p_2}^\dagger A_{p_3}^\dagger + A_{p_3}A_{p_2}A_{p_1}, \quad A_{p_1}^\dagger A_{p_2}^\dagger A_{p_3} + A_{p_3}^\dagger A_{p_2}A_{p_1}, \quad \dots ,\quad \left( \prod_p A_p \right)^\dagger + \prod_p A_p .
\end{equation}

\subsection{Exact geodesics}
As we showed above, the conjugate points associated to nonlocal directions are quite far from the identity due to the cost factor, while those associated with local directions occur at a time of $O(\text{poly}(S))$ since numerical experiments reveal the range of non-zero $\omega_p$ to be between $O(1/N)$ and $O(1)$, with a typical spacing of $1/N$.\footnote{It would be interesting to determine an analytic formula for these quantities, and it may be achievable since we are interested in the eigenvalues of a particularly simple random matrix $J_{ij}$: an $N \times N$ antisymmetric random matrix with Gaussian entries of mean zero and variance $1/N$ (for $\mathcal{J}=1$).} 
Therefore, as one might expect in the free theory, obstructions to complexity growth occur nearly immediately.
We would like to go beyond just identifying the location of such obstructions and actually find the new globally length-minimizing geodesics which replace the linear geodesic in the complexity calculation.

A general geometric strategy for finding these new geodesics will be to isolate relevant subalgebras of $\mathfrak{su}(2^{N/2})$ where the effect of the conjugate point can be completely understood.
While technically there are conjugate points associated with 2-local operators which occur sooner, it is illustrative to begin with the family of points corresponding to a single ladder operator $A_p^\dagger$.
Again, there is a two-fold degeneracy of these conjugate points which arises due to the $A_p$.
To understand the behavior of the conjugate point at $t_* = \pi/\omega_p$, we must at least study the algebra generated by $A_p$ and $A_p^\dagger$.
Furthermore, whatever our choice of subalgebra, we must also include the relevant terms in the Hamiltonian, namely the projection of $H$ to our subalgebra.
The smallest possible subalgebra that fits our needs is just a copy of $\mathfrak{su}(2)$, generated by
\beq
[J^{(p)}_+,J^{(p)}_-] = J^{(p)}_3 , 
\eeq
\beq
[J_3^{(p)}, J_+^{(p)}]= 2J_+^{(p)} ,
\eeq
\beq
[J_3^{(p)}, J_-^{(p)}]= -2J_-^{(p)} .
\eeq
While $J_+$ and $J_-$ are not Hermitian, they are traceless, and we may Hermiticize them by taking linear combinations to obtain valid $\mathfrak{su}(2)$ generators.
This $\mathfrak{su}(2)$ is essentially a copy of the Pauli algebra, and exponentiates to an $SU(2)$ subgroup within our total manifold $SU(2^{N/2})$.
Since $SU(2)$ is simply $S^3$, we have an immediate interpretation of the conjugate points at $t_* = \pi/\omega_p$.
The path traced by $e^{-itH}$ in $SU(2^{N/2})$ begins at the north pole of this $S^3$, and the conjugate point sits at the south pole.

There is a corresponding algebraic avenue for understanding this result. 
Because $J_3^{(p)}$ has eigenvalues $\pm 1$ in the energy eigenbasis, and we have $[J_3^{(p_1)}, J_3^{(p_2)}] = 0$, the time evolution operator splits as
\begin{equation}
    e^{-itH} = \prod_{p=1}^{N/2} e^{-i\omega_p t J_3^{(p)}} .
\end{equation}
The conjugate point occurs at $t_* = \pi/\omega_p$ because it is precisely at this time that we have the equivalence
\begin{equation}
    e^{-i\omega_p t_* J_3^{(p)}} = e^{i\omega_p t_* J_3^{(p)}} .
\end{equation}
Thus, we are led to conclude that the new geodesics should be written by simply modifying the coefficients in the Hamiltonian in the appropriate way to trace out the other half of the great circle and to return to the north pole on $S^3$ at time $2t_*$.

We must simply reverse the velocity in the $J_3^{(p)}$ direction and decrease the coefficient appropriately so that it returns to zero as we return to the north pole of the relevant $SU(2)$.
The appropriate operation which achieves this, and takes into account the necessary changes when encountering all other conjugate points in the family $t_* = \pi \mathbb{Z} / \omega_p$, can be neatly written as
\begin{equation}
    \omega_p t J_3^{(p)} \to -i \log (e^{i \omega_p t}) J_3^{(p)} ,
\label{eq:1-local-replacement}
\end{equation}
where we always take the principal branch of the logarithm with a cut on $(-\infty,0)$.
This function effectively computes $\omega_p t$ modulo $2\pi$, with a result in the range $(-\pi, \pi]$.
The new velocity, including the effects of all conjugate points associated to 1-local operators, is
\begin{equation}
    Ht \to -i \sum_p \log (e^{i\omega_p t}) J_3^{(p)} .
\end{equation}
This velocity is still constant (i.e., $s$-independent) and purely local, so it is still a geodesic.
This family of geodesics was found in \cite{Atia:2016sax} as fast-forwarding Hamiltonians for free fermion time evolution.
However, as we will now see, there is more structure in the free theory which allows for a constant factor improvement over the above construction.

Before dealing with the conjugate points corresponding to 2-local operators, we note that the family of geodesics just described induces a self-averaging behavior for the complexity in precisely the way which was understood in a toy model developed in \cite{Balasubramanian:2019wgd}.
The toy model made use of the single-qubit complexity geometry, which is simply $SU(2)$.
An $N$-member ensemble of single-qubit Hamiltonians was defined, and the ensemble-averaged complexity in that situation behaved in precisely the manner we have just described for the free SYK model with $N$ Majorana fermions.
This is a quantitative instance of self-averaging, an effect which is generally difficult to understand analytically.

With that being said, there are effects at times of $O(\text{poly}(S))$ in the $N$-Majorana theory which are ``intrinsically quantum", and do not arise from self-averaging.
These are the conjugate points like $t_* = \pi \mathbb{Z} / (\omega_{p_1} + \omega_{p_2})$ associated with 2-local operators, which have no analog in the ensemble average toy model.
We will see explicitly why this is the case by again analyzing these conjugate points from both geometric and algebraic viewpoints.
On the geometric side, we search for a subalgebra which includes the operators generating the (two-fold degenerate) conjugate point, any other operators necessary for the subalgebra to close, and the projection of the Hamiltonian to this subspace.
It turns out that we can again manage with just a single $\mathfrak{su}(2)$ subalgebra generated by
\begin{equation}
    A_{p_1}^\dagger A_{p_2}^\dagger, \quad A_{p_1} A_{p_2}, \quad J_3^{(p_1)} + J_3^{(p_2)} .
\end{equation}
It may seem like the third operator is not actually a projection of the Hamiltonian, since $H$ involves a weighted sum of $J_3^{(p)}$'s with different coefficients.
The point here is that we are projecting $H$ along a particular direction which involves operators from both the $p_1$ and the $p_2$ subalgebras discussed in the 1-local case.
In other words, we rewrite
\begin{equation}
    \omega_{p_1} J_3^{(p_1)} + \omega_{p_2} J_3^{(p_2)} = \frac{\omega_{p_1} + \omega_{p_2}}{2} (J_3^{(p_1)} + J_3^{(p_2)}) + \frac{\omega_{p_1} - \omega_{p_2}}{2} (J_3^{(p_1)} - J_3^{(p_2)}) ,
\end{equation}
and project along the $J_3^{(p_1)} + J_3^{(p_2)}$ direction.
This subalgebra exponentiates to a copy of $SU(2)$, and we again have an interpretation of the conjugate point as the south pole of an $S^3$.

The algebraic viewpoint is a bit more instructive in this case as opposed to the 1-local situation.
In that case, we observed that $i J_3^{(p)}$ had eigenvalues $\pm i$, and so its matrix exponential was $2\pi$-periodic, which led to a conjugate point at $\pi/\omega_p$ where $\omega_p$ was the coefficient of $J_3^{(p)}$ in $H$.
However, in general, for sums of different $J_3^{(p)}$, the most we can say is that the eigenvalues are integers.
Luckily, for the sum of precisely two $J_3^{(p)}$, the eigenvalues are $\pm 2$ or zero.
Therefore, the matrix exponential of $i(J_3^{(p_1)} + J_3^{(p_2)})$ is actually $\pi$-periodic.
This explains why the conjugate point sits at $t_* = \pi / (\omega_{p_1} + \omega_{p_2})$ as opposed to $2 \pi / (\omega_{p_1} + \omega_{p_2})$.

With an understanding of these conjugate points, we can write the new velocity.
Again we simply make the replacement
\begin{equation}
    \omega_{p_1} t J_3^{(p_1)} + \omega_{p_2} t J_3^{(p_2)} \to -i \log ( e^{i (\omega_{p_1} + \omega_{p_2}) t} ) \frac{J_3^{(p_1)} + J_3^{(p_2)}}{2} + \frac{\omega_{p_1} - \omega_{p_2}}{2} t (J_3^{(p_1)} - J_3^{(p_2)}) ,
\label{eq:2-local-replacement}
\end{equation}
which handles all conjugate points in the family $t_* = \pi \mathbb{Z} / (\omega_{p_1} + \omega_{p_2})$.
When we encounter the conjugate point at $t_* = \pi / (\omega_{p_1} - \omega_{p_2})$, the same replacement will occur on the second term on the right hand side above.
Unfortunately, as much as we would like to write a single expression which incorporates the changes in the geodesic after all conjugate points associated to 1- and 2-local operators, we cannot accomplish this with our logarithm branch cut trick.
We will simply provide a description of the total velocity.

The linear geodesic begins with velocity $Ht$.
As we increase $t$, the endpoint of the geodesic moves, and we may encounter a conjugate point.
To keep track of these changes, we keep a table of coefficients $c_p(t)$, and we will periodically update these with $t$ so that the velocity of the globally minimal geodesic is always (before times of order $\mu$)
\begin{equation}
    V = \sum_p c_p(t) J_3^{(p)} .
\label{eq:global-velocity}
\end{equation}
Initially, at very small times $t$, we have $c_p(t) = \omega_p t$, and these coefficients will always locally increase linearly with $\omega_p t$.
To know when we should update a particular $c_p(t)$, we keep track of three types of quantities: $c_{p_1}(t)+c_{p_2}(t)$, $c_{p_1}(t)-c_{p_2}(t)$, and $c_p(t)$ themselves.
We will update $c_p(t)$ so that all such quantities are in the range $(-\pi,\pi]$.
Whenever one of the first or second type increases beyond $\pi$, we rewrite the velocity as in \eqref{eq:2-local-replacement} and update $c_{p_1}(t)$ and $c_{p_2}(t)$.
In the case of the first type, their sum is updated to be in the range $(-\pi,\pi]$ but their difference is unchanged.
For the second type, their difference is updated but their sum is unchanged.
Similarly, when one of the third kind increases beyond a multiple of $\pi$, we simply update the individual $c_p(t)$ to be in the range $(-\pi,\pi]$.
Notice that in the first two cases we needed a second linear relationship (keeping one of the sum or difference fixed) in order to update both $c_{p_1}(t)$ and $c_{p_2}(t)$.

\begin{figure}[t]
    \centering
    \includegraphics[scale=.8]{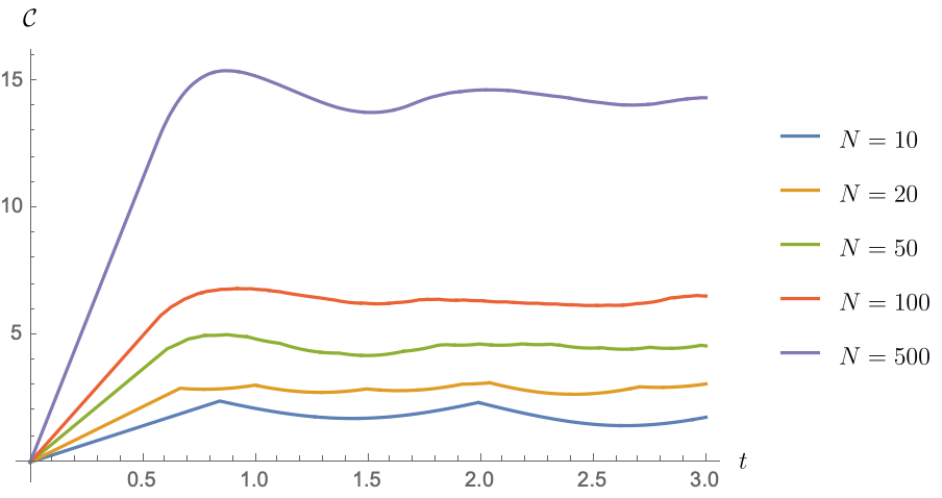}
    \caption{A plot of complexity $\mathcal{C}(t)$ for instances of the $N=10$, $N=20$, $N=50$, $N=100$, and $N=500$ free SYK model with $\mathcal{J}=1$.  The $\omega_p/2$ which control the growth of the coefficients $c_p(t)$ in \eqref{eq:global-velocity} are the positive eigenvalues of the antisymmetric coupling matrix $J_{ij}$ whose entries are independent Gaussian random variables with mean zero and variance $\sigma^2 = \mathcal{J}^2/N$.
    }
    \label{fig:free-n100}
\end{figure}

It is this ``quantum" effect which separates the exact $N$-Majorana free theory from the ensemble average of single-qubit theories.
Indeed, the quantum interference effects of the conjugate points related to 2-local operators actually prevent us from reaching the conjugate points associated with 1-local operators: if we ever had $c_i(t) = \pi$ for some $c_i(t)$, we would certainly have some sum or difference of $c_i(t)$ equal to $\pi$ already, unless all the other $c_{j\neq i}(t)$ are zero, which is quite finely tuned.
The fact that we are able to understand all the globally minimizing geodesics before exponential times as a function of $t$ by studying only conjugate points on the linear geodesic is due to the geometric description of all local operator conjugate points as south poles of 3-spheres.
A plot of complexity for the free SYK model is shown in Fig.~\ref{fig:free-n100}.
As all $O(N)$ terms in the diagonalized Hamiltonian are upper-bounded by $\pi$ due to the local conjugate points and corresponding geodesic loops, there is a hard upper bound on the free complexity of $O(\sqrt{N})$.
The conjugate points associated with non-local operators are not relevant for this discussion because they occur at far later times of $O(\mu \sim e^{\alpha S})$.
Thus, we have essentially determined the full structure of geometric complexity in the free theory at sub-exponential times, up to the existence of geodesic loops which are not signaled by conjugate points.

We can make progress on this front by ruling out at least one simple class of potential geodesic loops which are not signaled by conjugate points.
Though we have demonstrated that conjugate points corresponding to nonlocal (3-local and higher) operators occur at times of order $\mu$, and are thus not relevant for complexity growth below such times, we may wonder if a similar algebraic effect as \eqref{eq:2-local-replacement} can occur for e.g. a sum of three $J_3^{(p)}$'s even without a conjugate point.
It is clear that there is an algebraic relationship which would allow such a replacement: the sum of three or more $J_3^{(p)}$'s is still integer valued, so the matrix exponential will be at most $2\pi$-periodic.
This would be a geodesic loop that occurs without a conjugate point in the free theory.
However, this cannot occur, because of the way the coefficients scale.
In general, a sum of $m$ $J_3^{(p)}$'s has a half-periodicity (which was the conjugate point location for $m=1$ and $m=2$) when
\begin{equation}
    \sum_{i=1}^m c_i(t) = 4^{\lfloor \frac{m-1}{2} \rfloor} \pi .
\end{equation}
Notice that for $m=1$ and $m=2$, the right hand side is $\pi$, and this is led to our update rules for the $c_i(t)$.
However, for $m=3$ it is $4\pi$, which means the average value of the $c_i(t)$ is $4\pi / 3$, which is greater than $\pi$.
We cannot reach this regime, because the $c_i(t)$ are all valued in $(-\pi,\pi]$ due to effects of the 1- and 2-local operators.
For $m=4$, the average value is $\pi$, but this also cannot occur because (since $c_i(t) \leq \pi$) we must have $c_i(t) = \pi$ for all $i$ for the average of them to be $\pi$.
This violates the conditions placed by the 2-local operators, namely that the sum of any two $c_i(t)$ is less than or equal to $\pi$.
A similar story holds for all $m > 4$.
So, no periodicity effects arise for this number of $J_3^{(p)}$'s, and indeed there are no conjugate points associated with such effects.

Throughout this discussion, we have assumed that $k=2$, or in other words that 3-local and greater operators are considered nonlocal from the perspective of the complexity metric.
However, the classification and locations of conjugate points at arbitrary $\mu$ that we described in Claims 1.(i) and 1.(ii) in Sec.~\ref{sec:general_criteria}, and then applied to the free theory, does not actually depend on this assumption.
The reason our analysis cannot be extended to $k>2$ is more subtle.
Let us consider $k=3$ for concreteness.
By Claim 1.(i), there is a conjugate point family at $\pi \mathbb{Z} / (\omega_1 + \omega_2 + \omega_3)$, associated with operators like $A_1^\dagger A_2^\dagger A_3^\dagger + A_3 A_2 A_1$.
The next step to understand these conjugate points is to analyze this operator and the Hamiltonian projection from the geometric or algebraic perspective.
From the geometric perspective, the situation is significantly more complicated than the 1- and 2-local cases because the relevant subalgebra is no longer $\mathfrak{su}(2)$.
The two Hermitian operators associated to the conjugate point and the Hamiltonian projection do not close under the Lie bracket, and more operators must be added to ensure closure.
Moreover, beyond $S^3$, none of the higher-dimensional spheres are Lie groups, so the geometric interpretation of the conjugate point will no longer simply be arrival at the south pole of a sphere.
The algebraic perspective has an analogous difficulty: the sum of three or more $J_3^{(p)}$'s can certainly have an eigenvalue of $\pm 1$ or $\pm 2$, which is less than the multiplicity we would need to explain the appearance of the conjugate point so soon by some periodicity condition on the matrix exponential.

In a certain sense, this result is not surprising.
The 3-local and higher operators do not have such simple interpretations because physically they represent ``shortcuts through chaos" which generate free time evolution faster than the free system itself.
That is, after the linear geodesic (corresponding to time evolution with respect to the free Hamiltonian) is replaced by a new globally minimizing geodesic at a non-local conjugate point, the shorter trajectory along the new global minimizer can be thought of as Hamiltonian evolution with respect to a different, chaotic effective Hamiltonian.
These shortcuts would be interesting to understand, as they utilize chaos in a structured way.\footnote{It is conceivable that the 3-local deformation added to the free Hamiltonian, which makes the total effective Hamiltonian $H_{\text{eff}} = H + \epsilon \delta V/t$, may not be chaotic for a finite range of values $\epsilon > 0$.  We do not have concrete arguments against this, but it is unlikely that the flow in the space of paths generated by a 3-local $\delta V$ will remain in the local subspace, since closure of the relevant subalgebra will introduce even more non-local operators which may enter the effective Hamiltonian of the new length-minimizing geodesic. This would lead to a theory which involves many non-local interactions, which is likely chaotic. It would be interesting to confirm this intuition.} In other words: ``Chaos isn't a pit. Chaos is a ladder."

Of course, it could be that the chaotic deformation ``wraps around" a submanifold in the same way as the conjugate points we were able to understand above, and leaves us with a globally minimizing velocity that does not actually involve 3-local or more terms.
This observation does not change our conclusion that there are special chaotic deformations which allow speedups for free time evolution; it only means these speedups are not optimal.

\subsection*{Summary}

We found all conjugate points along the linear geodesic in the complexity metric, and we determined the associated geodesic loops.
To find the conjugate points, we determined all eigenvectors of the super-operator \textbf{Y}$_\mu$ at arbitrary $\mu$.
Using this information, we constructed a geodesic (as a function of $t$) which is globally length-minimizing from the identity to $e^{-iHt}$, up to the existence of possible geodesic loops which were not associated with any conjugate points.
The length, and therefore the complexity, was bounded at $O(\sqrt{N})$.


\section{Integrable theories and deformations}\label{sec:perturbation}

In Sec.~\ref{sec:free}, we studied obstructions to complexity growth along the linear geodesic associated with time evolution in the free SYK model. In this section, we will study a class of interacting-but-integrable Hamiltonians. To this end, consider adding a quartic interaction $H_1$ to the free (quadratic) Hamiltonian $H_0$ which preserves integrability.
An example of such an interaction is a term which is quadratic in the $J_3^{(p)}$, so that the total Hamiltonian is
\beq
H = H_0 + \epsilon H_1 = \sum_i \omega_i J_3^{(i)} + \frac{\epsilon}{4}  \sum_{ij} M_{ij} J_3^{(i)} J_3^{(j)} .
\label{eq:perturbed-hamiltonian}
\eeq
Since $[J_3^{(i)},J_3^{(j)}]=0$, we may take $M_{ij}=M_{ji}$ to be a symmetric matrix. 
In order to avoid introducing a nonzero trace, we take $M_{kk} = 0$. Since $H$ commutes with all the $J^{(i)}_3$'s, this interaction term preserves integrability. 
It is important to note that since our full Hamiltonian is now quartic, we will treat $k \leq 4$-local operators as easy in the complexity metric.

The analysis of conjugate points for the above integrable Hamiltonian is somewhat involved, and so we will approach it via perturbation theory in the coupling $\epsilon$. However, we note that this analysis becomes much simpler if we modify our gate set slightly by allowing ourselves access to one new elementary operation. To see this, observe that the adjoint eigenvectors of the Hamiltonian $H$ are given by 
\beq \label{gateset}
|m\rangle\langle n| = c_{p,q}\; A_{i_1}^{\dagger}\cdots A_{i_p}^{\dagger}P_0\,A_{i_1}\cdots A_{i_q} ,\;\;\; P_0 = |0\rangle\langle 0|,
\eeq
where there are $p$ Dirac excitations in $\ket{m}$ and $q$ Dirac excitations in $\ket{n}$, and $c_{p,q}$ is a constant. Note that these operators are almost like ``local'' operators built out of products of individual fermions, except  
for the inclusion of the projector $P_0$ in the product above. In principle, we could consider a gate set where operators of the form \eqref{gateset} with $(p+q)\leq k$ are treated as local/simple, while the rest are treated as hard. We can then again use Claim 1 from Sec.~\ref{sec:general_criteria} to compute the locations of all conjugate points in this case. These will be given by
\beq
t_* = \frac{2\pi}{E_m-E_n}\mathbb{Z} 
\eeq
for the simple operators, and 
\beq
t_* = \frac{2\pi(1+\mu)}{E_m-E_n}\mathbb{Z}
\eeq
for the hard operators, where $E_m$ are the eigenvalues of $H$. At any rate, we will not consider this choice of gate set any further in this work, instead focusing on the more standard choice with $k\leq 4$-local operators being treated as simple.

\subsection{Perturbative conjugate points}

To begin with, we will study the effect of the interaction term on the location of conjugate points perturbatively in the coupling constant $\epsilon$.\footnote{Readers who do not wish to follow the detailed perturbative calculations may skip ahead to the summary at the end of this section, and proceed to Sec.~\ref{sec:integ_loops}.} Since the general perturbative analysis is very complicated, we focus specifically on the conjugate points of $H_0$ associated with Jacobi fields that are 1-local operators, such as $A_p^\dagger + A_p$.
Recall that the conjugate points associated with these operators correspond to zero modes of the super-operator $\SY$ which appear at certain times, and in general they are eigen-operators of $\SY$ (defined using $H_0$) with eigenvalue
\begin{align}
    \SY (A_p^\dagger) & = \frac{e^{i\omega_p t}-1}{i \omega_p t} A_p^\dagger , \\
    \SY (A_p) & = \frac{e^{-i\omega_p t}-1}{-i\omega_p t} A_p .
\end{align}
Since these zero modes (and all others in the free theory) are two-fold degenerate, we must employ degenerate perturbation theory.
In fact, since we have expanded our definition of easy operators to include up to 4-local terms, there are additional 3-local operators which have the same eigenvalues under $\SY$.
These are e.g. $A_p^\dagger J_3^{(q)}$ and $A_p J_3^{(q)}$ for $p \neq q$, and these 3-local operators lead to conjugate points at the same times as the 1-local operators above, so the degeneracy is enhanced.

We proceed by perturbing Jacobi fields and conjugate point times in response to the perturbation of the Hamiltonian \eqref{eq:perturbed-hamiltonian},
\beq
\delta V(s) = \delta V^{(0)} (s) + \epsilon\delta V_{(1)} (s) + \cdots ,
\eeq
\beq
t_* = t_*^{(0)}+ \epsilon t^{(1)}_*+ \cdots .
\eeq
We reproduce here the equations governing the Jacobi equation and the super-operator, in which we will make the above replacements and expand:
\beq
\frac{d}{ds}\delta V_L(s) = -it\mu \left[H, \delta V_{NL}(s)\right]_{L} ,
\eeq
\beq
(1+\mu)\frac{d}{ds}\delta V_{NL}(s) = -it\mu \left[H, \delta V_{NL}(s)\right]_{NL},
\eeq
\beq
\mathbf{Y}(\delta V(0)) = U^{-1}\delta U(1) = \int_0^1 ds\,e^{istH} \delta V(s) e^{-istH} = 0.
\eeq

Subsequently, we will proceed order by order to see the effect of the perturbation on the locations of conjugate points.

\subsection*{Zeroth order} 
At $O(\epsilon^0)$, the total Hamiltonian is the free Hamiltonian $H_0$, so we can pick $\delta V_0(0)$ to be any linear combination of the form:
\beq
\delta V^{(0)}(0) = z_i A_i + \bar{z}_i A_i^{\dagger} + \sum_{j\neq i}  J_3^{(j)} (z_{j} A_i + \bar{z}_{j} A_i^{\dagger}),
\eeq
where $z_i$ and $z_{j\neq i}$ are complex numbers. We then obtain a corresponding conjugate point family at 
\begin{equation} 
t_*^{(0)} = \frac{\pi}{\omega_i}\mathbb{Z} .
\end{equation} 
Note that in this case $\delta V_0(s) = \delta V_0(0)$, and lies entirely along the easy directions. Importantly, we assume $\mu > 0$ here: if $\mu =0$, then there is a much larger degeneracy due to operators of the form $A_i J_3^{(j_1)}\cdots J_3^{(j_p)}$ (with $j_1 \neq \dots \neq j_p \neq i$) and the above ansatz needs to be modified.

\subsection*{First order}
At $O(\epsilon^1)$, the Jacobi equation reads
\beq
\frac{d}{ds}\delta V^{(1)}_L(s) = -it\mu \left[H_0, \delta V^{(1)}_{NL}(s)\right]_{L}
\eeq
\beq
(1+\mu)\frac{d}{ds}\delta V^{(1)}_{NL}(s) = -it\mu \left[H_0, \delta V^{(1)}_{NL}(s)\right]_{NL},
\eeq
Since $H_0$ is quadratic, it does not mix between local and non-local directions. Thus, the solutions are
\beq
\delta V^{(1)}_{L}(s) = \delta V^{(1)}_{L}(0),\;\;\delta V^{(1)}_{NL}(s) = e^{-i\frac{\mu ts}{1+\mu}H_0}\delta V^{(1)}_{NL}(0)e^{i\frac{\mu ts}{1+\mu}H_0}.
\eeq
From here, we can compute the perturbative terms in the super-operator,
\beqn
U^{-1}\delta U(1) &=& \int_0^1 ds e^{is(t^{(0)}_*+\epsilon t^{(1)}_*) (H_0+\epsilon H_1)}\left(\delta V^{(0)}(s)+\epsilon\delta V^{(1)}(s)+\cdots\right) e^{-is(t^{(0)}_*+\epsilon t^{(1)}_*) (H_0+\epsilon H_1)}\nonumber\\
&=&\epsilon\int_0^1 ds \,is t_*^{(1)}e^{ist^{(0)}_*H_0} \left[H_0, \delta V^{(0)}(0)\right]e^{-ist^{(0)}_*H_0}\nonumber\\
&+&\int_0^1 ds e^{ist^{(0)}_* H_0}\left[\delta V^{(0)}(0)+\epsilon \left(\delta V_L^{(1)}(0)+e^{-i\frac{\mu t^{(0)}_*s}{1+\mu}H_0}\delta V^{(1)}_{NL}(0)e^{i\frac{\mu t^{(0)}_*s}{1+\mu}H_0}\right) \right]e^{-ist_{\ast}^{(0)} H_0}\nonumber\\
&+&i\epsilon t^{(0)}_* \int_0^1 ds \,s\,e^{ist^{(0)}_* H_0}\left[H_1, \delta V^{(0)}(0)\right] e^{-ist^{(0)}_* H_0} + O(\epsilon^2),
\eeqn
where we have assumed $[H_0, H_1] = 0$ as is the case for the particular integrable deformation \eqref{eq:perturbed-hamiltonian}. In order to extract the change in the location of the conjugate points, we set the $O(\epsilon)$ term to zero so that zero modes of the super-operator with respect to the free Hamiltonian remain zero modes of the perturbed Hamiltonian:
\beqn
& & \int_0^1 ds \,is t_*^{(1)}e^{ist^{(0)}_*H_0} \left[H_0, \delta V^{(0)}(0)\right]e^{-ist^{(0)}_*H_0}\nonumber\\
&+&\int_0^1 ds e^{ist_*^{(0)} H_0}\left[ \left(\delta V_L^{(1)}(0)+e^{-i\frac{\mu t_*^{(0)}s}{1+\mu}H_0}\delta V^{(1)}_{NL}(0)e^{i\frac{\mu t_*^{(0)}s}{1+\mu}H_0}\right) \right]e^{-ist_*^{(0)} H_0}\nonumber\\
&+&i t_*^{(0)} \int_0^1 ds \,s\,e^{ist_*^{(0)} H_0}\left[H_1, \delta V^{(0)}(0)\right] e^{-ist_*^{(0)} H_0} = 0. 
\eeqn
In order to make further progress, we project this equation into the local and non-local directions:
\beqn\label{loc}
&\text{Local}: &\;\; \int_0^1 ds \,is t_*^{(1)}e^{ist^{(0)}_*H_0} \left[H_0, \delta V^{(0)}(0)\right]e^{-ist^{(0)}_*H_0}+\int_0^1 ds e^{ist_*^{(0)} H_0}\left[\delta V_L^{(1)}(0) \right]e^{-ist_*^{(0)} H_0}\nonumber\\
&+&i t_*^{(0)} \int_0^1 ds \,s\,e^{ist_*^{(0)} H_0}\left[H_1, \delta V^{(0)}(0)\right]_L e^{-ist_*^{(0)} H_0} = 0. 
\eeqn
\beqn\label{nonloc}
&\text{Non-local}:&\;\; \int_0^1 ds  e^{i\frac{ t_*^{(0)}s}{1+\mu}H_0}\delta V^{(1)}_{NL}(0)e^{-i\frac{ t_*^{(0)}s}{1+\mu}H_0} \nonumber\\
&+&i t_*^{(0)} \int_0^1 ds  \,s\,e^{ist_*^{(0)} H_0}\left[H_1, \delta V^{(0)}(0)\right]_{NL} e^{-ist_*^{(0)} H_0} = 0. 
\eeqn
Here we have again used the fact that $H_0$ is quadratic, and so does not mix between local and non-local operators. 
Plugging in the Hamiltonian deformation $H_1$ and the ansatz for $\delta V^{(0)}(0)$, we find
\beq
[H_1, \delta V^{(0)}(0)] = \sum_k M_{ik}(-z_iA_i+\bar{z}_i A_i^{\dagger})J_3^{(k)} 
+ \sum_k M_{ik} \sum_{j\neq i}(-z_jA_i+\bar{z}_j A_i^{\dagger})J_3^{(j)} J_3^{(k)}.
\eeq
In taking the local projection, we need to be careful because $(J_3^{(p)})^2 = 1$. So the local projection becomes
\beq
[H_1, \delta V^{(0)}(0)]_L = \sum_k M_{ik}(-z_iA_i+\bar{z}_i A_i^{\dagger})J_3^{(k)}
+\sum_{j\neq i}M_{ij}(-z_j A_i + \bar{z}_j A_i^{\dagger}).
\eeq
Now going back to the local constraint:
\beqn
&\text{Local}: &\;\; \int_0^1 ds \,is t_*^{(1)}e^{ist^{(0)}_*H_0} \left[H_0, \delta V^{(0)}(0)\right]e^{-ist^{(0)}_*H_0}+\int_0^1 ds e^{ist_*^{(0)} H_0}\left[\delta V_L^{(1)}(0) \right]e^{-ist_*^{(0)} H_0}\nonumber\\
&+&i t_*^{(0)} \int_0^1 ds \,s\,e^{ist_*^{(0)} H_0}\left[H_1, \delta V^{(0)}(0)\right]_L e^{-ist_*^{(0)} H_0} = 0,
\eeqn
we take its overlap with $A_i$ and $A_i J_3^{(j)}$ respectively. This kills the $\delta V_L^{(1)}$ term above, and we get
\beq
\frac{t^{(1)}_*}{t^{(0)}_*}z_i +\frac{1}{2\omega_i}\sum_{j\neq i}M_{ij}z_j=0.
\eeq
\beq
\frac{t^{(1)}_*}{t^{(0)}_*}z_j +\frac{1}{2\omega_i}M_{ij}z_i=0.
\eeq
Here taking $i=1$ suffices to show the general structure. The equations then can be written in the matrix form $X\vec{z}=0$, where 
\beq
X = \left(\begin{array} {ccccc} 
\frac{t^{(1)}_*}{t^{(0)}_*} & \frac{1}{2\omega_1}M_{12} & \frac{1}{2\omega_1}M_{13} & \frac{1}{2\omega_1}M_{14} &  \cdots \\ 
\frac{1}{2\omega_1} M_{12}  & \frac{t^{(1)}_*}{t^{(0)}_*} & 0 & 0 & \cdots\\
\frac{1}{2\omega_1} M_{13}  & 0 &  \frac{t^{(1)}_*}{t^{(0)}_*}& 0  & \cdots\\
\vdots &  &  & & \end{array}\right), \;\;\; \vec{z}= (z_1, z_2,\ldots ).
\eeq
There are only nontrivial solutions when the determinant of $X$ vanishes. The determinant can be written
\beq
\mathrm{det}(X) = \tilde{t}^{N/2-2}\left(\tilde{t}^2- \sum_{j\neq 1}\frac{1}{4\omega_1^2}M_{1j}^2\right),
\eeq
where $\tilde{t} = \frac{t^{(1)}_*}{t^{(0)}_*}$. \emph{So we find $N/2-2$ pairs of conjugate points do not move, while the remaining two pairs move to the new locations:}
\beq
t^{(1)}_* = \pm \frac{t^{(0)}_*}{2\omega_1}\sqrt{\sum_{j\neq 1} M_{1j}^2}.
\eeq
The $\vec{z}$'s which correspond to the non-trivial displacement at first order are given by
\beq
\vec{z}_{(\pm)} = \left(\mp \sqrt{\sum_{j\neq 1}M_{1j}^2}, M_{12},M_{13},\cdots, M_{1\frac{N}{2}}\right).
\eeq
The $\vec{z}$'s corresponding to $t^{(1)}_*=0$ all have $z_1=0$ and satisfy $\sum_{j\neq 1}M_{1j} z_j =0$, with solutions of the form
\beqn
\vec{z}_{(3)} &=& (0, -M_{13},M_{12},0, \cdots,0),\nonumber\\
& & \vdots\nonumber\\
 \vec{z}_{(N/2)} &=& (0, -M_{1\frac{N}{2}},0,\cdots,0, M_{12}).
\eeqn
In addition, we need to also work out $\delta V^{(1)}(0)$. From \eqref{loc} it follows that $\delta V_L^{(1)}(0)$ should be of the general form: 
\beq
\delta V^{(1)}_L(0) = w_i A_i + \bar{w}_i A_i^{\dagger} + \sum_{j\neq i}   (w_{j} A_i + \bar{w}_{j} A_i^{\dagger})J_3^{(j)},
\eeq
but the coefficients $w_i$ and $w_j$ are not determined at this order. A priori, we could have had other operators appearing in this expansion, but their coefficients must be zero by \eqref{loc}. On the other hand, we can solve for $\delta V_{NL}^{(1)}$ from equation \eqref{nonloc}. Note that
\beq
\left[H_1, \delta V^{(0)} (0)\right]_{NL} = \sum_{j\neq i}\sum_{k\neq j}M_{ik} \left(-  z_{j} A_i + \bar{z}_{j} A_i^{\dagger}\right)J_3^{(j)} J_3^{(k)},
\eeq
which suggests the following ansatz for $\delta V_{NL}^{(1)}(0)$:
\beq
\delta V^{(1)}_{NL} (0) = \sum_{j\neq i}\sum_{k\neq i} \left( c_{jk} A_i + \bar{c}_{jk} A_i^{\dagger}\right) J_3^{(j)} J_3^{(k)},
\eeq
where the $c_{jk}$'s are some complex coefficients to be determined. Substituting this into equation \eqref{nonloc}, we find $c_{jj} = 0$, while for $j \neq k$,
\beq
c_{jk}= \frac{t^{(0)}_*\phi'(-2\omega_it^{(0)}_*)}{\phi(-\frac{2\omega_i t_*^{(0)}}{1+\mu})}M_{i(k}z_{j)}= \frac{-1}{2\omega_i\phi(-\frac{2\omega_i t_*^{(0)}}{1+\mu})}M_{i(k}z_{j)}, \;\;\; \phi(x) = \frac{e^{ix}-1}{ix}.
\eeq

\subsection*{Second order}
At second order in $\epsilon$, the Jacobi equations are given by
\beq
\frac{d}{ds}\delta V^{(2)}_{L}(s) = -it\mu \left[H_1, \delta V_{NL}^{(1)}(s)\right]_L ,
\eeq
\beq
(1+\mu)\frac{d}{ds}\delta V^{(2)}_{NL}(s) =-it\mu \left[H_0, \delta V_{NL}^{(2)}(s)\right]_{NL} -it\mu \left[H_1, \delta V_{NL}^{(1)}(s)\right]_{NL}.
\eeq
We will only need to know the explicit form of $\delta V_{L}^{(2)}$, which is given by
\beqn
\delta V^{(2)}_{L}(s) &=&\delta V^{(2)}_{L}(0)  -it\mu \int_0^s ds'\,\left[H_1, \delta V_{NL}^{(1)}(s')\right]_L\nonumber\\
&=&\delta V^{(2)}_{L}(0)  -it\mu \int_0^s ds'\,e^{-\frac{i\mu t s'}{1+\mu}H_0}\left[H_1, \delta V_{NL}^{(1)}(0)\right]_Le^{\frac{i\mu t s'}{1+\mu}H_0} .
\eeqn

In order to study the displacement of conjugate points at second order, we now compute $U^{-1}\delta U$:
\beq
U^{-1}\delta U = \int_0^1 ds\,e^{ist(H_0 + \epsilon H_1)}\left(\delta V^{(0)} (s)+ \epsilon \delta V^{(1)}(s) + \epsilon^2 \delta V^{(2)}(s)\right) e^{-ist(H_0 + \epsilon H_1)}.
\eeq
Then, we must substitute $t = t^{(0)}_* + \epsilon t^{(1)}_* + \frac{1}{2}\epsilon^2 t^{(2)}_*$, and extract the second order terms. In doing this, we should be careful to keep in mind that $\delta V$ also depends on $t$. 
\beqn
U^{-1}\delta U\Big|_{\epsilon^2} &=&\int_0^1ds\,\left[\frac{1}{2}ist^{(2)}_*e^{ist^{(0)}_*H_0}[H_0,\delta V^{(0)}(0)]e^{-ist^{(0)}_*H_0}+ist^{(1)}_*e^{ist^{(0)}_*H_0}[H_1,\delta V^{(0)}(0)]e^{-ist^{(0)}_*H_0}\right]\nonumber\\
&+&\int_0^1ds\,\frac{(is)^2}{2}e^{ist^{(0)}_*H_0}\left[(t^{(1)}_*H_0+ t^{(0)}_*H_1), \left[(t^{(1)}_*H_0+ t^{(0)}_*H_1),\delta V^{(0)}(0)\right]\right]e^{-ist^{(0)}_*H_0}\nonumber\\
&+&\int_0^1ds\,is\,e^{ist^{(0)}_*H_0} \left[(t^{(1)}_*H_0+ t^{(0)}_*H_1),\delta V^{(1)}_L(0)\right]e^{-ist^{(0)}_*H_0}\nonumber\\
&+&\int_0^1ds\,is\,e^{i\frac{st^{(0)}_*}{1+\mu}H_0} \left[\left(\frac{t^{(1)}_*}{1+\mu}H_0+ t^{(0)}_*H_1\right),\delta V^{(1)}_{NL}(0)\right]e^{-i\frac{st^{(0)}_*}{1+\mu}H_0}\nonumber\\
&+&\int_0^1ds\,e^{ist^{(0)}_*H_0}\left( \delta V^{(2)}_{L}(0)-i\mu t_*^{(0)}\int_0^s ds'\,e^{-\frac{i\mu t_*^{(0)} s'}{1+\mu}H_0}\left[H_1, \delta V_{NL}^{(1)}(0)\right]_Le^{\frac{i\mu t_*^{(0)} s'}{1+\mu}H_0}\right)e^{-ist^{(0)}_*H_0}\nonumber\\
&+&\int_0^1ds\,e^{ist^{(0)}_*H_0}\;\delta V^{(2)}_{NL}(s)\;e^{-ist^{(0)}_*H_0}.
\eeqn
As in the first order case, the displacement of the conjugate points is determined by taking the overlap of this equation with the local directions, in particular with $A_i$ and $A_i J_3^{(j)}$ (for $j\neq i$). The terms proportional to $\delta V^{(2)}$ drop out of these overlaps, and so we do not need to explicitly compute $\delta V^{(2)}$ at this stage. 

In order to simplify the computation, we will only track the conjugate points which do not already move at first order, i.e., which have $t_*^{(1)}=0$. For these points, we have
\beqn
U^{-1}\delta U\Big|_{\epsilon^2} &=&\int_0^1ds\,\frac{1}{2}ist^{(2)}_*e^{ist^{(0)}_*H_0}[H_0,\delta V^{(0)}(0)]e^{-ist^{(0)}_*H_0}\nonumber \\
&+&\int_0^1ds\,\frac{(is)^2}{2}e^{ist^{(0)}_*H_0}\left[t^{(0)}_*H_1, \left[t^{(0)}_*H_1,\delta V^{(0) }(0)\right]\right]e^{-ist^{(0)}_*H_0}\nonumber\\
&+&\int_0^1ds\,is\,e^{ist^{(0)}_*H_0} \left[t^{(0)}_*H_1,\delta V^{(1)}_L(0)\right]e^{-ist^{(0)}_*H_0}+\int_0^1ds\,is\,e^{i\frac{st^{(0)}_*}{1+\mu}H_0} \left[ t^{(0)}_*H_1,\delta V^{(1)}_{NL}(0)\right]e^{-i\frac{st^{(0)}_*}{1+\mu}H_0}\nonumber\\
&+&\int_0^1ds\,e^{ist^{(0)}_*H_0}\left( \delta V^{(2)}_{L}(0)-i\mu t_*^{(0)}\int_0^s ds'\,e^{-\frac{i\mu t_*^{(0)} s'}{1+\mu}H_0}\left[H_1, \delta V_{NL}^{(1)}(0)\right]_Le^{\frac{i\mu t_*^{(0)} s'}{1+\mu}H_0}\right)e^{-ist^{(0)}_*H_0}\nonumber\\
&+&\int_0^1ds\,e^{ist^{(0)}_*H_0}\;\delta V^{(2)}_{NL}(s)\;e^{-ist^{(0)}_*H_0}.
\eeqn
So, we need to compute
\beqn
\left[H_1,\left[H_1,\delta V^{(0)}(0)\right]\right]&=&\sum_{k,\ell}M_{ik}M_{i\ell}(z_iA_i+\bar{z}_iA_i^{\dagger})J_3^{(k)} J_3^{(\ell)}\nonumber\\
&+&\sum_{k,\ell}\sum_{j\neq i} M_{ik}M_{i\ell}(z_jA_i+\bar{z}_jA_i^{\dagger})J_3^{(j)} J_3^{(k)} J_3^{(\ell)}\nonumber\\
&=&\sum_k M_{ik}^2\left((z_iA_i+\bar{z}_iA_i^{\dagger})+\sum_{j\neq i}(z_jA_i+\bar{z}_jA_i^{\dagger})J_3^{(j)}\right)\nonumber\\
&+&2\sum_{j\neq i}\sum_{k \neq j}M_{ij}M_{ik}(z_jA_i+\bar{z}_jA_i^{\dagger})J_3^{(k)} + \text{non-local terms}. 
\eeqn
In addition, we also need 
\beq
\left[H_1, \delta V^{(1)}_{NL}(0)\right]=\sum_{j\neq i}\sum_{k\neq i}\left(-c_{jk}A_i+\bar{c}_{jk}A_i^{\dagger}\right)\left(M_{ij}J_3^{(k)} +M_{ik} J_3^{(j)} \right) + \text{non-local terms}.
\eeq
Finally, it is easy to check that the displacement projected along $A_i$ vanishes for the conjugate points which do not move at first order, if we take $\sum_jM_{ij}w_j = 0$. For these, the displacement projected along $A_i J_3^{(j)}$ is given by:
\beqn
0 &=&-\omega_i\phi'(-2\omega_i t^{(0)}_*)\,t^{(2)}_*z_j-t^{(0)}_*\phi'(-2\omega_i t^{(0)}_*)M_{ij}w_i  \\
&+&\left[\frac{1}{2}(t^{(0)}_*)^2\phi''(-2\omega_i t^{(0)}_*)+\frac{t_*^{(0)}}{2\omega_i} \frac{\phi'\left(\frac{-2\omega_i t^{(0)}_*}{(1+\mu)}\right)}{\phi\left(\frac{-2\omega_i t^{(0)}_*}{1+\mu}\right)}-\frac{1+\mu}{4\omega_i^2}\right]\left(\sum_kM_{ik}^2-2M_{ij}^2\right)z_j\nonumber
\eeqn
Using $\phi(x) = e^{ix/2} \frac{\sin(x/2)}{x/2}$, one can check that \emph{the imaginary parts inside the square brackets precisely cancel}. Therefore, these equations take the form:
\beq
\left[\frac{1}{2t^{(0)}_*} t^{(2)}_* + \alpha\left(\sum_k M_{ik}^2-2M_{ij}^2\right)\right]z_j = - \frac{1}{2\omega_i}M_{ij}w_i.
\eeq
where $\alpha $ is the real constant equal to the term in brackets above. Assuming that the $M_{ij}$ are all different, then we can solve these equations for the $z_j$:
\beq
z_j = \frac{-M_{ij}w_i}{2\omega_i\left[\frac{1}{2t^{(0)}_*} t^{(2)}_* + \alpha\left(\sum_kM_{ik}^2-2M_{ij}^2\right)\right]}.
\eeq
Finally, we need to impose the constraint $\sum_{j\neq i} M_{ij}z_j = 0$ assuming $w_i\neq 0$, which translates to
\beq
\sum_{j\neq i}\frac{M^2_{ij}}{\left[\frac{1}{2t^{(0)}_*} t^{(2)}_* + \alpha\left(\sum_kM_{ik}^2-2M_{ij}^2\right)\right]}=0.
\eeq
Defining $\tau = (\frac{t^{(2)}_*}{4\alpha t^{(0)}_*} +  \frac{1}{2}\sum_k M_{ik}^2)$, we can then write this equation as
\beq
f(\tau) = \sum_{j\neq i}\frac{M_{ij}^2}{(\tau -  M_{ij}^2)} = 0. \label{eq:secondorderzeros}
\eeq
\emph{Therefore, the second order displacements of the $N/2-2$ conjugate points are generically nonzero and can be obtained from the zeros of the complex function $f(\tau)$}. These zeroes are always real, as can be checked by explicitly substituting $\tau = x+iy$ into \eqref{eq:secondorderzeros}. Note that if any two $M_{ij}$ coincide, then we lose a zero, and that zero corresponds to looking for a solution with $w_i = 0$. We will not consider these additional special cases here. 

\subsection{Integrable geodesic loops} \label{sec:integ_loops}
\begin{figure}
    \centering
    \includegraphics[scale=.7]{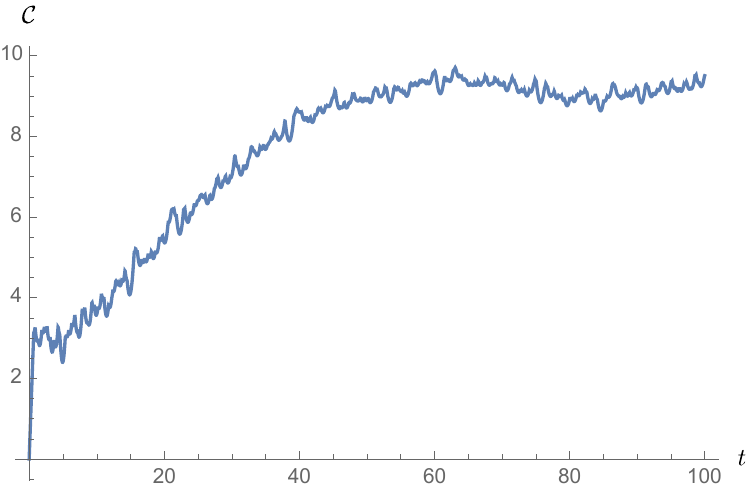}
    \caption{Complexity bound \eqref{eq:integrable-bound} for an $N=20$ instance of the integrable Hamiltonian \eqref{eq:perturbed-hamiltonian} where $\epsilon M_{ij}/4$ is drawn from the $q=4$ SYK distribution with $\mathcal{J}=1$. The initial sharp linear growth is due to the combined initial linear growth of both terms in \eqref{eq:integrable-bound}, and the small fluctuations are due to the frequent geodesic loops in $e^{-iH_0t}$.  The larger fluctuations, and the coarse-grained shape of the function itself, are controlled by the geodesic loops in $e^{-i\epsilon H_1 t}$ that we have included in defining $d_{ij}(t)$.  The plateau is clearly $O(N)$; its height without the integrable perturbation would be less than the height of the initial sharp rise, which is at most $O(\sqrt{N})$.}
    \label{fig:integrable}
\end{figure}

While we have focused on perturbation theory for the conjugate point locations, it is also possible to find certain geodesic loops in this model analytically.
Recall that in the free theory, we found many geodesic loops, each of which came from a conjugate point associated with an easy operator.
In the deformed theory, we do not have an exact analytic handle on conjugate point locations, but the same sorts of loops can occur because the two terms in \eqref{eq:perturbed-hamiltonian} commute.
This means that the time evolution operator splits as
\begin{equation}
    e^{-iHt} = e^{-iH_0t} e^{-i\epsilon H_1 t} .
\end{equation}
The loops we found for the operator $e^{-iH_0t}$ in Sec.~\ref{sec:free} also apply here.
Furthermore, since the product $J_3^{(i)} J_3^{(j)}$ also has eigenvalues $\pm 1$, there are additional loops associated with the $\pi/2$ half-periodicity of the coefficients $\epsilon M_{ij} t/4$.
The individual coefficients have this half-periodicity because $M_{ij}=M_{ji}$, so there is an extra factor of 2 in the total coefficient of $J_3^{(i)}J_3^{(j)}$.
To take these into account, we follow \eqref{eq:global-velocity} and define coefficients $d_{ij}(t)$ with bounded range
\begin{equation}
    d_{ij}(t) \equiv \epsilon M_{ij} t/4 \mod \pi ,
\end{equation}
where we define the $\pi$ modulus to take values in $(-\pi/2,\pi/2]$.
Then, using the global velocities \eqref{eq:global-velocity} for $H_0$, a bounded-length path to $e^{-iHt}$ is
\begin{equation}
    V = \sum_p c_p(t) J_3^{(p)} + \sum_{i,j} d_{ij}(t) J_3^{(i)} J_3^{(j)} .
\end{equation}
The complexity is upper-bounded by the length of this path:
\begin{equation}
    \mathcal{C}(t) \leq \sqrt{\sum_p c_p(t)^2 + \sum_{i<j} (2d_{ij}(t))^2} .
\label{eq:integrable-bound}
\end{equation}
An instance of this function is shown in Fig.~\ref{fig:integrable}.
Qualitatively, we may conclude that the complexity reaches a plateau here as well, but with greater height than the free case.
The free complexity is upper-bounded by $O(\sqrt{N})$ since there are $N/2$ coefficients $c_p(t)$ with maximum value $\pi$, but the integrable perturbation allows for $N(N-1)/2$ more terms in the $d_{ij}(t)$, which leads to an upper bound of $O(N)$.
A strict upper bound in this case is in fact
\begin{equation}
    \mathcal{C}(t) \: < \: \sqrt{\frac{N}{2}\pi^2 + \frac{N}{4}\left(\frac{N}{2}-1\right) \pi^2 } \: \sim \: \frac{\pi N}{2\sqrt{2}}  \qquad \text{as} \qquad N \to \infty ,
\end{equation}
where we simply took the upper limits $c_p = \pi$ and $d_{ij} = \pi/2$.
We have been careful to say upper-bounded in this discussion because we have not exactly located the conjugate points in this model, and there may be some which are closer to the identity than any of the geodesic loops we considered here.
Of course, as in the free case, we also do not have analytic control over every geodesic loop.
This and other integrable interacting models could furnish interesting examples of geodesic loops in complexity geometry which are not signaled by a conjugate point in a straightforward way.

The above construction is clearly generalizable to the case where the Hamiltonian perturbation is 
\begin{equation}
H_{c-1} = \frac{1}{2(c!)} \sum_{i_1, \dots , i_c} \lambda_{i_1 \dots i_c} J_3^{(i_1)} \dots J_3^{(i_c)} ,
\label{eq:hamiltonian-generalization}
\end{equation}
where we require $k \geq 2c$ so that $H_{c-1}$ is an easy operator in the complexity metric, and $\lambda$ is symmetric in all indices and vanishes when $i_j = i_\ell$ for any $j \neq \ell$ (so it is strictly $2c$-local).
Following the same procedure as before, the complexity of $e^{-i(H_0+\epsilon H_{c-1})t}$ is upper-bounded by
\begin{equation}
    \mathcal{C}(t) \: < \: \sqrt{ \frac{N}{2}\pi^2 + \binom{N/2}{c} \pi^2 } \: \sim \: \frac{\pi N^{c/2}}{2^{c/2} \sqrt{c!}}  \qquad \text{as} \qquad N \to \infty .
\label{eq:integrable-generalization}
\end{equation}
Thus, we have a family of integrable models with complexity of time evolution that is upper-bounded by a polynomial $O($poly$(N))$ that depends on the order of the interaction $c$.

\subsection*{Summary}

We calculated the first and second order shifts in location of the conjugate points associated with 1-local operators in the free theory under the integrable deformation \eqref{eq:perturbed-hamiltonian}.
At first order, all but two of the $N/2$ degenerate conjugate points remain fixed, and the two which move do so by a distance which depends on the perturbation couplings $M_{ij}$ but not on the cost factor $\mu$.
At second order, the $N/2-2$ points which did not move at first order begin to move, and are shifted by a distance which is sensitive to $\mu$.
As this shift can become large for $\mu \gg 1$, the perturbation theory may break down.
We also found geodesic loops which were analogous to certain loops found in the free theory, but for which we did not find associated conjugate points.
These represent potential examples of geodesic loops which are not signaled by conjugate points.

The perturbative results suggest that the complexity grows linearly for a long time as the conjugate points we studied move to later times as $\mu$ is increased; however, the existence of these geodesic loops shows otherwise. There are also other conjugate points associated to operators of higher locality which may be independent of $\mu$, the existence of which will be suggested by our numerical results in Sec.~\ref{sec:syk-numerical}.
Due to the geodesic loops, an upper bound of $O(N)$ can be placed on the complexity of $e^{-iHt}$ for the integrable $H$ in \eqref{eq:perturbed-hamiltonian}.
More generally, if the perturbation term commutes with the free Hamiltonian, our results will carry over, with a possibly greater upper bound on complexity.
An example of this more general result is the bound \eqref{eq:integrable-generalization}, which is $O($poly$(N))$ and specifically $O(N^{c/2})$, on the complexity of $e^{-i(H_0+\epsilon H_{c-1})t}$ with the $2c$-local integrable perturbation $H_{c-1}$ given in \eqref{eq:hamiltonian-generalization}.


\section{Impact parameter and local conjugate points in chaotic theories}\label{sec:local-chaotic}
We now turn to the interesting case of chaotic Hamiltonians. In \cite{Balasubramanian:2019wgd} it was argued that in a chaotic model, the super-operator $\mathbf{Y}_{\mu}$ takes a simple form in the energy eigen-operator basis:
\beq
\mathbf{Y}_{\mu}(|m\rangle\langle n|) = \phi\left(\frac{(E_m-E_n)t}{1+\mu}\right)|m\rangle\langle n| + \cdots,
\eeq
where under appropriate assumptions the Frobenius norm of the correction term $\cdots$ was shown to be exponentially small. Thus, the diagonal entries of the super-operator $\mathbf{Y}_{\mu}$ in the $|m\rangle\langle n|$ basis are $O(1)$ for $t \ll \frac{(1+\mu)}{E_m-E_n}$. Since the off-diagonal entries are small, we thus expect that the eigenvalues will also be bounded away from zero, and given that $\mu$ scales exponentially with $S$, we conclude that conjugate points do not occur at sub-exponential times. However, there is a caveat: while the off-diagonal elements of $\mathbf{Y}_{\mu}$ are suppressed, at the same time there are an exponentially large number of such off-diagonal entries. So although ``almost all'' of the eigenvalues of $\mathbf{Y}_{\mu}$ will be $O(1)$ for sub-exponential times, we cannot be certain that a small number of zero modes cannot occur. In fact, \emph{local conjugate points} (see Claim 2 in Sec.~\ref{sec:general_criteria}) are prime suspects at sub-exponential times, as their locations do not depend on the cost factor $\mu$. In Claim 2, we re-formulated such conjugate points in terms of zero modes of the positive semi-definite matrix $M_{\alpha\beta}$, which is the matrix of infinite temperature thermal two-point functions between time-averaged simple operators:
\beq
M_{\alpha\beta}(t) = \int_0^1ds\int_0^1ds'\,\mathrm{Tr}\,\left(T_{\alpha}e^{i(s-s')tH}T_{\beta}e^{-i(s-s')tH}\right).
\eeq
We will now argue that in chaotic systems, zero modes of $M_{\alpha\beta}$ -- and hence local conjugate points -- can only potentially arise at exponential times. Our strategy will be to show that the minimum eigenvalue $\lambda_{\text{min}}(t)$ of $M_{\alpha\beta}$ is exponentially large for $t< e^S$, and becomes small only thereafter. We will refer to $\lambda_{\text{min}}$ as the \emph{impact parameter} (see Sec.~\ref{sec:general_criteria}).  

By expanding in the energy eigenbasis and evaluating the integrals, the matrix $M_{\alpha\beta}$ can be written as:
\beq \label{M}
M_{\alpha\beta}(t) = \sum_{m,n} \langle m|T_{\alpha}|n\rangle \langle n|T_{\beta} |m\rangle g(t(E_m-E_n)),\;\; g(x) = \left(\frac{\sin(x/2)}{x/2}\right)^2,
\eeq
where $|m\rangle,\;|n\rangle$ are energy eigenstates with energies $E_m,\;E_n$. With the above formula for $M_{\alpha\beta}(t)$, we can now estimate the time $t_*$ at which we expect a zero mode by using intuition from random matrix theory and the Eigenstate Thermalization Hypothesis (ETH).
In this context, we assume ETH is satisfied for the $k$-local operators that we consider easy in the complexity metric.\footnote{This may not always be a safe assumption, as the precise degree of locality and the particular operators for which ETH is expected to hold are not always clear. But for our purposes, we can take this as the definition of a chaotic system.} First, notice that for times less than the inverse maximum energy difference $1/(E_{\text{max}}-E_{\text{min}})$, we have $g(t(E_m-E_n)) \approx 1$.
If we were to make this replacement in $M_{\alpha\beta}$, we would find
\begin{equation}
    M_{\alpha\beta} \approx \sum_{m,n} \bra{m} T_\alpha \ket{n} \bra{n} T_\beta \ket{m} = e^S \delta_{\alpha\beta},\qquad\quad t\ll \frac{1}{(E_{\text{max}}-E_{\text{min}})}.
\end{equation}
This diagonal result appears because the projectors $\ket{n}\bra{n}$ sum to the identity operator, and then we are left with the trace $\tr (T_\alpha T_\beta)$. The generators $\{T_{\alpha}\}$ are orthogonal, and we have chosen the norm to be 
\begin{equation} 
e^S = \text{dim}\,\mathcal{H} = 2^{N/2} , 
\end{equation}
since in the SYK model the $T_\alpha$ are traceless, Hermitian products of Majorana fermions which square to the identity operator.
This matrix clearly has no zero modes.
Going back to the exact expression in equation \eqref{M}, the sum over $m$ and $n$ is modified by the presence of the function $g$, but the diagonal (i.e., $m=n$) terms in the sum are unaffected by $g$:
\begin{equation}
    M_{\alpha\beta} = \sum_n \bra{n} T_\alpha \ket{n} \bra{n} T_\beta \ket{n} + \sum_{\substack{m,n\\m \neq n}} \bra{m} T_\alpha \ket{n} \bra{n} T_\beta \ket{m} g(t(E_m-E_n)) .
\end{equation}
We can replace these diagonal terms with $2^{N/2}\delta_{\alpha\beta}$ by rearranging the above equation as:
\begin{equation}\label{Mab1}
    M_{\alpha\beta} = 2^{N/2}\delta_{\alpha\beta} + \sum_{\substack{m,n\\m \neq n}} \bra{m} T_\alpha \ket{n} \bra{n} T_\beta \ket{m} \left( g(t(E_m-E_n)) - 1 \right) .
\end{equation}
Now our basic strategy will be to argue that for $t< e^S$, (i) the diagonal entries of $M_{\alpha\beta}$ are $O(e^S)$, while (ii) the off-diagonal entries of $M_{\alpha\beta}$ are $O(1)$. Since the matrix is polynomial in size (as the $\alpha,\beta$ indices run over simple operators), this then implies that the eigenvalues will all be $O(e^S)$. On the other hand, when $t\gg e^S$, the diagonal entries can become $O(1)$, and thus the impact parameter, i.e., the minimum eigenvalue of $M_{\alpha\beta}$, can become small, and zero modes could potentially arise.

\noindent\textbf{Diagonal elements}: In general, the sum over $m,n$ in the second term above for $\alpha \neq \beta$ involves a sum of many $g$ functions along with incommensurate complex numbers $\bra{m} T_\alpha \ket{n}$ and $\bra{n} T_\beta \ket{m}$.
However, the diagonal of $M_{\alpha\beta}$ obeys
\begin{equation}
    M_{\alpha\alpha} = 2^{N/2} + \sum_{\substack{m,n\\m \neq n}} |\bra{m} T_\alpha \ket{n}|^2  \left( g(t(E_m-E_n)) - 1 \right) ,
\end{equation}
and so the sum of $g$ functions appears here with all strictly non-negative coefficients.
At this point, we invoke ETH, which in this context states that (for $m \neq n$)
\begin{equation}
    |\bra{m} T_\alpha \ket{n}|^2 \sim 2^{-N/2} |r_{\alpha,mn}|^2 ,
\label{eq:diagonal-ETH-estimate}
\end{equation}
where $r_{\alpha,mn}$ is a random matrix with entries of $O(1)$ magnitude whose squared elements $|r_{\alpha,mn}|^2$ are all roughly equal and $O(1)$.
What this means is that the sum
\begin{equation}
    \sum_{\substack{m,n\\m \neq n}} \left( g(t(E_m-E_n)) - 1 \right) ,
\end{equation}
must become $O(2^N)$ before the diagonal entries $M_{\alpha\alpha}$ can vanish.
This will only occur when almost all of the $g$ functions are close to zero, which can only happen when $t \gg e^S$.

More quantitatively, let us try to approximate the timescale at which this occurs.
Notice that we can expand the sum above to include $m = n$, since these terms have $g(0) = 1$.
Then, we must determine when the sum $\sum_{m,n} g(t(E_m-E_n))$ becomes small, i.e., $O(1)$.
At large $N$, we can approximate the double sum as a double integral over two copies of the spectral density $\rho(E)$:
\begin{equation}
    M_{\alpha\alpha} \approx 2^{-N/2} \int dE_m \int dE_n\; \rho(E_m) \rho(E_n) \left( \frac{\sin (t(E_m-E_n)/2)}{t(E_m-E_n)/2} \right)^2 .
\end{equation}
Strictly speaking, we should use the SYK spectral density for the function $\rho(E)$.\footnote{If we consider the SYK ensemble, we should use the density-density correlator $\avg{\rho(E_m)\rho(E_n)}$.  In random matrix theory, there are additional contributions to this object which include a contact term and the sine kernel.  However, for our purposes it is sufficient to approximate this quantity as the product of two densities at large $N$. }
However, we expect that our conclusions about conjugate points should apply to other chaotic systems as well.
The key feature of the spectral density for $q=4$ SYK is that there is an exponential number of states, $e^S = 2^{N/2}$, within a polynomial size window $-N \leq E \leq N$.
The precise size of the window is not important for the argument, only that it is polynomial in $N$.
Similarly, the relevant information about the exact height of the spectral density is that it is exponential in $N$.
These properties also hold in e.g. a microcanonical ensemble of black hole microstates, where the window is actually $O(1)$ in size with $O(e^N)$ states.
Since we are only interested in these very coarse features of the spectral density, we may approximate $\rho(E)$ above by a constant distribution on $-N \leq E \leq N$:
\begin{equation}
    \rho(E) \approx \frac{2^{N/2}}{2N} .
\end{equation}
Of course, for sufficiently abnormal models, this density will not be a good approximation, but for chaotic SYK or a black hole microstate ensemble it is sufficient.
The result of the integrals is
\begin{equation}
    M_{\alpha\alpha} \approx \frac{2^{N/2}}{2N^2 t^2} \left( 2N t\; \text{Si}(2N t) + \cos (2Nt) - 1 \right) ,
\label{eq:diag-integral-approx}
\end{equation}
where Si$(x) \equiv \int_0^x dz \sin(z)/z$.
The above estimate is generically an underestimate because the ansatz of a constant spectral density gives additional support to pairs of eigenvalues $E_m$ and $E_n$ which have separation larger than $O(e^{-S})$.
The most important feature of \eqref{eq:diag-integral-approx} is that the function Si$(x) \approx \pi$ for $x \gg 1$, so $M_{\alpha\alpha}$ is bounded away from zero by roughly $\frac{2^{N/2}}{N t}$ at large $N$.
This quantity is exponential in $N$ for any $t \sim \text{poly}(N)$, and becomes $O(1)$ only when
\begin{equation}
    t \sim \frac{2^{N/2}}{N} = O(e^S) .
\end{equation}
Therefore, the diagonal $M_{\alpha\alpha}$ is $O(e^S)$ until an exponential time $t \sim e^S$, at which point it becomes $O(1)$.
It is clear that this conclusion holds when the spectral width is any $O(\text{poly}(N))$, instead of exactly $2N$, as long as the spectral height is $O(e^S)$.

Notice that we did not assume anything about the structure of the matrix $r_{\alpha,mn}$ in making this argument.
We only needed the entries to be distributed so that the squares $|r_{\alpha,mn}|^2$ took roughly the same $O(1)$ value for any $m$ and $n$, but the entries themselves did not need to be independent random variables.
This is less than the usual statement about the ETH ensemble, where the variance of any given $r_{\alpha,mn}$ is not only fixed, but the $r_{\alpha,mn}$ themselves are all independent random variables.

\noindent\textbf{Off-diagonal elements}: Having understood the rough order of magnitude for the diagonal entries of $M_{\alpha\beta}$, we now turn to the off-diagonal pieces.
For these, we have again a sum of $g$ functions from equation \eqref{Mab1}, but now the coefficients in the sum can be negative.
We can get some rough intuition for the order of this quantity by again invoking ETH on the local operator matrix elements.
\begin{equation}
    \bra{m} T_\alpha \ket{n} \bra{n} T_\beta \ket{m} \approx e^{-S/2} r_{\alpha,mn} r_{\beta,nm} ,
\end{equation}
If the $r_{\alpha,mn}$ were drawn from independent Gaussian distributions with mean zero and $O(1)$ variance, it would be straightforward to compute the typical (expectation) value of the above expression.
We would simply find zero for the typical value since $r_\alpha$ and $r_\beta$ are independent matrices and have mean zero.
To ensure that the fluctuations of this quantity are not excessively large, we could also estimate the variance, which involves a calculation of $\avg{r_{\alpha,mn} r_{\beta,nm} r_{\alpha,m'n'} r_{\beta,n'm'}}$ in the aforementioned ensemble.
Since $r_{\alpha,mn}$ and $r_{\beta,mn}$ are independent, this four-point function factorizes into a product of two-point functions.
ETH would then tell us that these two-point functions $\avg{r_{\alpha,mn}r_{\alpha,pq}}$ are proportional to $\delta_{mp}\delta_{nq}$ since the entries of $r_{\alpha,mn}$ are supposed to be \textit{independent} Gaussian random variables. Going back to \eqref{Mab1}, we thus conclude that the off-diagonal entries of $M_{\alpha\beta}$ are always $O(1)$.
However, it cannot be precisely correct to employ ETH in this manner for \textit{any} choice of eigenstates $\ket{m}$ and $\ket{n}$ because the operators $T_\alpha$ have a known spectrum (all eigenvalues are $\pm 1$) which greatly differs from the spectrum of a random matrix with independent Gaussian random entries at large $N$.
So, we will need a different sort of ensemble to get a consistent estimate of the mean and variance of $M_{\alpha\beta}$ for $\alpha \neq \beta$.

One candidate which is consistent with all constraints on the matrices $T_\alpha$ is the Haar ensemble of unitary matrices employed in the following manner.
We pick some fixed basis $\ket{i}_P$ (for instance, the Pauli basis) in which the form of $T_\alpha$ is known by construction to be relatively sparse or simple.
Then, we assume that the eigenvectors $\ket{n}_E$ of the chaotic Hamiltonian $H$ can be roughly thought of as a Haar random unitary rotation of this basis via\footnote{A similar ensemble was used to model a microcanonical window of states in quantum gravity in \cite{Pollack:2020gfa}, although in that context the ensemble had a physical interpretation as the dual of a gravitational path integral in the spirit of \cite{Saad:2019lba}.  Here, by contrast, we use the Haar ensemble to extract information about the typical value and variance of certain matrix elements with the understanding that we are really studying the expected behavior of a quantum chaotic system with fixed Hamiltonian, such as a single instance of the SYK model.}
\begin{equation}
    \ket{n}_E = \sum_i U_{ni} \ket{i}_P .
\end{equation}
The off-diagonal terms in $M_{\alpha\beta}$ are given by
\begin{equation}
    M_{\alpha\beta} = \sum_{\substack{m,n\\m\neq n}} \bra{m} T_\alpha \ket{n}_E \bra{n} T_\beta \ket{m}_E (g(t(E_m-E_n))-1), \qquad \alpha \neq \beta .
\end{equation}
We would like to get an estimate for the mean value of the quantity
\begin{equation}
    \bra{m} T_\alpha \ket{n}_E \bra{n} T_\beta \ket{m}_E = \sum_{i,j,k,\ell} \bra{i} U^\dagger_{im} T_\alpha U_{nj} \ket{j}_P \bra{k} U^\dagger_{kn} T_\beta U_{m\ell} \ket{\ell}_P .
\end{equation}
To compute the typical value, we integrate this expression over the Haar ensemble for $U$ by making use of
\begin{equation}
\int dU\; U_{nj} U_{m\ell} U^\dagger_{im} U^\dagger_{kn} = \frac{1}{e^{2S}-1} \left( \delta_{nm} \delta_{ij} \delta_{k\ell} + \delta_{jk} \delta_{i\ell} \right) - \frac{1}{e^S(e^{2S}-1)} \left( \delta_{ij} \delta_{k\ell} + \delta_{nm} \delta_{jk} \delta_{i\ell} \right) .
\end{equation}
The asymptotic forms \cite{Weingarten:1977ya} and exact expressions \cite{collins2003moments} for such integrals are well known.
With an eye toward the sums over $m$ and $n$ in $M_{\alpha\beta}$, we notice that any term with $\delta_{nm}$ must vanish in the full expression since $m \neq n$.
We obtain (writing $\avg{\cdot}_H$ for the Haar expectation)
\begin{equation}
\avg{M_{\alpha\beta}}_H = \sum_{\substack{m,n\\m\neq n}} \left( \frac{1}{e^{2S}-1} \tr (T_\alpha T_\beta) - \frac{1}{e^S(e^{2S}-1)} \tr T_\alpha \tr T_\beta  \right) (g(t(E_m-E_n))-1), \quad \alpha \neq \beta .
\label{eq:off-diagonal-haar-average}
\end{equation}
The Haar integration has given us the typical value of $M_{\alpha\beta}$ in terms of traces of the operators $T_\alpha$ and $T_\beta$.
By construction, we have $\tr T_\alpha = 0$ and $\tr (T_\alpha T_\beta) = 0$ for $\alpha \neq \beta$, so the typical value in a chaotic Hamiltonian ensemble defined this way is
\begin{equation}
\avg{M_{\alpha\beta}}_H = 0, \qquad \alpha \neq \beta .
\end{equation}
Incidentally, this calculation also shows that the diagonal terms $\alpha = \beta$ have a Haar average of order $e^S$ until exponential times.
By setting $\alpha = \beta$ in the large parentheses of \eqref{eq:off-diagonal-haar-average}, we conclude
\begin{equation}
    \avg{|\bra{m} T_\alpha \ket{n}_E|^2}_H \sim e^{-S} , 
\end{equation}
which is consistent with our estimate that relied on the ETH ensemble \eqref{eq:diagonal-ETH-estimate}, so the conclusions from that discussion concerning $M_{\alpha\alpha}$ match the results of the Haar ensemble.

If the off-diagonal elements of $M_{\alpha\beta}$ are all approximately zero for a given chaotic Hamiltonian, the only way a zero mode can arise is by the vanishing of a diagonal element, which we have shown does not occur until $t \sim e^S$.
To be complete, we should also study the variance $\avg{M^2_{\alpha\beta}}_H$ and ensure it is not too large.
A small $O(1)$ variance will ensure that fluctuations in the off-diagonal elements are small relative to the diagonal magnitude.

The variance can be estimated by computing  $\avg{M_{\alpha\beta}^2}_H$, where
\begin{equation}
    M^2_{\alpha\beta} = \sum_{\substack{m,n\\m\neq n}} \sum_{\substack{m',n'\\m'\neq n'}} \bra{m} T_\alpha \ket{n}_E \;
    \bra{n} T_\beta \ket{m}_E \;
    \bra{m'} T_\alpha \ket{n'}_E \;
    \bra{n'} T_\beta \ket{m'}_E \; (g(\Delta_{mn}t)-1)(g(\Delta_{m'n'}t)-1) ,
\label{eq:M2ab-full}
\end{equation}
where $\alpha \neq \beta$ and $\Delta_{mn} \equiv E_m-E_n$.
The basic quantity which we would like to integrate against the Haar measure is
\begin{equation}
    \bra{m} T_\alpha \ket{n}_E \;
    \bra{n} T_\beta \ket{m}_E \;
    \bra{m'} T_\alpha \ket{n'}_E \;
    \bra{n'} T_\beta \ket{m'}_E ,
\label{eq:stddev-Mab-relevant}
\end{equation}
which can be converted to the Pauli basis $\ket{i}_P$ by
\begin{equation}
    \sum_{i,j,k,\ell,p,q,r,s} \bra{p} U^\dagger_{pm} T_\alpha U_{ni} \ket{i}_P \;
    \bra{q} U^\dagger_{qn} T_\beta U_{mj} \ket{j}_P \;
    \bra{r} U^\dagger_{rm'} T_\alpha U_{n'k} \ket{k}_P \;
    \bra{s} U^\dagger_{sn'} T_\beta U_{m'\ell} \ket{\ell}_P .
\end{equation}
The relevant Haar integral is
\begin{equation}
    \int dU \; U_{ni} U_{mj} U_{n'k} U_{m'\ell} U^\dagger_{pm} U^\dagger_{qn} U^\dagger_{rm'} U^\dagger_{sn'} .
\end{equation}
On general grounds, the overall result for the Haar expectation of \eqref{eq:stddev-Mab-relevant} will be written in terms of traces or products of traces of the operators $T_\alpha$, $T_\beta$, $T_\alpha$, and $T_\beta$.
There are three such combinations which can be nonzero:
\begin{equation} 
\tr T_\alpha^2 \tr T_\beta^2 = 2^N , \quad \tr (T_\alpha^2 T_\beta^2) = 2^{N/2}, \quad \tr (T_\alpha T_\beta T_\alpha T_\beta) = - 2^{N/2} .
\label{eq:haar-combinations}
\end{equation}
We deal with each of these three case by case.

The first combination in \eqref{eq:haar-combinations}, the double trace factor yielding $2^N$, is produced by certain products of delta functions from the Haar integral
\begin{equation}
    \delta_{ir} \delta_{js} \delta_{kp}  \delta_{\ell q} \left( O(e^{-4S}) \delta_{nm'}\delta_{mn'} + O(e^{-5S}) (\delta_{mn'} + \delta_{nm'} ) + O(e^{-6S})  \right) ,
\label{eq:double-trace-haar}
\end{equation}
where we have only kept terms which contribute at leading order to $\avg{M^2_{\alpha\beta}}_H$.
Notice that all of these terms actually contribute at $O(1)$. 
For example, the $O(e^{-4S})$ term comes with two delta functions that cancel two of the four sums over $n,m,n',m'$ in \eqref{eq:M2ab-full}, which leads to a sum over $e^{2S}$ terms of order $O(e^{-2S})$ since the $e^{-4S}$ suppression can absorb the double trace factor $e^{2S}$.
Similarly, the $O(e^{-5S})$ terms come with one delta function to cancel one of the $n,m,n',m'$ sums in \eqref{eq:M2ab-full}, and again contributes at $O(1)$.
Finally, the $O(e^{-6S})$ term comes without any delta function constraints, but is suppressed enough to absorb all four sums in \eqref{eq:M2ab-full} (each over $e^{S}$ elements) where all elements have magnitude of order the trace contribution $e^{2S}$, and ends up at $O(1)$.

The second combination in \eqref{eq:haar-combinations} can be formed with a variety of delta function combinations appearing from the Haar integral.
Fortunately, because the trace factor is only $e^S$ in this case, the only possible dangerous term which may contribute beyond $O(1)$ must take the form
\begin{equation}
    O(e^{-4S}) \delta_{iq} \delta_{jp} \delta_{ks} \delta_{\ell r} ,
\label{eq:dangerous}
\end{equation}
which is the unique term that appears at $O(e^{-4S})$ Haar suppression without any additional delta functions which would cancel the sums in \eqref{eq:M2ab-full}.
However, a term of this form does not lead to $\tr (T_\alpha^2 T_\beta^2)$, but instead leads to $\tr (T_\alpha T_\beta) \tr (T_\alpha T_\beta)$, which vanishes.
So, the leading contribution of the Haar integral to the coefficient of the second term in \eqref{eq:haar-combinations} is $O(e^{-5S})$, and this is enough to absorb the $e^S$ trace factor appearing in all $e^{4S}$ terms of the four sums in \eqref{eq:M2ab-full}, yielding an at most $O(1)$ contribution to $\avg{M^2_{\alpha\beta}}_H$.

The third and final combination in \eqref{eq:haar-combinations} must also contribute at most an $O(1)$ result to $\avg{M^2_{\alpha\beta}}_H$, as the same argument concerning the unique form of the possible dangerous term holds in this case as well, since the trace factor is again only $O(e^S)$.

\begin{figure}[!h]
\centering
\begin{tabular}{c c}
\includegraphics[height=5.5cm]{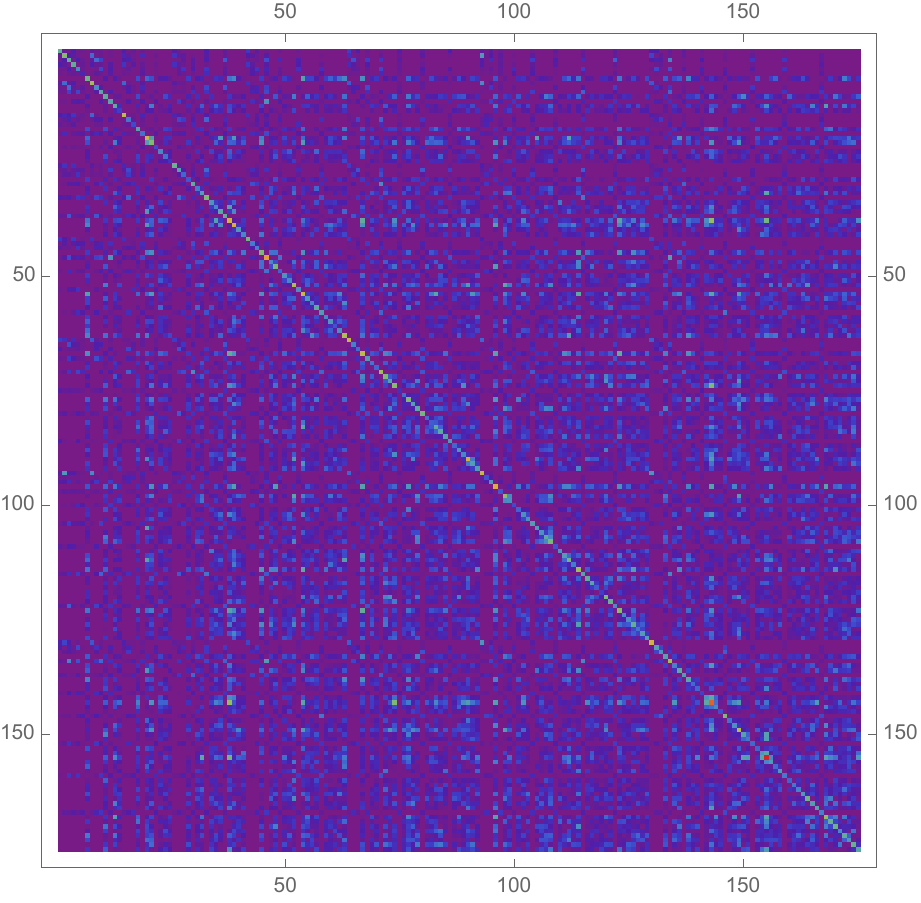}     &  \includegraphics[height=5.5cm]{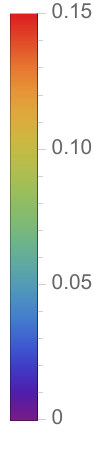}
\end{tabular}
\caption{An array plot of the matrix $e^{-S}| M_{\alpha\beta}|$ for $N=10,\;q=3,\;k=3,\;\mathcal{J}=1$ SYK and time $t=50$. We note that most of the off-diagonal elements are smaller than $e^{-S}$, while many diagonal matrix elements are also $O(e^{-S})$ at such late times.}
\label{fig:Mab}
\end{figure}

Putting it all together, we have shown that the Haar average of $M^2_{\alpha\beta}$, assuming the eigenvectors of our chaotic Hamiltonian are related to some simple basis by a Haar-random unitary transformation, is
\begin{equation}
    \avg{M^2_{\alpha\beta}}_H \sim O(1) , \qquad \alpha \neq \beta .
\end{equation}
An analogous argument shows that the diagonal variance is similar, 
\begin{equation} 
\avg{M^2_{\alpha\alpha}}_H - \avg{M_{\alpha\alpha}}_H^2 \sim O(1) , 
\end{equation}
where the dangerous term \eqref{eq:dangerous} actually makes an important $O(e^{2S})$ contribution to $\avg{M^2_{\alpha\alpha}}_H$ in order to cancel the leading term from $\avg{M_{\alpha\alpha}}_H^2$.  
The next-to-leading term from the trace factor generated by \eqref{eq:dangerous} actually contributes $O(1)$ to the diagonal variance rather than $O(e^S)$, since we will have $O(e^{-5S})$ Haar suppression along with at least one delta function to cancel one sum in the analog of \eqref{eq:M2ab-full} for $\alpha = \beta$. 
This is because \eqref{eq:dangerous} is the unique permutation leading to $\tr (T_\alpha T_\beta) \tr (T_\alpha T_\beta)$, which is the first of only two new non-vanishing trace factors when $\alpha = \beta$, without any such delta functions.
The permutation leading to the second new pairing $\tr (T_\alpha T_\beta) \tr (T_\beta T_\alpha)$, where the first $T_\alpha$ is multiplied instead with the second $T_\beta$ in \eqref{eq:M2ab-full}, comes with two delta functions $\delta_{nn'}\delta_{mm'}$ at $O(e^{-4S})$ and one delta function at $O(e^{-5S})$, just as in \eqref{eq:double-trace-haar}, so there are only $O(1)$ contributions due to this trace factor.
The other non-vanishing trace factors for $\alpha = \beta$ are all captured by the three cases in \eqref{eq:haar-combinations}, and the suppression arguments we made for those when $\alpha \neq \beta$ also apply when $\alpha = \beta$.
Thus, all diagonal variance contributions are $O(1)$ as claimed.
Note that the diagonal variance may have some mild dependence on $t$; here we have only argued that it is $O(N^0)$.
The numerical structure of $M_{\alpha\beta}$ at large $t$ is shown in Fig.~\ref{fig:Mab}.

\begin{figure}[!h]
\centering
\begin{tabular}{c c}
\includegraphics[height=4cm]{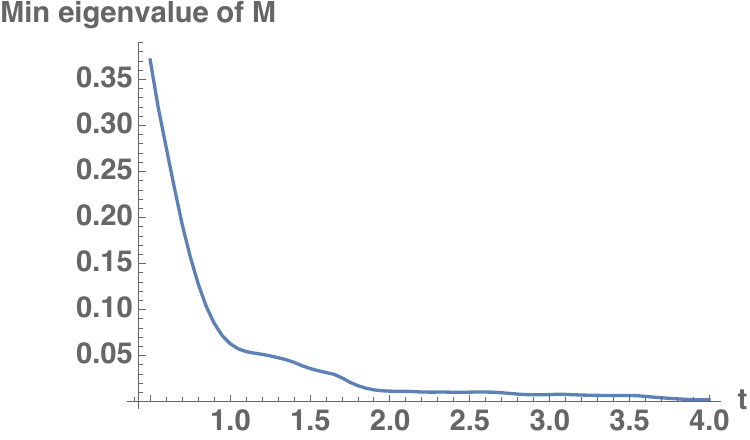}     &  \includegraphics[height=4cm]{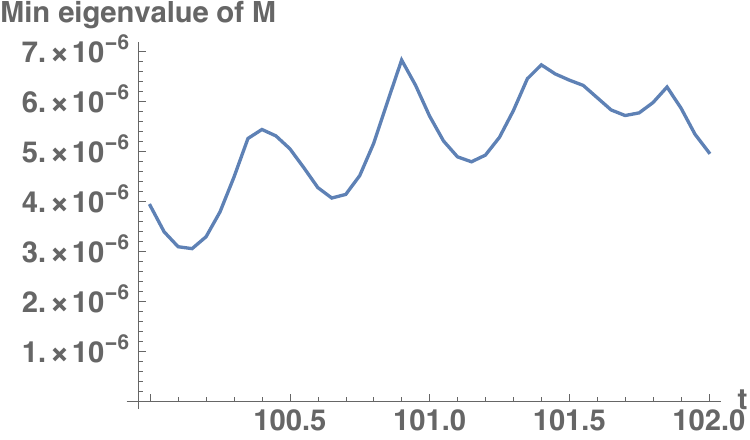}
\end{tabular}
\caption{(Left) The minimum eigenvalue of $e^{-S}M_{\alpha\beta}$, i.e., the impact parameter, for an SYK Hamiltonian with $N=10,\;q=3,\;k=3,\;\mathcal{J}=1$ at small times. (Right) The minimum eigenvalue past exponential time becomes very small. }
\label{fig:MabEig}
\end{figure}

As we discussed above, this estimate is sufficient to argue that there should be no zero mode of $M_{\alpha\beta}(t)$ before times $t \sim e^S$, as the diagonal of the matrix is overwhelmingly large compared to the off-diagonal elements, and in addition, the matrix size scales as $\text{poly}(S)$. Thus, for $t< e^S$, the impact parameter will be $O(e^S)$. On the other hand for $t > e^S$, the diagonal elements of $M_{\alpha\beta}(t)$ are $O(1)$ and in particular of the same order as the off-diagonal elements; we thus expect the impact parameter to become small (see Fig.~\ref{fig:MabEig}).  

\subsection*{Summary}
We have argued in this section that the minimum eigenvalue of $M_{\alpha\beta}(t)$ must be $O(e^S)$ for $t< e^S$, but becomes small for $t> e^S$. This implies that local conjugate points in chaotic theories can potentially occur only beyond exponential time. Even if exact zero modes of $M_{\alpha\beta}$ do not occur, we expect the minimum eigenvalue $\lambda_{\text{min}}$ of $M_{\alpha\beta}$ to become very small after $t\sim e^S$ (see Fig.~\ref{fig:MabEig}). Physically, this means that for $t \gg e^S$, it is possible to find an infinitesimally nearby curve with a local initial velocity $V(0) = Ht + \epsilon \delta V(0)$ which satisfies the geodesic equation up to $O(\epsilon^2)$, such that the end point is very close to the target unitary $e^{-itH}$:
\beq
|| U(1) - e^{-itH}||^2_F = \epsilon^2\, e^{-S}\lambda_{\text{min}}||\delta V(0)||^2_F+ O(\epsilon^3) \ll 1\;\;\;\;\cdots(t \gg e^S).
\eeq
Thus, it becomes possible to approximate time evolution by an infinitesimally nearby geodesic with a small error after exponential time. If the impact parameter $\lambda_{\text{min}}$ is exactly zero for some $t_* > e^S$, then we have a conjugate point at that location, and then we can find a shorter geodesic path to $e^{-itH}$ exactly, with no error. 

Our arguments in this section were based on ETH and random matrix theory. A fairly similar story was told for the complete super-operator under the Eigenstate Complexity Hypothesis assumption in \cite{Balasubramanian:2019wgd} (also explored in Sec.~\ref{sec:ECH}), but there are two key differences here.
First, since there are only polynomially many entries in $M_{\alpha\beta}$, we need not worry about the off-diagonal entries ``backreacting" on the diagonal to force an unexpected zero mode at early times.
Instead, a zero mode can only occur when a significant portion of the diagonal becomes suppressed at the same order as the off-diagonal entries, and this does not occur until $t_* \sim e^S$.
Second, the zero modes which arise in this way are actually independent of $\mu$, and thus are fixed obstructions to the complexity growth of even maximally chaotic systems.
We speculate further on the implications in Sec.~\ref{sec:disc}.


\section{Numerical analysis of conjugate points}\label{sec:syk-numerical}

We now present numerical calculations of conjugate point locations for free, interacting integrable, and chaotic SYK models. The general method that we use is to explicitly construct a matrix representation of the super-operator and compute its smallest eigenvalue (i.e., the eigenvalue with the smallest absolute value) using the Arnoldi (iterative) algorithm \cite{Arnoldi}. This gives us a concrete, albeit numerical, method to study obstructions to complexity growth. We will limit ourselves to systems up to $N=8$ (four qubits) for computational feasibility, but in principle this method is not limited to small $N$. 

We first reproduce the general expression for the super-operator from previous sections for reference,
\begin{equation}
\begin{split}
    \textbf{Y}_{\mu}(\delta V(0)) = & \int_0^1 ds e^{iHts} \biggl[ \delta V_L(0) - i \mu t \sum_{\dot{\alpha}} \frac{\exp \left( \frac{-i \mu t \lambda_{\dot{\alpha}} s}{1+\mu}  \right)-1}{\frac{-i \mu t \lambda_{\dot{\alpha}} }{1+\mu}} \delta \tilde{V}^{\dot{\alpha}}(0) [H,\tilde{T}_{\dot{\alpha}}]_L \\
    & + \sum_{\dot{\alpha}} \exp \left( \frac{-i \mu t \lambda_{\dot{\alpha}} s}{1+\mu}  \right) \delta \tilde{V}^{\dot{\alpha}}(0) \tilde{T}_{\dot{\alpha}} \biggr] e^{-iHts} .
\end{split}
\label{eq:super-operator2}
\end{equation}
There are two key observations that allow us to represent the super-operator more efficiently. The first is that the integrand simplifies immensely if we construct the super-operator in the basis of $\{T_{\alpha}, \tilde{T}_{\dot{\alpha}}\}$ where the $T_{\alpha}$ are a basis for the local subspace and the $\tilde{T}_{\dot{\alpha}}$ are the basis for the nonlocal subspace that diagonalizes $[H,\:\cdot\,]_{NL}$ with eigenvalues $\lambda_{\dot{\alpha}}$. In that case, the sums disappear and we need only consider the first term or the last two terms depending on the column of the matrix representation that we wish to construct. The second observation is that the integral can be done analytically provided that we express the basis $\{T_{\alpha}, \tilde{T}_{\dot{\alpha}}\}$ in the energy eigenbasis $|m\rangle\langle n|$. Note that this is not the same thing as writing the super-operator in the energy eigenbasis, which would not respect the split into local and nonlocal terms. In the energy basis these operators have coefficients:
\begin{align}
    T_{\alpha} = \sum_{m,n} c_{mn} |m\rangle\langle n|, \qquad c_{mn} = \langle m | T_{\alpha} |n\rangle.
\end{align}
The $c_{mn}$, as well as the energy eigenstates $|m\rangle$, their corresponding eigenvalues, and the diagonalization of $[H,\:\cdot\,]_{NL}$, can all be precomputed before constructing the super-operator. 

Now we construct the matrix representation of the super-operator $\textbf{Y}_{ij} = \text{tr}(\mathcal{O}_i^{\dagger}\textbf{Y}_{\mu}(\mathcal{O}_j) )$ as follows, letting $\mathcal{O}_i = \sum_{m,n} c_{mn}^{(i)} |m\rangle\langle n|$ index $\{T_{\alpha}, \tilde{T}_{\dot{\alpha}}\}$, $\phi(x) = (\exp(ix) - 1)/(ix)$, $\Delta_{mn} = E_m - E_n$ be the difference of energy eigenvalues, and $M_{\dot{\alpha}}  = \frac{\mu \lambda_{\dot{\alpha}}}{1+\mu}$,
\begin{align}
    \textbf{Y}_{ij} &= \text{tr} \begin{cases} 
    \int_0^1 ds \mathcal{O}_i^{\dagger} e^{iHts}   \mathcal{O}_j e^{-iHts}  ,\qquad  &\mathcal{O}_j \in \{T_{\alpha}\} \\
    \int_0^1 ds \mathcal{O}_i^{\dagger} e^{iHts} \biggl[- i \mu t \phi(-\frac{\mu t \lambda_{\dot{\alpha}} s}{1+\mu}) [H,\mathcal{O}_j]_L  +  \exp \left( \frac{-i \mu t \lambda_{\dot{\alpha}} s}{1+\mu}  \right)  \mathcal{O}_j \biggr] e^{-iHts}  ,\qquad  &\mathcal{O}_j \in \{ \tilde{T}_{\dot{\alpha}}\}
    \end{cases} \\
    &=  \begin{cases} 
    \sum_{m,n} c_{nm}^{(i)} c_{mn}^{(j)} \phi( t\Delta_{mn}), &\mathcal{O}_j \in \{T_{\alpha}\} \\
     \sum_{m,n} c_{nm}^{(i)} c_{mn}^{(j)} \left[\phi(t(\Delta_{mn} - M_{\dot{\alpha}} )) + \mu\left(\Delta_{mn} - \lambda_{\dot{\alpha}}\right)\left(\frac{\phi(t(\Delta_{mn} - M_{\dot{\alpha}} )) - \phi(t\Delta_{mn})}{M_{\dot{\alpha}} } \right) \right],  &\mathcal{O}_j \in \{ \tilde{T}_{\dot{\alpha}}\}.
    \end{cases}
\end{align}
In writing the second equality we have used the fact that $[H,\tilde{T}_{\dot{\alpha}}]_L = [H,\tilde{T}_{\dot{\alpha}}] - \lambda_{\dot{\alpha}} \tilde{T}_{\dot{\alpha}}$, evaluated the Hamiltonian on the energy eigenstates, and used cyclicity of the trace to remove some sums. By precomputing the energy spectrum and the coefficients $c_{mn}$, $\textbf{Y}_{ij}$ can therefore be more efficiently constructed without costly numerical integration or matrix products. In terms of the dimension $d=2^{N/2}$ of the Hilbert space, computing the coefficients $c_{mn}$ naively requires $O(d^2)$ operations for each of the $d^2 = 2^N$ operators for a total of $O(d^4)$ complexity to compute the $c_{mn}$. Similarly, at each fixed $i,j$ one must sum up $O(d^2)$ function evaluations for each of the $d^4$ matrix elements $\textbf{Y}_{ij}$, so evaluating the matrix representation of the super-operator is $O(d^6)$. In practical terms this means that at $N=8$ fermions constructing the super-operator requires a reasonable $O(10^7)$ operations \emph{at each time point}, with a number of time points that is typically on the order of $d^2$, while at $N=10$ one already requires $O(10^9)$ operations at each time point. For this reason, we restrict to $N\leq 8$ in the numerical results below. The plots of minimum eigenvalue versus time below with $N=6$ take one to two minutes per curve to generate on a four-core desktop while at $N=8$ they take one to two hours per curve.

\subsection{Free SYK}

We first recall the key point of Sec.~\ref{sec:free} regarding the location of conjugate points in the free model. When $H$ is quadratic, the adjoint action of the Hamiltonian does not mix the local and nonlocal operator subspaces. Consequently, we can diagonalize $[H,\:\cdot\,]$ in the local and nonlocal subspaces independently with corresponding eigenvalues $\lambda_{\alpha}$ and $\lambda_{\dot{\alpha}}$. The super-operator then reduces to the simpler expression
\begin{align}
\SY &= \int_0^1 ds\,\left[e^{istH}\delta V_L(0) e^{-istH}+ e^{\frac{istH}{1+\mu}}\delta V_{NL}(0) e^{-\frac{istH}{1+\mu}}\right] \\
&= \sum_{\alpha} \frac{1}{it\lambda_{\alpha}} (e^{it \lambda_{\alpha}} -1)\delta V_{\alpha}(0) +  \sum_{\dot{\alpha}} \frac{1+\mu}{it\lambda_{\dot{\alpha}}} (e^{\frac{it \lambda_{\dot{\alpha}}}{1+\mu}} - 1)\delta V_{\dot{\alpha}} (0)  .
\end{align}
In the basis of operators $\{T_{\alpha}, \tilde{T}_{\dot{\alpha}}\}$ the super-operator is therefore diagonal with eigenvalues given by the coefficients above. Consequently, we have two\footnote{The degeneracy can be larger than two when multiple $\lambda_{\alpha}$ or $\lambda_{\dot{\alpha}}$ coincide, as discussed in Sec.~\ref{sec:perturbation}.} zero modes whenever
\begin{equation}
    t = \frac{2\pi}{\lambda_{\alpha}} \mathbb{Z} \qquad \text{or} \qquad t =  \frac{2\pi (1+\mu)}{\lambda_{\dot{\alpha}}} \mathbb{Z} . \label{eq:freeconjpts}
\end{equation}
So in the free model, every conjugate point is associated with an individual easy or hard operator.
The easy conjugate points never move with $\mu$, while the hard ones are occur at exponential times when $\mu$ is $O(e^{\alpha S})$.

\begin{figure}[h!]
    \centering
    \includegraphics[width=.75\textwidth]{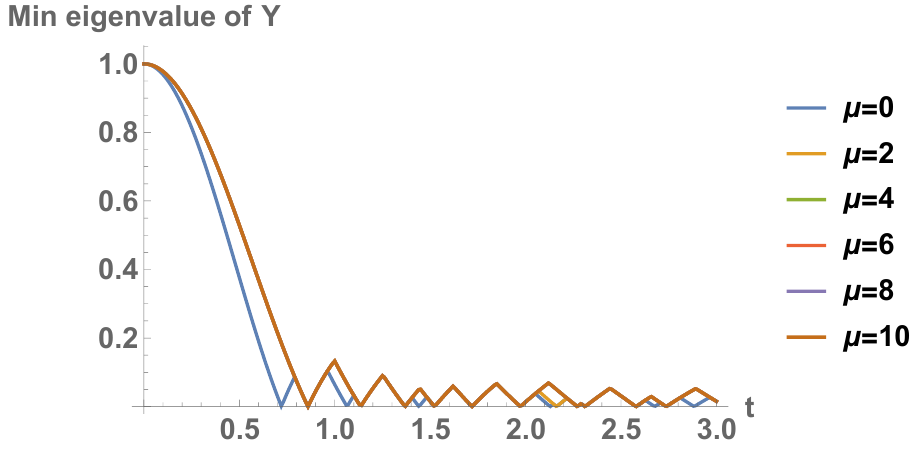}
    \caption{The smallest eigenvalue of the super-operator $\textbf{Y}_\mu$ for the $N=6$ free fermion model for various $\mu$ with the various $\omega$ equal to 2.06238, 1.59206, and 0.703448. The plots for $\mu > 2$ overlap for nearly all of the displayed values of $t$.}
    \label{fig:freeSYKconjpts}
\end{figure}

As computed numerically, the minimum eigenvalue of the $N=6$ free theory is shown for several values of $\mu$ in Fig.~\ref{fig:freeSYKconjpts}, where we take $k=2$ to match the locality of the Hamiltonian. Conjugate points occur where the minimal eigenvalue of $\SY$ touches the $x$ axis. The conjugate point locations as displayed in Fig.~\ref{fig:freeSYKconjpts} exactly match the analytic expression in \eqref{eq:freeconjpts}. One can clearly see the shifting of several conjugate points as $\mu$ is increased; for example, the first conjugate point near $t = 0.7$ at $\mu = 0$ gets shifted to three times its value, near $t = 2.1$, when $\mu = 2$, and subsequently moves off the right end of the figure for larger values of $\mu$. Most of the curves overlap for $\mu > 2$ since once $\mu$ is sufficiently large only the local conjugate points, whose locations are not functions of $\mu$, remain in a bounded-time region.

\subsection{Integrable and chaotic models}

\begin{figure}[h!]
    \centering
    \includegraphics[width=.44\textwidth]{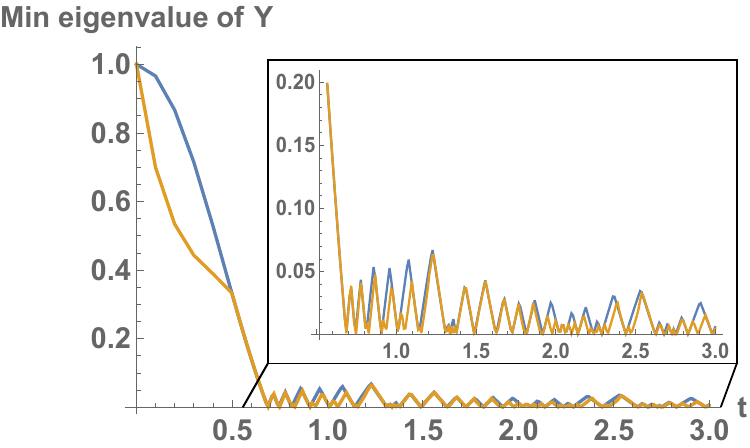}
    \includegraphics[width=.54\textwidth]{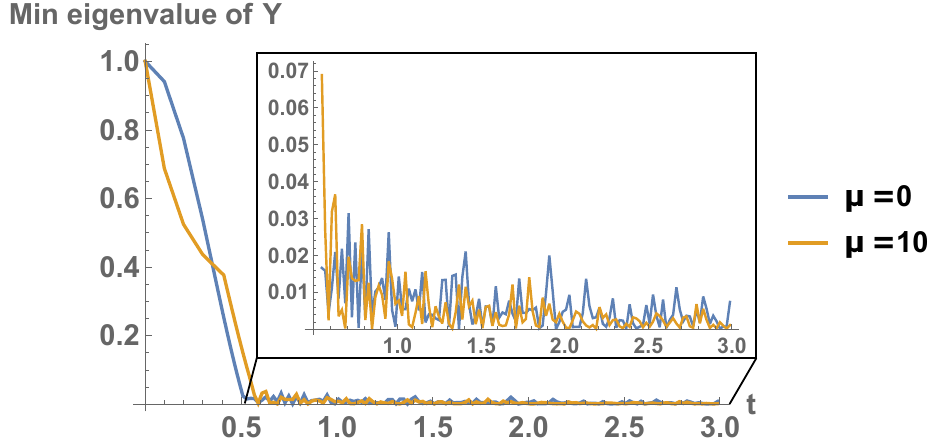}
    \caption{The smallest eigenvalue of the super-operator $\textbf{Y}_\mu$ for the SYK model with the chaotic 3-body deformation $\delta H_3$, with $N=6$ (left) and $N=8$ (right) for $\mu = 0$ (blue) and $\mu = 10$ (gold). The values for $t \gtrsim 0.5$ are displayed inset for visual clarity.}
    \label{fig:chaos3conjpts}
\end{figure}

We now compute the locations of conjugate points where we deform the free Hamiltonian as in Sec.~\ref{sec:perturbation} by $H = H_0 + \epsilon\, \delta H$. Since the numerics are not restricted to taking $\epsilon$ to be perturbative we will consider $\epsilon = 1.0$ in all plots in this section to illustrate large effects of each type of interaction.\footnote{One reason to keep the $H_0$ term is that the algebra generated by $\text{ad}_H$ acting on a basis of operators has too trivial a structure at small $N$ when $H$ is chosen to be only a single $q$-local term, which can cause unwanted numerical coincidences.} We will consider three different choices of $\delta H$ with the same base Hamiltonian $H_0$ considered across all cases at fixed $N$,
\begin{align}
    \delta H_1 &= \sum_{ij} M_{ij} J_3^{(i)} J_3^{(j)} \quad &&\text{integrable 4-body}, \label{eq: integrable 4-body} \\
    \delta H_2 &= \sum_{1\leq i <j <k < \ell \leq N} J_{ijk\ell} \psi^i \psi^j \psi^k \psi^{\ell} \quad &&\text{chaotic 4-body}, \\
    \delta H_3 &= i\sum_{1\leq i < j < k \leq N} J_{ijk} \psi^i \psi^j \psi^k \quad &&\text{chaotic 3-body}. \label{eq: chaotic 3-body} 
\end{align}
Notably, $\delta H_2$ and $\delta H_3$ are effectively the maximally chaotic SYK$_4$ and SYK$_3$ theories \cite{Maldacena:2016hyu} while $\delta H_1$ is the integrable interaction from Sec.~\ref{sec:perturbation}.\footnote{We drop the numerical prefactor of $1/4$ on $\delta H_1$ that was written in Sec.~\ref{sec:perturbation}, but draw $M_{ij}$ rather than $\epsilon M_{ij} / 4$ from the $q=4$ SYK distribution with $\mathcal{J}=1$ in this section, so the numerical results in each section are on equal footing.} The results for $\delta H_1$, $\delta H_2$, and $\delta H_3$ are displayed below in Figs.~\ref{fig:integconjpts}, \ref{fig:chaos4conjpts}, and \ref{fig:chaos3conjpts}, respectively. We compare the plots of minimum eigenvalue versus time for $\mu = 0$ and $\mu = 10$, where we have chosen $\mu = 10$ as a numerically feasible upper bound to illustrate the large $\mu$ regime.\footnote{Note that for $N=6$, the large $\mu$ scale is $e^S \sim 2^{N/2} = 8$, while for $N=8$ it is only $2^4 = 16$.} Clearly, in all three cases, a number of conjugate points remain fixed, corresponding to \emph{local} eigen-operators of the super-operator (which were the main subject of Sec.~\ref{sec:local-chaotic} for large $N$ chaotic theories), while those corresponding to eigen-operators with some nonlocal component will tend to shoot off with some $\mu$-dependent speed. One can see this by examining the zeros of the below plots, which correspond to zero modes of the super-operator and thus conjugate points. Zeros that remain fixed in location between $\mu = 0$ and $\mu = 10$ are independent of $\mu$ and therefore correspond to local eigen-operators, while those that move do not.

\begin{figure}[h!]
    \centering
    \includegraphics[width=.43\textwidth]{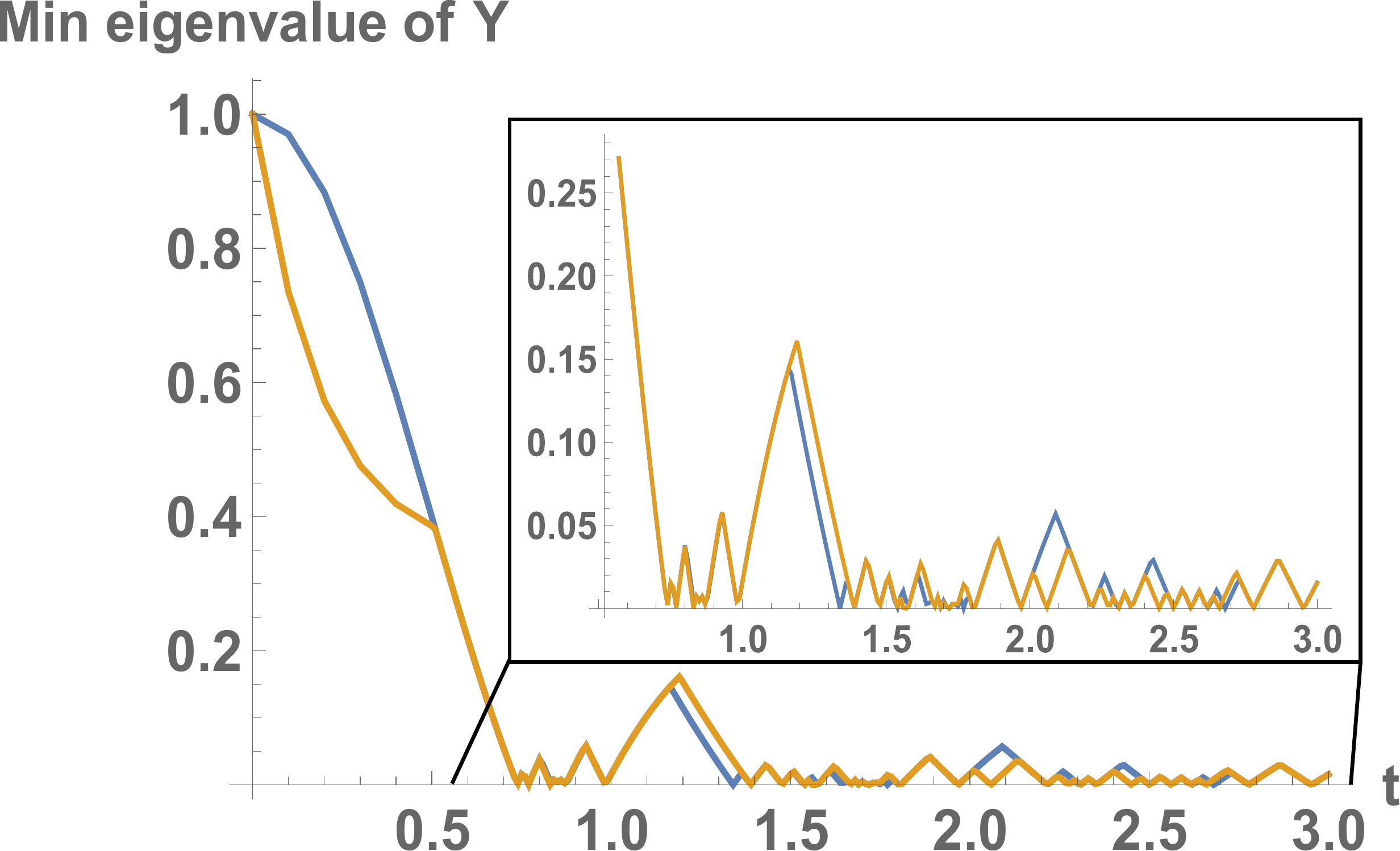}
    \includegraphics[width=.55\textwidth]{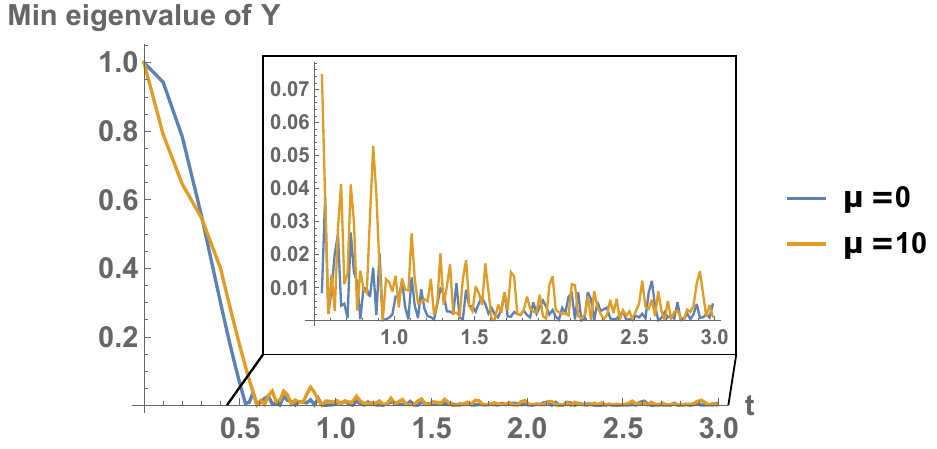}
    \caption{The smallest eigenvalue of the super-operator $\textbf{Y}_\mu$ for the SYK model with the integrable 4-body deformation $\delta H_1$, with $N=6$ (left) and $N=8$ (right) for $\mu = 0$ (blue) and $\mu = 10$ (gold). The values for $t \gtrsim 0.5$ are displayed inset for visual clarity.}
    \label{fig:integconjpts}
\end{figure}

\begin{figure}[h!]
    \centering
    \includegraphics[width=.41\textwidth]{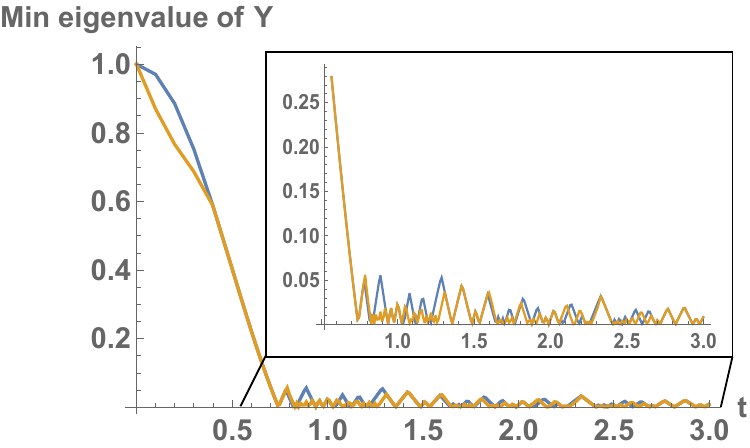}
    \includegraphics[width=.57\textwidth]{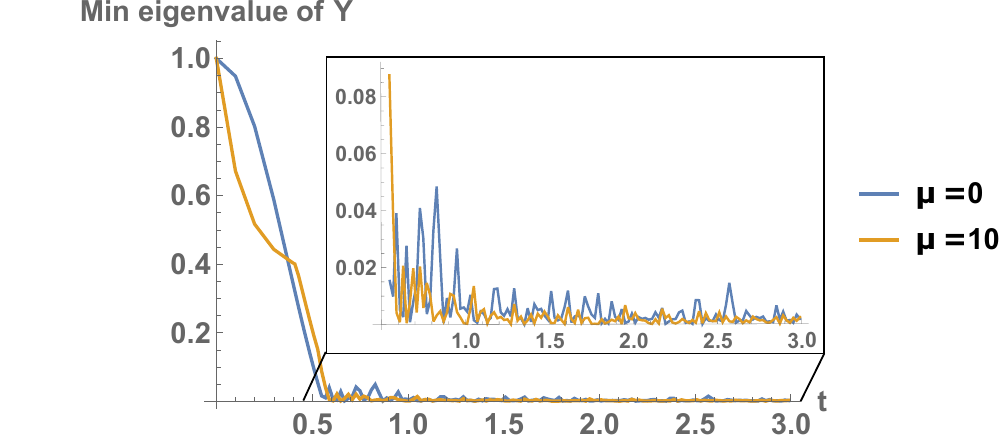}
    \caption{The smallest eigenvalue of the super-operator $\textbf{Y}_\mu$ for the SYK model with the chaotic 4-body deformation $\delta H_2$, with $N=6$ (left) and $N=8$ (right) for $\mu = 0$ (blue) and $\mu = 10$ (gold). The values for $t \gtrsim 0.5$ are displayed inset for visual clarity.}
    \label{fig:chaos4conjpts}
\end{figure}

\begin{figure}[h!]
    \centering
    \includegraphics[width=.75\textwidth]{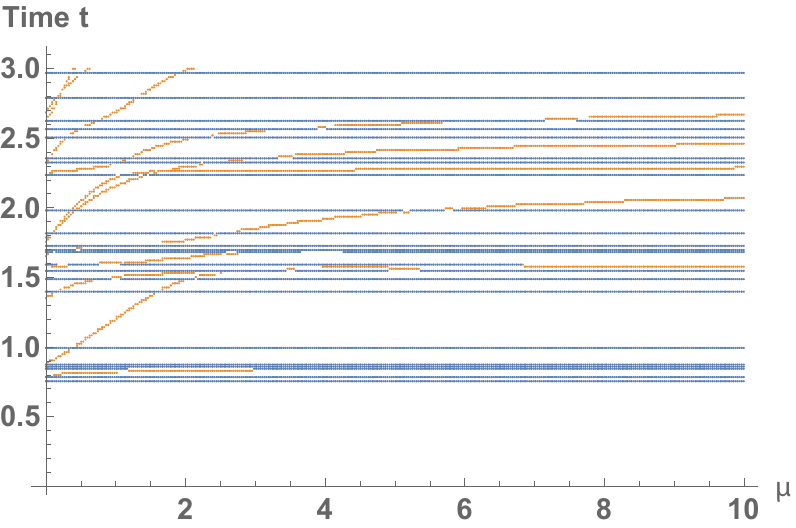}
    \caption{The times of all conjugate points with the integrable 4-body deformation $\delta H_1$ with $N=6$ fermions. This plot and Figs.~\ref{fig:AllconjptvsmuChaos4} and \ref{fig:AllconjptvsmuChaos3} are made by identifying all of the zeros of $\textbf{Y}_{\mu}$ at each fixed $\mu$ point and sampling a lattice of time values, rather than by tracking the motion of individual conjugate points. Consequently, it may be difficult to distinguish conjugate points that lie within a lattice spacing of each other for certain values of $\mu$. Some of the more easily distinguishable nonlocal conjugate points are highlighted in orange by hand, while the blue lines correspond to $\mu$-independent local conjugate points.}
    \label{fig:AllconjptvsmuInteg}
\end{figure}

\begin{figure}[h!]
    \centering
    \includegraphics[width=.75\textwidth]{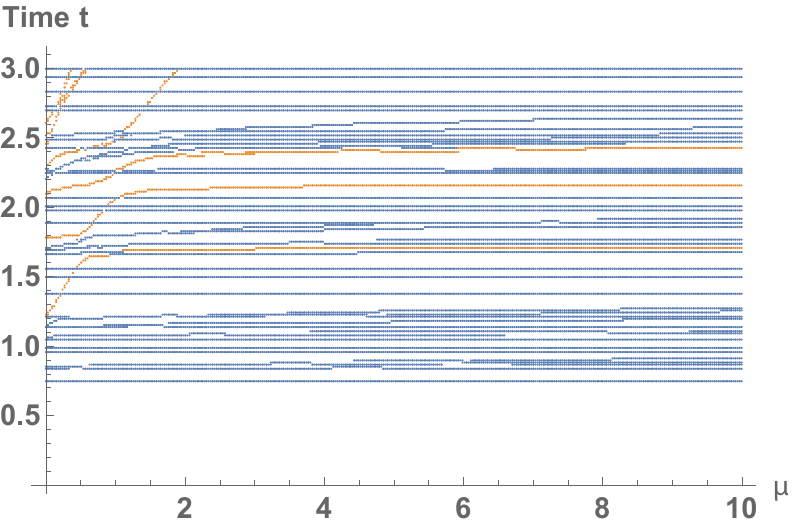}
    \caption{The times of all conjugate points with the chaotic 4-body deformation $\delta H_2$ with $N=6$ fermions. For ease of visibility, some of the easily distinguishable nonlocal conjugate points are highlighted in orange, while the blue lines correspond to $\mu$-independent local conjugate points.}
    \label{fig:AllconjptvsmuChaos4}
\end{figure}

\begin{figure}[h!]
    \centering
    \includegraphics[width=.75\textwidth]{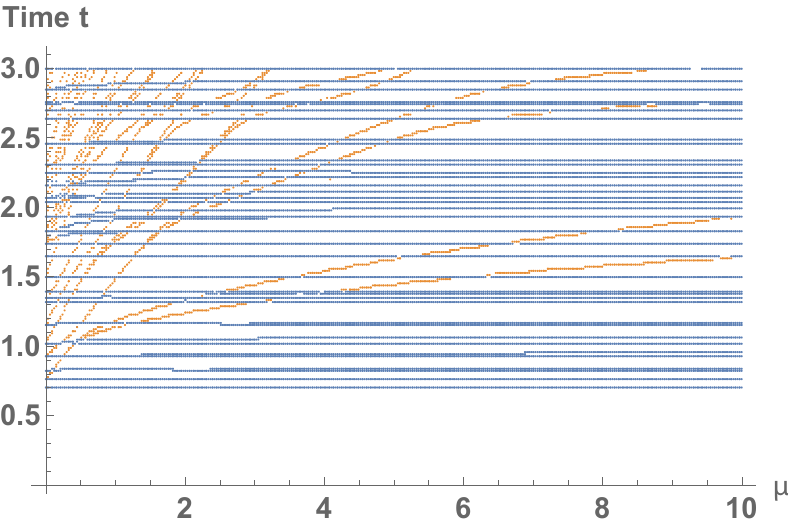}
    \caption{The times of all conjugate points with the chaotic 3-body deformation $\delta H_3$ with $N=6$ fermions. For ease of visibility, some of the easily distinguishable nonlocal conjugate points are highlighted in orange, while the blue lines correspond to $\mu$-independent local conjugate points.}
    \label{fig:AllconjptvsmuChaos3}
\end{figure}

In Figs.~\ref{fig:AllconjptvsmuInteg}, \ref{fig:AllconjptvsmuChaos4}, and \ref{fig:AllconjptvsmuChaos3}, we illustrate how the times of various conjugate points behave as $\mu$ is increased for all three types of interactions. The results clearly illustrate the following conclusions:

\begin{enumerate}[(a)]
    \item At very early times for $N=6,8$ (3, 4 qubits) the linear geodesic encounters a conjugate point and $e^{-iHt}$ fails to be the globally minimizing geodesic. This is remarkable because the complexity geometry corresponds to the manifolds $SU(8)$ and $SU(16)$ which are already very complicated fiber bundles over spheres. What this illustrates is that even in such highly nontrivial geometries conjugate points can become relevant for complexity growth almost immediately. This fact is also interesting since it implies that interacting qubit Hamiltonians can be fast-forwarded at early times.
    
    \item There exist a family of conjugate points whose times are independent of $\mu$. These correspond to local eigen-operators of the super-operator, since the expression \eqref{eq:super-operator} for the super-operator shows that eigen-operators with a nonlocal component will generically have eigenvalues that are functions of $\mu$. These nonlocal eigen-operators clearly correspond to the conjugate points that move to later times as $\mu$ is increased. The existence of the local eigen-operators is surprising; as the subspace of local operators is small compared to the space of all operators, one might have expected that the typical eigen-operator generically contained nonlocal pieces.
    
    \item The size of the nonlocal subspace controls the density of \emph{nonlocal} conjugate points and possibly also the speed at which they approach later times as $\mu$ is increased. This is visible in the greatly increased density of conjugate points at early times in Fig.~\ref{fig:AllconjptvsmuChaos3}, where the degree of locality is $k=q=3$ in contrast to the other two cases that take $k=q=4$. It appears that many of these conjugate points rapidly shoot off to late times whereas in the other two cases many of the nonlocal conjugate points appear to level off quickly.
    
    Notably, there is not a large distinction between the integrable interaction $\delta H_1$ in Fig.~\ref{fig:AllconjptvsmuInteg} and the $4$-body chaotic interaction $\delta H_2$ in Fig.~\ref{fig:AllconjptvsmuChaos4} with regard to the behavior of the conjugate points. It appears that the degree of the locality of the Hamiltonian is the most significant factor in these small $N$ plots.
    
    We note here that we have chosen $k$ to be the same order as the locality of the Hamiltonian, both so that the linear geodesic $V(s)=Ht$ is explicitly a solution to the geodesic equations as well as so that the energies of the system are an ``easy" observable to measure. However, we could have also chosen $k$ to be smaller, since $k=2$ is sufficient for the geometric complexity to approach the true quantum circuit complexity \cite{Nielsen2007}. In this case, the nonlocal subspace of operators is substantially enlarged and we expect that at even smaller values of $\mu$ the nonlocal conjugate points occur at late times.
    
    Although we take $N$ to be small for computational feasibility, we emphasize that in the large $N$ limit the size of the nonlocal subspace vastly exceeds the size of the local subspace. The size of the nonlocal subspace scales as $O(e^N)$ while the size of the local subspace scales as $O(\text{poly}(N))$ regardless of $k$. In this limit, there will be many more nonlocal conjugate points than local, in contrast to what we see in the $N=6$ plots.
    
    \item The first conjugate point is rapidly followed by many more conjugate points, both local and nonlocal. This substantiates arguments detailed in Sec.~\ref{sec:free} that the occurrence of the first conjugate point is rapidly followed by the end of the linear regime for complexity, even though many conjugate points may be required to reach the plateau regime.

\end{enumerate}


\begin{figure}[h!]
    \centering
    \includegraphics[width=.75\textwidth]{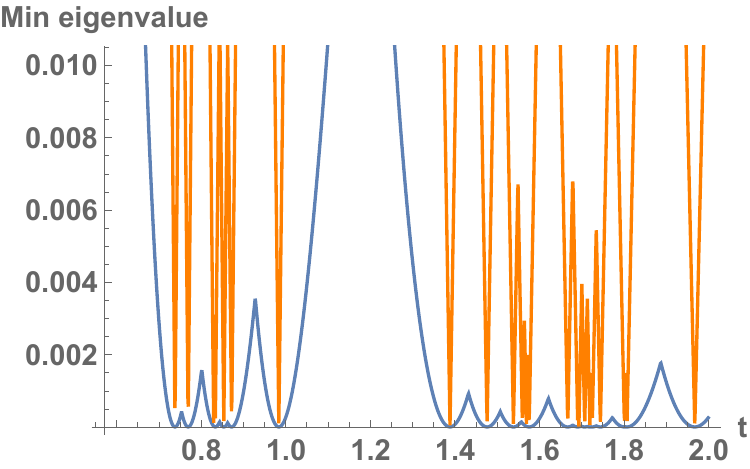}
    \caption{The minimum eigenvalue of the super-operator (orange) and of the matrix $M_{\alpha \beta}$ (blue) for $N=6$ fermions and the integrable interaction $\delta H_1$ with the same coupling strengths as chosen above.}.
    \label{fig:local_comparison}
\end{figure}

That the local conjugate points occur at times which are independent of $\mu$ might seem \emph{prima facie} to be at tension with the result in \cite{Balasubramanian:2019wgd} that the first conjugate point should not occur until times of order $\mu \sim e^{\alpha S}$ for Hamiltonians respecting the Eigenstate Complexity Hypothesis (to be discussed in Sec.~\ref{sec:ECH}). However, there is no real contradiction. Firstly, the results of \cite{Balasubramanian:2019wgd} apply to large $N$ systems, and the present numerical studies are at small $N$. Secondly, as we increase $N$, the $\mu$-independent time scale at which such local conjugate points occur cannot be sub-exponential in chaotic systems. Indeed, in Sec.~\ref{sec:local-chaotic}, we gave a general argument for chaotic Hamiltonians based on the Eigenstate Thermalization Hypothesis \cite{Rigol2008} and random matrix theory ideas that this time scale should be exponential, resolving the apparent tension. This argument need not apply to integrable Hamiltonians, which may still encounter conjugate points at early times in the large $N$ limit. 

We also note that Claim 2 in Sec.~\ref{sec:general_criteria} shows that in order to identify the locations of \emph{local} conjugate points, we need not compute the full super-operator. Instead, we can compute a smaller matrix of polynomial size,
\begin{align}
    M_{\alpha\beta}(t) &= \int_0^1 ds\int_0^1ds'\,\mathrm{Tr}\,[e^{i(s-s')tH}T_{\alpha}e^{-i(s-s')tH}T_{\beta}] \\
                         &= \sum_{m,n} c_{mn}^{(\alpha)}c_{nm}^{(\beta)} g(t(E_m-E_n)),\;\; g(x) = \left(\frac{\sin(x/2)}{x/2}\right)^2,
\end{align}
where $\alpha,\beta$ only index the \emph{local} operators. The zero modes of this matrix $M_{\alpha \beta}$ then identify the times of \emph{local} conjugate points with substantially increased computational efficiency. In Fig.~\ref{fig:local_comparison} we have demonstrated this by plotting the minimum eigenvalues of the super-operator and of $M_{\alpha \beta}$. The zero modes of each clearly coincide, though the matrix $M_{\alpha \beta}$ also appears to be more numerically stable in the sense that the precision of the numerical zero modes locates them closer to exactly zero.\footnote{At such small times, there are almost no nonlocal conjugate points of the super-operator present.}

\subsection*{Summary}

We computed the zero modes of the super-operator numerically for $N=6,8$ fermions by applying the Arnoldi algorithm to compute the minimum eigenvalue of its matrix representation, for various choices of Hamiltonian including free, integrable interacting, and chaotic SYK models. We demonstrated that a large class of conjugate points corresponding to local eigen-operators of the super-operator remain fixed, while those which have nonlocal components occur at later times for larger $\mu$. The density of nonlocal conjugate points and the times at which they occur appears to be controlled by the size of the subspace of nonlocal operators, which becomes very large when $N$ is large. Lastly, we confirmed that the matrix $M_{\alpha \beta} (t)$ defined in Sec.~\ref{sec:general_criteria} correctly identifies the locations of local conjugate points with greatly improved efficiency. This indicates that there are no conjugate points obstructing the linear growth of complexity until time scales of order $O(e^N)$ when the Hamiltonian is chaotic, based on the arguments of Sec.~\ref{sec:local-chaotic}. This result is complementary to the results of Sec.~\ref{sec:free} and Sec.~\ref{sec:perturbation}, which showed respectively that the free fermion complexity growth ends at $O(\sqrt{N})$ time and that the complexity of an integrable Hamiltonian is upper bounded by $O(\text{poly}(N))$.


\section{Eigenstate Complexity Hypothesis} \label{sec:ECH}
From the geometric formulation of quantum computation, we have found evidence that there is a difference between chaotic and integrable models in the location of obstructions to complexity growth. Free models reach a complexity of order $O(\sqrt{N})$ at late times, while integrable models plateau at order $O(\text{poly}(N))$. Chaotic models, on the other hand, have complexity of order $O(e^N)$ at late times. Depending on the location and number of easy conjugate points, this linear growth may continue until exponential times.

To argue for this behavior of the complexity of time evolution in chaotic theories, \cite{Balasubramanian:2019wgd} found that a certain matrix $R_{mn}$ of operator matrix elements in the energy eigenbasis was relevant:
\begin{equation}
    R_{mn}=\frac{\sum_{\alpha} |\langle m| T_{\alpha} |n\rangle|^2 }{ \sum_{\alpha} |\langle m| T_{\alpha} |n\rangle|^2+ \sum_{\dot{\alpha}} |\langle m| T_{\dot{\alpha}} |n\rangle|^2} .
\label{eq:ECH-matrix}
\end{equation} 
The $T_{\alpha}$ are the easy operators (at most $k$-local), $T_{\dot{\alpha}}$ are the hard operators (at least $(k+1)$-local), and $\ket{m}$ and $\ket{n}$ are energy eigenstates. 
In order for chaotic Hamiltonians to show the expected complexity growth, individual off-diagonal matrix elements ($m \neq n$) should be suppressed by 
\begin{equation}
R_{mn} \sim e^{-2S}\text{poly}(S)r_{mn}
\label{eq:ECH} ,
\end{equation}
where $S$ is the log of the dimension of the Hilbert space and $r_{mn}$ are independent random numbers of order 1. 
In \cite{Balasubramanian:2019wgd}, the conjecture that \eqref{eq:ECH} should hold for chaotic Hamiltonians was called the Eigenstate Complexity Hypothesis (ECH), and we refer to $R$ in \eqref{eq:ECH-matrix} as the ECH matrix.
Intuitively, \eqref{eq:ECH} states that the eigenstates of a chaotic Hamiltonian are sufficiently different that acting with a local operator on an energy eigenstate will not yield a different one even approximately.

To make the polynomial factor in \eqref{eq:ECH} a bit more precise, we can compute the ECH matrix numerically in some simple chaotic systems; it is also interesting to compare to the integrable case.
We will find that the average $R_{mn}$ (with $m\neq n$) is in general suppressed by roughly the fraction $s(N,k)$ of operators that are $k$-local,
\begin{equation}
   s(N,k)=\frac{\text{(number of }k\text{-local operators)}} {(\text{total number of operators)}}.
\label{eq:ratio}
\end{equation} 
This holds regardless of whether the dynamics are chaotic or integrable, although higher moments of the distribution of $R_{mn}$ show more of a sensitivity to the dynamics.

For SYK, this means poly$(S)$ in \eqref{eq:ECH} is roughly $N^k$.
If there are extra symmetries in the problem, which is true in the case of $q=4$ SYK, some elements $R_{mn}$ could be slightly reinforced, but in general $\overline{R}$ (the average of all off-diagonal $R_{mn}$) should still be approximately equal to \eqref{eq:ratio}. 

\subsection{Pure SYK}

\begin{figure}[t]
\centering
\subfigure[] {\includegraphics[width=0.45\textwidth]{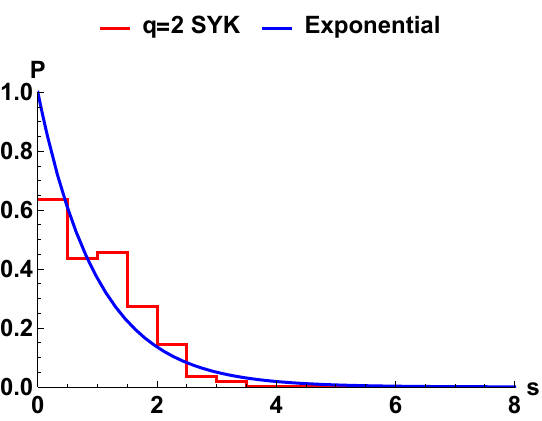}}
\subfigure[]{\includegraphics[width=0.45\textwidth]{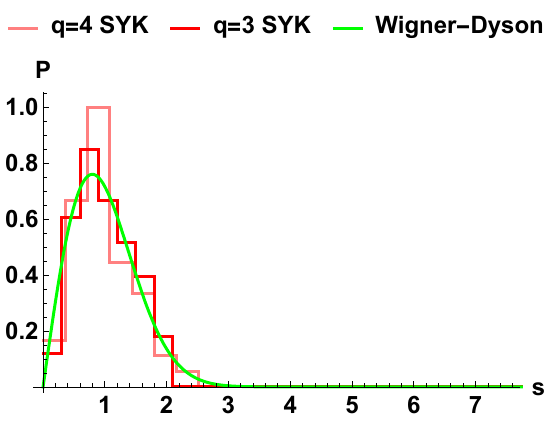}  }  
\caption{Level spacing for the (a) $q=2$ and (b) $q=3,4$ SYK models, with exponential and Wigner-Dyson level spacing distributions, respectively.}
\label{fig:syk levelspacing plots}
\end{figure}

Here we check the ECH criterion in a general SYK$_q$ model \eqref{eq:SYKHamil} for different choices of $N$ and $q$.
Recall that $N>q$, and we also take $k \geq q$ so that the Hamiltonian is a simple operator.
We first verify the integrability/non-integrability of the model using level spacing statistics: the integrable $q=2$ model shows a clear exponential distribution (i.e., Poisson distribution for the energies) while for $q=3,4$ the model is chaotic and shows Wigner-Dyson statistics (Fig.~\ref{fig:syk levelspacing plots}). These distributions match the expected level spacing distributions from random matrix theory \cite{doi:10.1080/00018732.2016.1198134}.

\begin{figure}[h!]
\centering
\subfigure[]
{\includegraphics[width=0.5\textwidth]{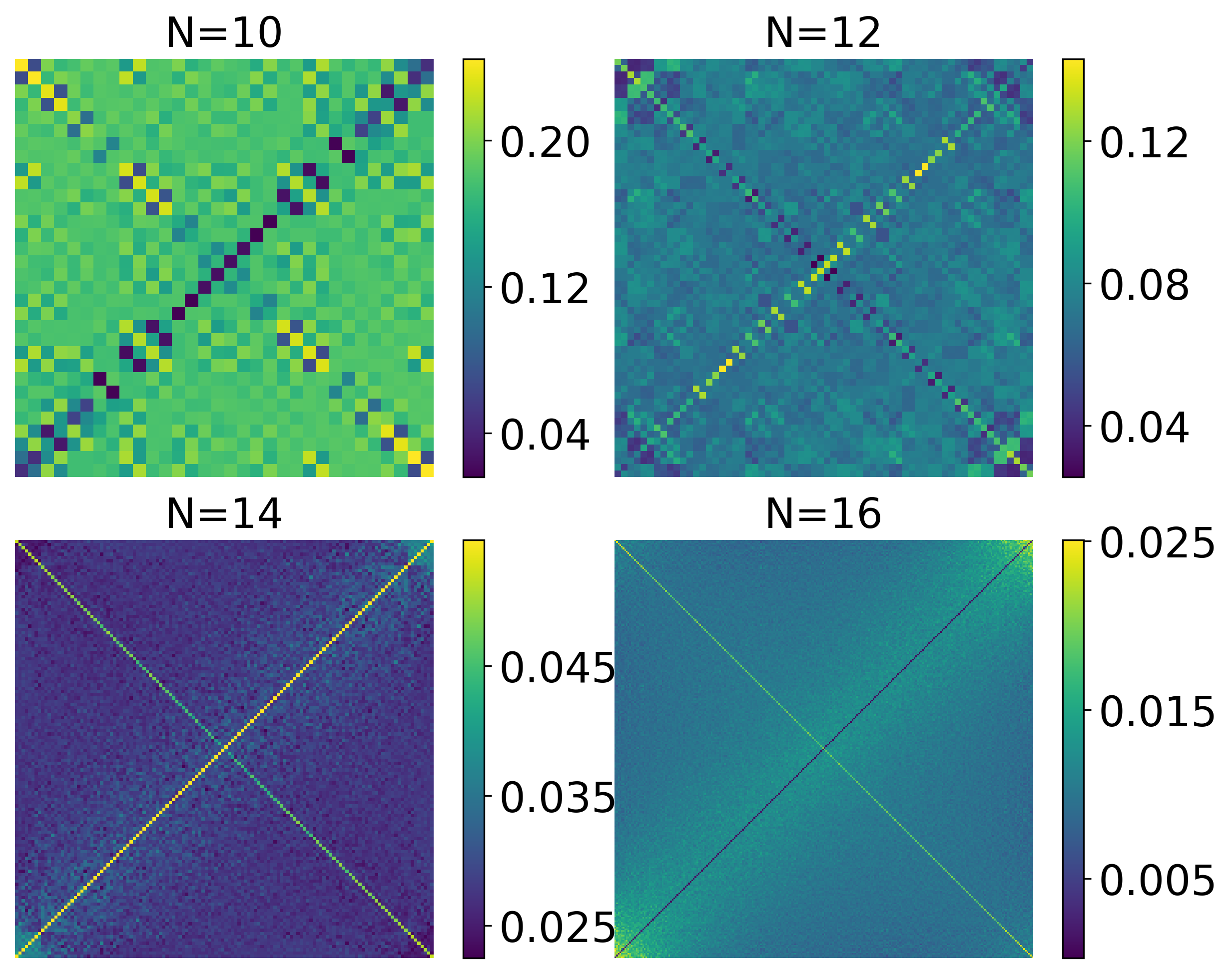}}
\caption{The $R_{mn}$ matrices of the $q=3$ SYK model with up to $k=3$-local operators taken to be easy. As one increases $N$, the overall $R_{mn}$ values are suppressed according to \eqref{eq:ratio}. }
\label{fig: ECH matrix q=3 syk}
\end{figure}

In the SYK model, the ratio \eqref{eq:ratio} is exactly
\begin{equation} 
s_{SYK}(N,k)=\frac{1}{2^N-1}\sum_{i=1}^{k} \binom{N}{i} .
\end{equation}
For example, for $N=14$ and $k=4$, this gives
\begin{equation}
    s_{SYK}(N=14,k=4) \approx 0.0897.
\label{eq:syk-ratio-n14-k4}
\end{equation}
For $q=2$, since the Hamiltonian is free, we would expect the ECH criterion to fail, i.e. the off-diagonal entries should not be suppressed by the ratio defined by \eqref{eq:ratio}.
\begin{figure}[t]
    \centering
    \includegraphics[width=0.45\textwidth]{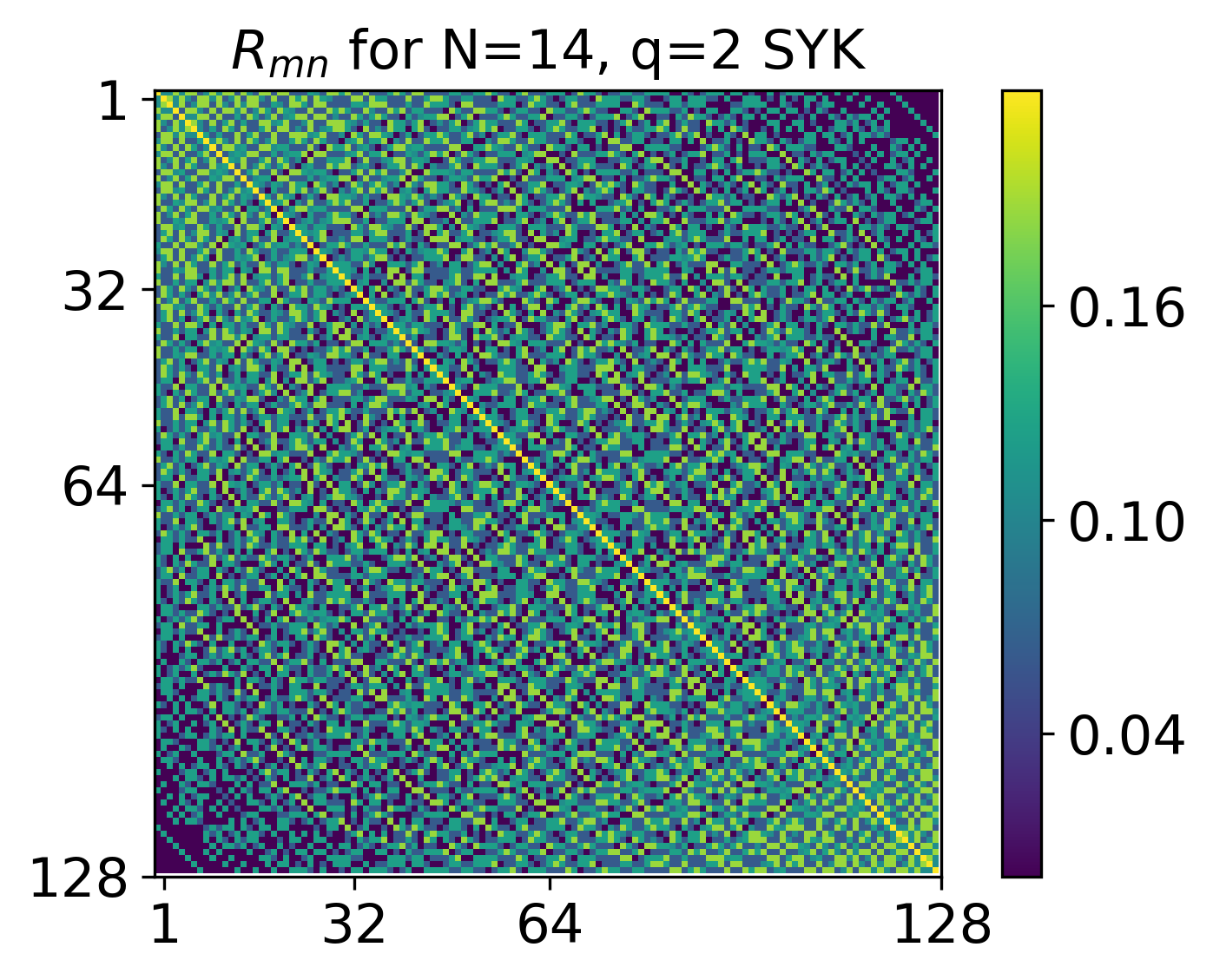}
    \caption{ECH matrix $R_{mn}$ for the $q=2$ SYK model with up to $k=4$-local operators considered easy.}
    \label{fig:syk-q=2-ech}
\end{figure}
Visualizing $R_{mn}$ for $q=2$ in Fig.~\ref{fig:syk-q=2-ech}, we see that there are indeed large off-diagonal elements that are comparable to the diagonal itself. By contrast, for the $q=3$ model (Fig.~\ref{fig: ECH matrix q=3 syk}), the off-diagonal elements are fairly uniformly suppressed with the exception of the anti-diagonal entries.

\begin{figure}[h!]
\centering
{\includegraphics[width=0.5\textwidth]{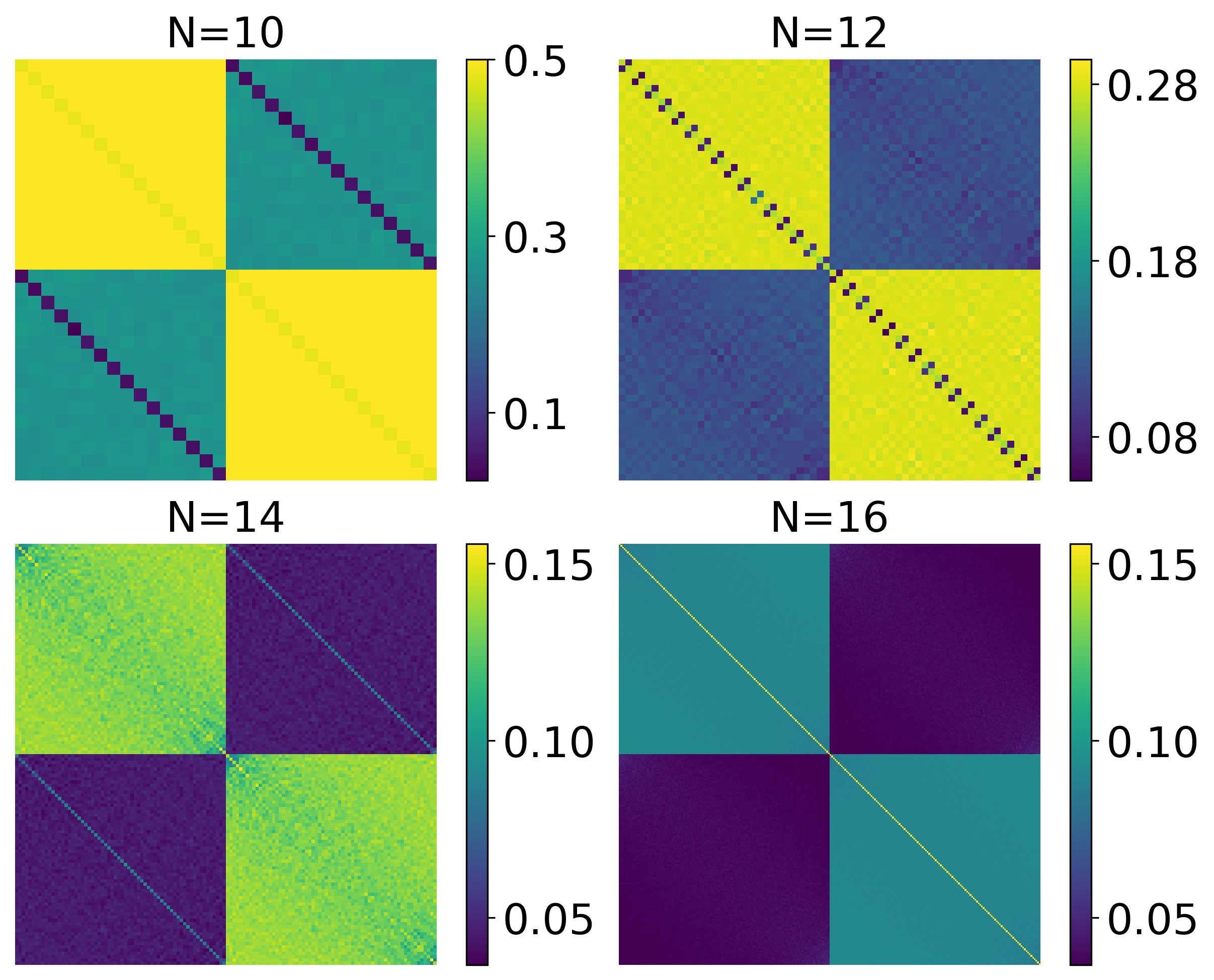}}
\caption{The $R_{mn}$ matrix for $q=4$ SYK model, featuring an extra fermion number symmetry $F$. The eigenvectors are organized into the even and odd symmetry sectors. The average value of the off-diagonal matrix elements becomes suppressed as $N$ increases. For $N=2,6$ mod $8$, the even and odd symmetry sectors are exactly equal (Appendix~\ref{sec:syk-ech-symmetries}).}
\label{fig: ECH matrix q=4 syk}
\end{figure}

In the higher $q$ models, the fermion number operator $F$ becomes relevant for interpreting our results.
\begin{equation}
    F \equiv i^{N/2} \prod_{j=1}^N \psi^j .
\end{equation}
For $q=4$, there is an extra fermion number symmetry preserved by the Hamiltonian, $[H,F]=0$, so the spectrum is 2-fold degenerate, and $F$ is a conserved charge with values $\pm 1$ (even and odd fermion number) on the eigenvectors of $H$. The conserved charge for $q=4$ leads to additional suppression and enhancement within the ECH matrix.
This can be explicitly seen in that there are two dominant colors in the ECH matrix across $N=10,12,14,16$ (Fig.~\ref{fig: ECH matrix q=4 syk}) when the eigenvectors are organized into even and odd symmetry sectors. 
The bright colors (larger magnitude) in the diagonal blocks indicate that two eigenstates within a given symmetry sector will have enhanced overlap with $k$-local operators. 
The darker colors (smaller magnitude) in the off-diagonal blocks indicate that only small overlaps can be achieved by acting with local operators between states in different symmetry sectors.


\begin{figure}[h!]
    \centering
    \includegraphics[width=0.8\textwidth]{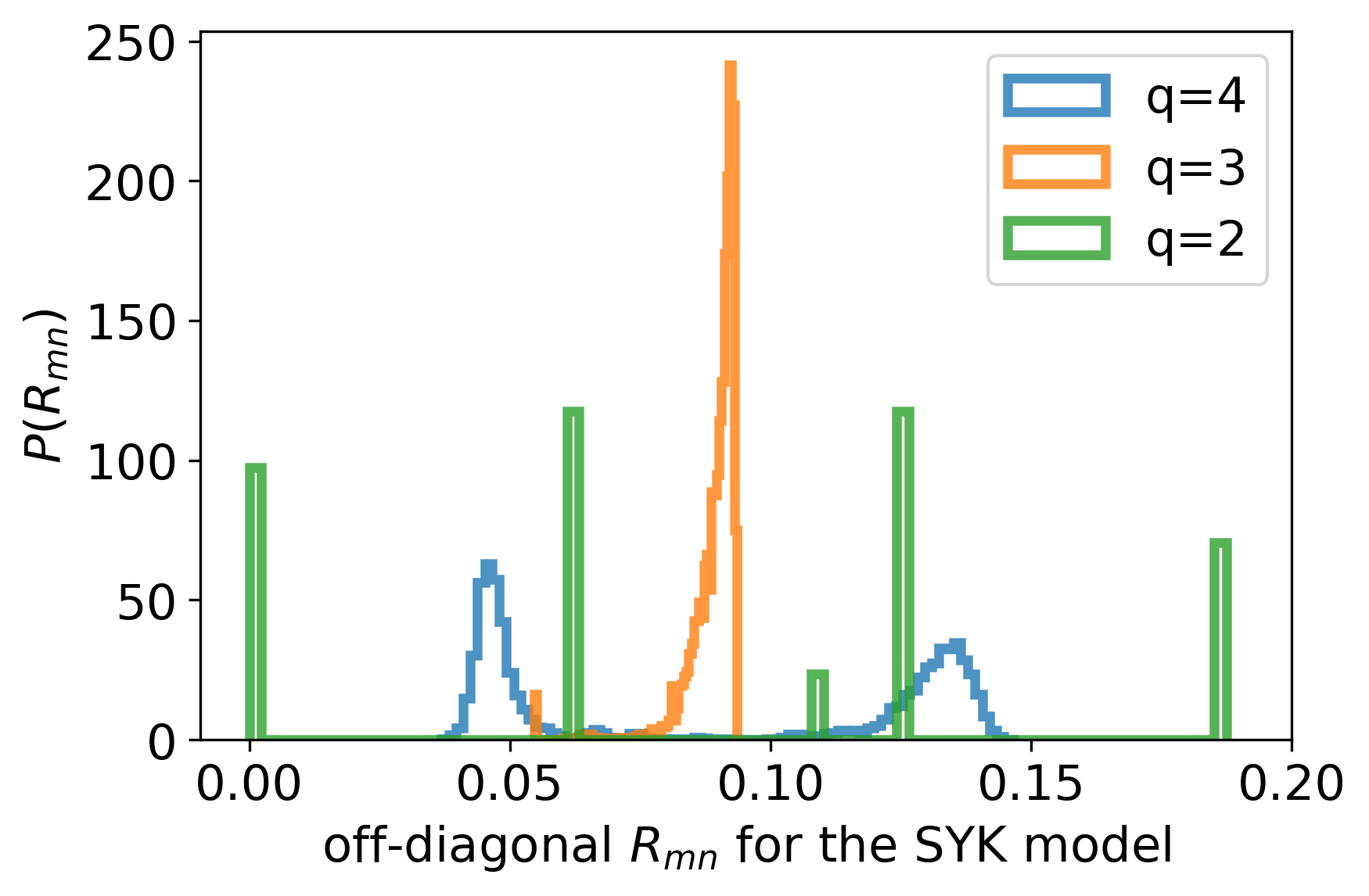}
    \caption{The distribution of off-diagonal $R_{mn}$ for the $q=2,3,4$ SYK model with $N = 14$. The degree of locality is controlled at $k=4$. Notice that the $q=3$ case has a small spike at less than the majority of the off-diagonal values. The spike comes from a more suppressed anti-diagonal values.}
    \label{fig:syk integrable vs chaotic}
\end{figure}

We can get a better quantitative picture of the differences between $q=2,3,4$ by studying the distribution of all of the off-diagonal entries in $R_{mn}$ (Fig.~\ref{fig:syk integrable vs chaotic}).
For $q=4$, we do see the two separate peaks corresponding to the two superselection sectors generated by the charge $F$, but the average of all off-diagonal entries (roughly the point in between the peaks, $\approx 8.8\times 10^{-2}$) is still suppressed by the ratio \eqref{eq:syk-ratio-n14-k4}.
For $q=3$, we have no additional symmetry, and we see that the off-diagonal values of the matrix $R_{mn}$ are all suppressed by approximately the same value (poly$(S)e^{-2S}$), following \eqref{eq:ECH}.

\begin{figure}[t]
\centering
\subfigure[] {\includegraphics[width=0.45\textwidth]{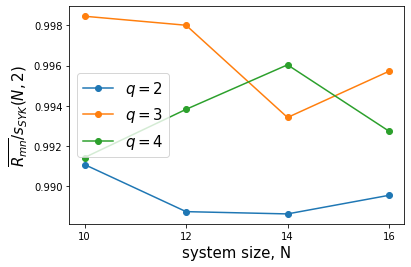} }
\subfigure[]{\includegraphics[width=0.45\textwidth]{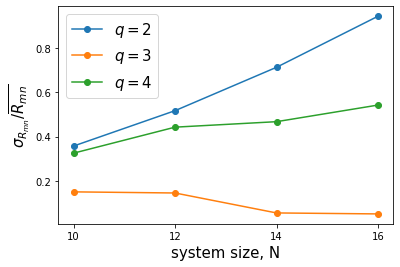} }
\caption{Comparison among the average and the standard deviation of the $R_{mn}$ matrix of the $q=2,3,4$ SYK model. The degree of locality is $k=4$. Each data point comes from a single realization of the model.}
\label{fig:ratio-stdev-scaling-SYK}
\end{figure}

Though we have claimed the ECH matrix should distinguish between integrable and chaotic theories by the magnitude of off-diagonal elements, one cannot discern the integrability of the model from the averaged value of $R_{mn}$ (Fig.~\ref{fig:ratio-stdev-scaling-SYK}). The off-diagonal values of both the free $q=2$ SYK model and the chaotic $q=3,4$ models are suppressed on average in Fig.~\ref{fig:syk integrable vs chaotic} by the ratio \eqref{eq:syk-ratio-n14-k4} with system size $N=14$ and the degree of locality $k=4$ fixed.
The key distinction is that in the $q=3$ case, the $R_{mn}$ are continuously distributed with a small standard deviation (Fig.~\ref{fig:ratio-stdev-scaling-SYK}). 
This is also true for $q=4$ within the same superselection sector of the charge $F$. 
The cluster centered at the larger averaged $R_{mn}$ in Fig.~\ref{fig:syk integrable vs chaotic} consists of the overlaps of the eigenvectors within the same superselection sector. The more suppressed $R_{mn}$ cluster comes from the overlaps between eigenvectors across different sectors.
By contrast, for $q=2$, $R_{mn}$ essentially has a delta function distribution at discrete points and a very large standard deviation consequently. The masses at different values suggest multiple symmetries in the $q=2$ SYK model, and the discrete support reflects the integrability of the free theory, where a very small number of parameters control the energies and correlations of the model.
The overlaps in the $q=2$ case with $N=14$ have delta functions at five major values (Fig.~\ref{fig:syk integrable vs chaotic}).
Notably in the free case, a significant number of the overlap matrix elements $R_{mn}$ are 0. This leads to smaller average value of $R_{mn}$ in Fig.~\ref{fig:ratio-stdev-scaling-SYK}. This together with the delta-function at outlying value indicates a large coefficient of variation at large $N$ (Fig.~\ref{fig:ratio-stdev-scaling-SYK}). Similar observations have been made regarding the matrix elements of local operators in the energy basis of an integrable system, $\bra{m} \mathcal{A} \ket{n}$, in the context of the Eigenstate Thermalization Hypothesis  \cite{doi:10.1080/00018732.2016.1198134,PhysRevA.80.053607}. In the case of the XXZ-chain, for example, the off-diagonal entries of $\bra{m} \mathcal{A} \ket{n}$ are generically either highly suppressed or very large, whereas the entries in the non-integrable case show a more uniform distribution.

We have focused on the SYK model here, but see Appendix~\ref{sec:ising-numerical} for a companion study of the ECH matrix for the mixed-field Ising chain of length $N=7,8,9,10$, which similarly features regimes of chaos and integrability.

\subsection{Deformations of free SYK}

\begin{figure}[t]
\centering
\subfigure[] {\includegraphics[width=0.45\textwidth]{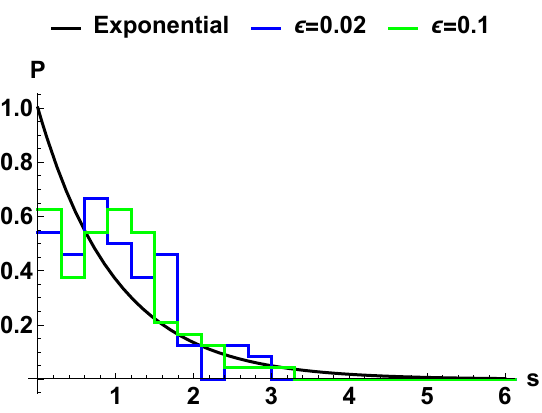}}
\subfigure[]{\includegraphics[width=0.45\textwidth]{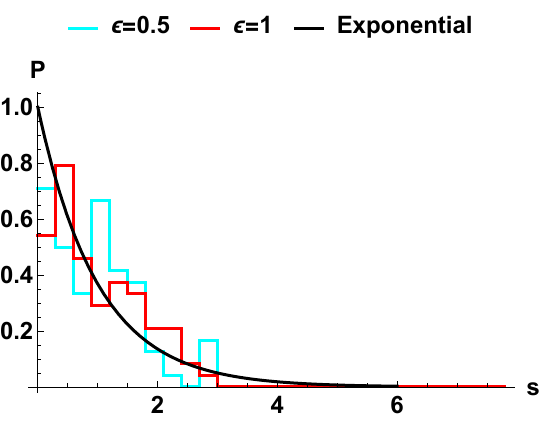}  }  
\caption{Level spacing for free SYK with an integrable quartic perturbation \eqref{eq: integrable 4-body} (a) when the coupling $\epsilon$ is small and (b) when the coupling is large.  Both small and large coupling show integrable statistics. }
\label{fig:syk levelspacing w int perturb}
\end{figure}

\begin{figure}[t]
\centering
\subfigure[] {\includegraphics[width=0.45\textwidth]{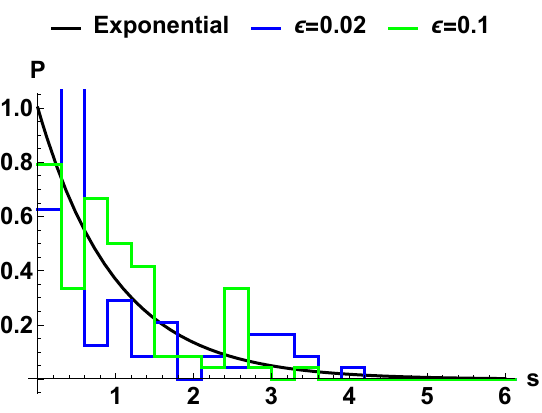}}
\subfigure[]{\includegraphics[width=0.45\textwidth]{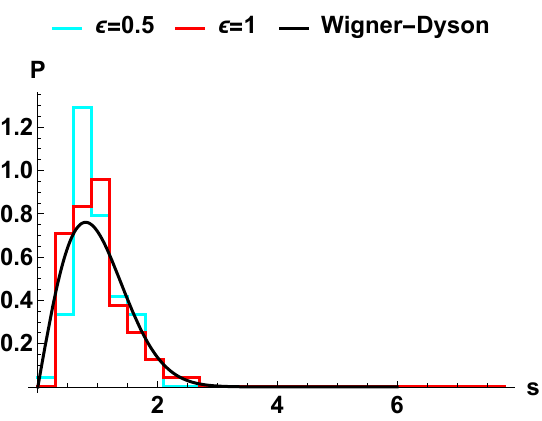}  }  
\caption{Level spacing for free SYK with an chaotic cubic perturbation \eqref{eq: chaotic 3-body} (a) when the coupling $\epsilon$ is small and (b) when the coupling is large.  As the chaotic term becomes comparable in magnitude to the free term, the level spacing statistics shift from exponential to Wigner-Dyson. }
\label{fig:syk levelspacing w chao perturb}
\end{figure}

In addition, we consider adding integrable or chaotic deformations to the free model as in Sec.~\ref{sec:perturbation} and Sec.~\ref{sec:syk-numerical}.
Adding an integrable term \eqref{eq: integrable 4-body} to the free Hamiltonian with Gaussian random coupling $M_{ij}$ does not change the level spacing distribution, regardless of the coupling strength $\epsilon$ (Fig.~\ref{fig:syk levelspacing w int perturb}).
By contrast, adding a 3-local chaotic perturbation \eqref{eq: chaotic 3-body} to the Hamiltonian affects the energy spacings.
As the coupling $\epsilon$ becomes large, the level spacing transitions to chaotic Wigner-Dyson statistics (Fig.~\ref{fig:syk levelspacing w chao perturb}).

\begin{figure}[h!]
    \centering
    \includegraphics[width=0.8\textwidth]{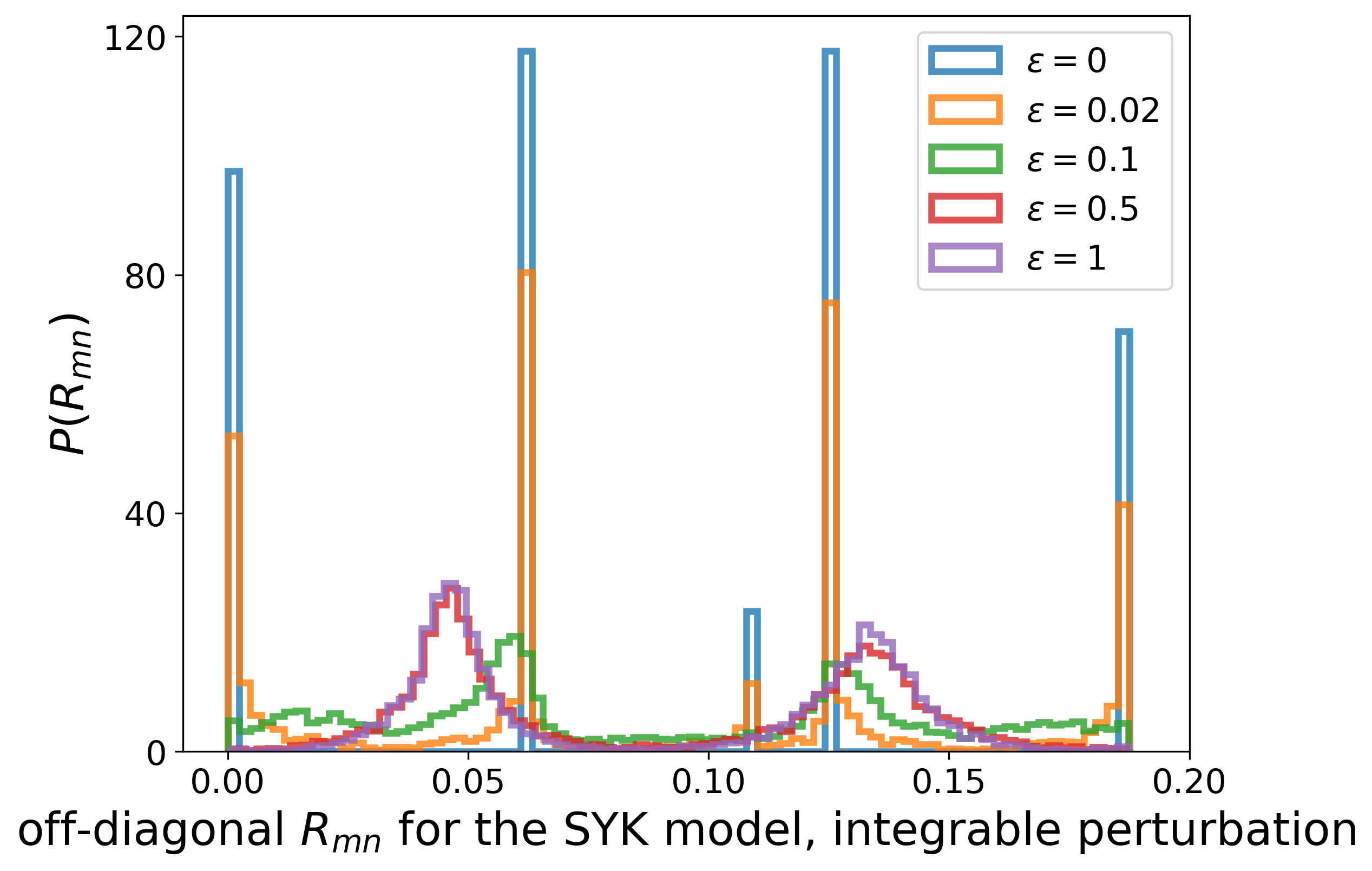}
    \caption{The distribution of off-diagonal $R_{mn}$ for the SYK model with an integrable perturbation \eqref{eq: integrable 4-body}. The number of Majorana fermions is 14 and the degree of locality is $k=4$.}
    \label{fig:syk ech w int perturb}
\end{figure}

\begin{figure}[h!]
    \centering
    \includegraphics[width=0.8\textwidth]{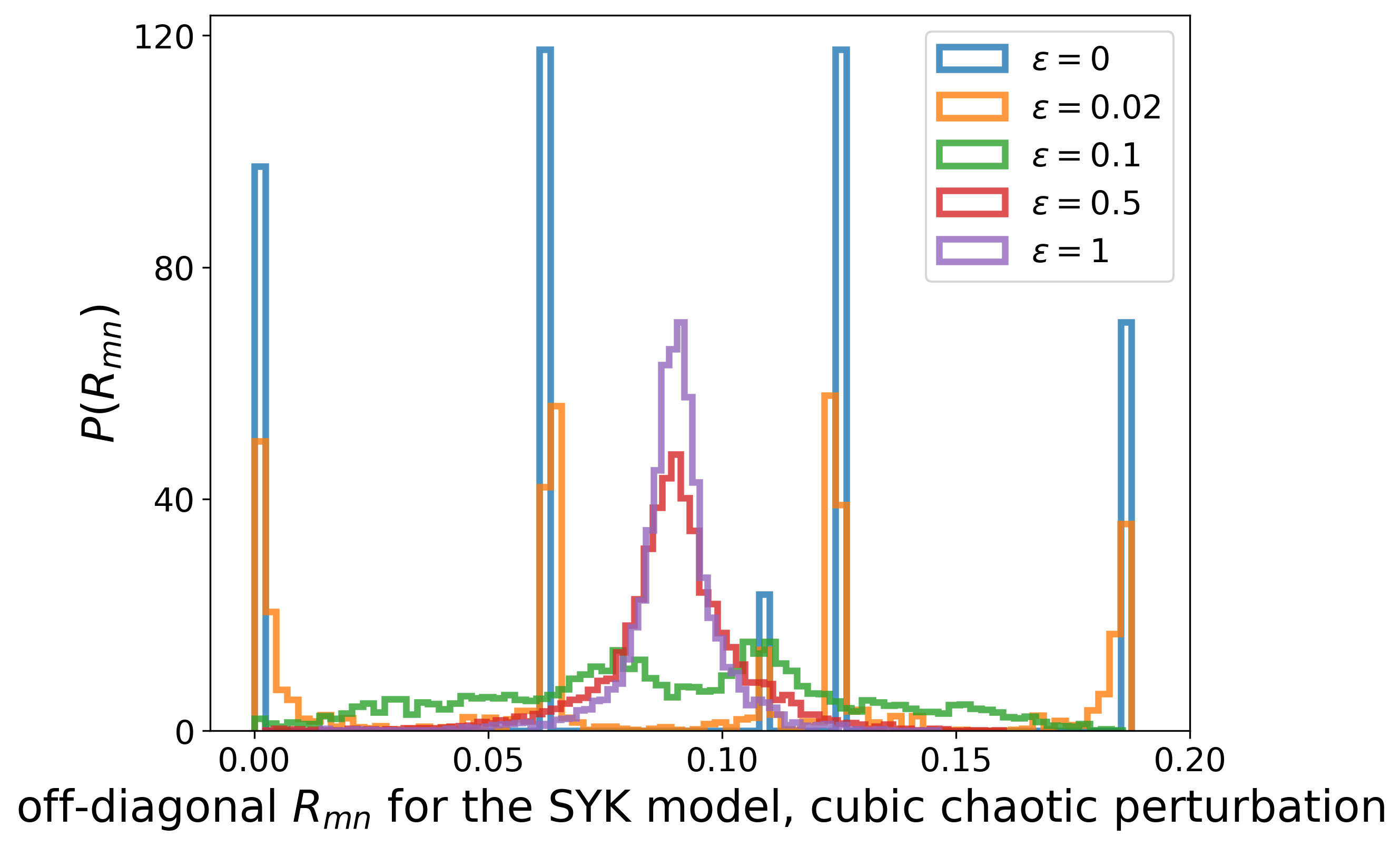}
    \caption{The distribution of off-diagonal $R_{mn}$ for the SYK model with a chaotic perturbation \eqref{eq: chaotic 3-body}. The number of Majorana fermions is 14 and the degree of locality is $k=4$.}
    \label{fig:syk ech w chao perturb}
\end{figure}

However, both the integrable and non-integrable perturbation to the Hamiltonian change the distribution of the off-diagonal entries of $R_{mn}$. At very small perturbation parameter, the peaks concentrated at $R_{mn}$ values of the free SYK model widen. When the perturbation parameter becomes large, the $R_{mn}$ form  distributions centered around new values that reflect the symmetry of the system (Figs.~\ref{fig:syk ech w int perturb} and \ref{fig:syk ech w chao perturb}). Since the integrable Hamiltonian commutes with the fermion number operator, its $R_{mn}$ distribution is bimodal corresponding to the two symmetry sectors; the cubic chaotic SYK model has no such symmetry as discussed before, so the distribution is unimodal. Apart from the existence of these symmetry sectors, the integrable and chaotic perturbations do not show qualitative differences in their distributions. Indeed, the $R_{mn}$ distribution shows a bigger contrast between free systems and interacting systems, rather than between integrable and non-integrable systems. Similar observations can be made in the $R_{mn}$ distribution of transverse Ising model (Appendix~\ref{sec:ising-numerical}), where the free spin chain also has a delta-function like probability distribution. The integrable but interacting spin chain, analogous to the SYK model perturbed by an integrable deformation, instead clusters around values that reflect the symmetry of the system.

The number of conserved quantities in an integrable theory scales extensively with the number of degrees of freedom, so we might have expected that the distribution of $R_{mn}$ in free and interacting integrable theories would be similar.  However, free and interacting integrable theories are distinct in that free Hamiltonians can be diagonalized into a sum of $O(N)$ independent terms whereas interacting integrable theories cannot in general. Because of this, the $2^N$ eigenstates of a free theory carry redundant information about the system, and this is why the values of the overlap matrix $R_{mn}$ show delta-function support. This reduction in the number of effective variables required to characterize $R_{mn}$ is already impossible for interacting integrable theories.

\subsection*{Summary}
From the simulations of the free, integrable, and chaotic SYK model with varying system sizes, we have  observed that the mean of the off-diagonal ECH matrix $R_{mn}$ is suppressed by the ratio \eqref{eq:ratio} in every type of system. The free systems have outliers compared to the mean but also a large number of zero entries that ``balance out" the large entries. However, the standard deviation in the free case scales linearly with the system size $N$ while it remains approximately constant in the chaotic cases. From the distribution of the $R_{mn}$, one can distinguish the free systems from interacting integrable and chaotic systems by the discreteness of their support, but the latter two are difficult to differentiate. Many characteristics of the ECH matrix have also been observed in the ETH matrix. The reason why interacting-integrable and chaotic theories have similar $R_{mn}$ distribution and the precise relation between the ECH and ETH matrices will be explored in future works.


\section{Discussion}\label{sec:disc}

In this paper, we studied conjugate points and geodesic loops in various SYK models. In the free model, we located all conjugate points and characterized the family of geodesic loops associated to the local conjugate points. This allowed us to exactly compute the complexity, which is bounded by $O(\sqrt{N})$, and to specify the fast-forwarding Hamiltonian at sub-exponential times. We studied the motion of conjugate points with $\mu$ under the addition of integrable or chaotic interactions both analytically and numerically. In the integrable case, we first showed how to set up perturbation theory for conjugate point locations in the strength of the coupling constant controlling interactions. We also described a family of geodesic loops which bound the complexity by $O(\text{poly}(N))$ for the class of integrable systems we considered. We then studied local conjugate points in chaotic theories. We argued based on the statistics of the matrix $M_{\alpha\beta}(t)$ that such local conjugate points do not occur in chaotic systems before exponential time, thus strengthening the arguments given in previous work \cite{Balasubramanian:2019wgd}. We then studied the locations of conjugate points non-perturbatively using numerics for SYK models up to $N=8$. Finally, we explored the Eigenstate Complexity Hypothesis (introduced in \cite{Balasubramanian:2019wgd}) in free, interacting-integrable and chaotic SYK Hamiltonians. We view these results as demonstrating a hierarchy of complexity growth between free, integrable, and chaotic models, and as a preliminary attempt at describing a more complete picture of the growth of complexity of time evolution, in which conjugate points play an essential role. Of course, global loops (which are not signaled by conjugate points) should also play an important part in this story, and perhaps they could even obstruct complexity growth before conjugate points. A more complete picture of complexity growth must therefore also include these.

\subsection*{Quantum error correction and AdS/CFT}

The modern picture of bulk reconstruction in AdS/CFT involves interpreting the bulk-to-boundary map as an isometry which encodes bulk ``logical" degrees of freedom within the set of CFT ``physical" degrees of freedom in an approximate quantum error correcting code (QECC) \cite{Almheiri:2014lwa}.
This picture sheds light on several subtle issues in bulk reconstruction, such as the fact that a single bulk operator can have multiple distinct boundary reconstructions on different subregions of the boundary.
However, just as there is some ambiguity in the definition of quantum complexity, there are some choices to be made in the definition of the QECC.
One such choice is the notion of the code subspace, which is usually taken to be (roughly speaking) the subspace of states which correspond to a bounded number of bulk operator insertions around a semiclassical background.
The QECC then reconstructs bulk operators within this Hilbert subspace, rather than on the full CFT Hilbert space. 
This allows the AdS bulk to incorporate, for instance, radial locality in the form of commutation between a bulk local operator and a boundary local operator \cite{Harlow:2018fse}.

It has always been relatively ambiguous what precisely the code subspace ought to be in a given situation.
There are known restrictions on, for instance, what fraction of black hole microstates in a single microcanonical window may be included in a code subspace while keeping the error in the approximate bulk reconstruction under control \cite{Hayden:2018khn}.
Relatedly, the choice of simple operators in geodesic complexity is somewhat ambiguous. One way to construct the code subspace from a CFT perspective is to start with some holographic state (say, the vacuum) and act on it with a few, not-too-heavy single-trace operators. The span of such states forms a subspace which one could regard as the code subspace. Taking inspiration from this idea, one could regard as simple operators (from a complexity perspective) the span of such not-too-heavy single-trace operators which generate the code subspace.  
This ties together the complexity-theoretic notion of locality and the error correction notion of locality.

One speculative way of operationalizing these ideas in the context of conjugate points is the following. Suppose we take a CFT state which corresponds to a small number of light operator insertions in some background state. In the bulk, this creates some particles near the boundary in some semiclassical geometry. Now one considers time evolution on the boundary by the (local) boundary Hamiltonian $H$. At very late boundary times $t$, we expect that the corresponding linear geodesic $e^{-iHt}$ encounters a conjugate point. After the conjugate point, a new globally minimizing geodesic takes over, which may correspond to evolution with a \emph{nonlocal} effective Hamiltonian as we have discussed in this paper. Meanwhile, in the bulk this time evolution is dual to a scattering process between the particles that were created near the boundary. Assuming this scattering process did not create a black hole, the local time evolution on the boundary does not take our initial state out of the code subspace. Since the new minimizing boundary geodesic at late times does not lie in the code subspace, we expect it corresponds to a time evolution in the bulk involving black holes, as it will take the initial state out of the code subspace. This suggests that the late-time out states of the scattering process in the bulk could have been reached more efficiently by a scattering process involving black holes in the bulk. 

\subsection*{Remarks about switchback effect}

The geometric complexity theory that we studied was constructed to be polynomially equivalent to quantum circuit complexity \cite{Nielsen2007}.
This required choosing the cost factor $\mu$ to be exponential in the system size, $\mu \sim e^S$.
However, there are arguments from AdS/CFT involving the so-called switchback effect \cite{Stanford:2014jda} which appear to imply that the notion of complexity which is relevant for holography is compatible with a more gradual weighting scheme \cite{Brown:2017jil}.\footnote{Although, see \cite{Caginalp:2020tzw} for a situation where the ordinary weighting scheme with $\mu \sim e^S$ appears to give the holographically expected results, at least for small $N$. }
The graduated scheme roughly involves taking operators below a locality threshold $k$ to have cost 1, and operators above this locality to have weight equal to the exponential of their degree of locality.
So, a $K$-local operator for $K>k$ would have weight $e^K$.

A potential issue with the graduated scheme is that it allows $(\log N)$-local gates to act with a polynomial cost.
In the more conventional formulation of complexity theory, a fixed upper bound $k \sim O(N^0)$ is chosen on the locality of polynomial cost gates.
However, if we interpret the graduated scheme as setting $k \sim c \log N$ instead of $k \sim O(N^0)$, it is in fact still polynomially equivalent to the standard scheme where only $O(N^0)$-local gates have polynomial cost.
To see this, notice that an arbitrary unitary operator acting on $c \log N$ qubits has complexity at most roughly $e^{c \log N} = N^c$, which is still polynomial in $N$.
Therefore, in a polynomial-length circuit formed using the graduated scheme, we may simply replace any $(\log N)$-local gates with polynomially many $O(1)$-local gates without changing the fact that the total circuit length is polynomial in $N$.

In fact, the graduated scheme assigns a cost of $N$ to a $(\log N)$-local gate, which is precisely the same as the maximum cost of a $(\log N)$-qubit unitary operator if we had used $O(1)$-local gates in the standard scheme.
Given this observation, it is not hard to imagine that the graduated and standard schemes are actually related by some $O(1)$ factor rather than only a polynomial.
Since there are differences in sectional curvature between the graduated and standard schemes \cite{Brown:2016wib}, it would be interesting to understand whether these differences really appear at the level of the complexity of time evolution. 
It may be that they are only related to $O(1)$ prefactors in that quantity, and more significant differences can only be seen for more complicated quantities like the complexity of a precursor \cite{Stanford:2014jda}.

\subsection*{Relation of ECH to ETH}

The behavior of the $R_{mn}$ is similar to the behavior of matrix elements of local observables in the energy basis that show up in the Eigenstate Thermalization Hypothesis (ETH) \cite{PhysRevA.43.2046, Rigol2008}. The ETH matrix in integrable systems features off-diagonal elements that are mostly either very small or zero, with a sparse number of large values. By contrast, the ETH matrix in chaotic models has more uniformly suppressed entries.  We have demonstrated analogous properties in the ECH matrix $R_{mn}$ for both the integrable and chaotic SYK$_q$ models (Fig.~\ref{fig:syk integrable vs chaotic}, \ref{fig:syk ech w int perturb}, \ref{fig:syk ech w chao perturb}) as well as the mixed-field Ising models in Appendix~\ref{sec:ising-numerical}.

Although the statistical properties of these matrices are similar, the physical settings behind the ECH and ETH statements are different. ECH is founded in ideas of circuit complexity whereas ETH is based on ideas in many-body physics and thermalization. Regardless, the mathematical expression of ECH involves the matrix elements of local operators in the energy basis that appear in ETH. Specifically, one might wonder if there is a relation between ECH and ETH applied to the local operator $\sum_\alpha T_\alpha$ which is the sum of all local operators. One approach that might be fruitful in understanding the statistics of the matrix elements of such an operator is to consider perturbing a Hamiltonian with a perturbation $\delta H = \sum_\alpha T_\alpha$. In this case, it may be possible to understand the statistics by relating the eigenstates of the perturbed and unperturbed Hamiltonian since the corrections to the energies and eigenstates generically involve matrix elements like $\langle m| \delta H | n\rangle$.  A related thought experiment investigating the statistics of matrix elements of such perturbations was carried out in \cite{PhysRevA.43.2046}.

In Sec.~\ref{sec:local-chaotic} we showed that in a chaotic theory complexity will grow linearly until times $O(e^N)$ provided we assume that the energy eigenstates of the chaotic theory are essentially Haar random rotations of a fixed ``simple'' basis.  There is a related expectation in ETH, i.e., that all  the eigenstates of a thermalizing Hamiltonian ``look'' thermal \cite{Rigol2008}.  Indeed,  Deutsch showed that for a real, symmetric Hamiltonian, a thermalizing perturbation leads to energy eigenstates that are Gaussian random linear combinations of the unperturbed eigenstates \cite{PhysRevA.43.2046}. This averaging suppresses  the variance of observables by factors of $e^S$  just like in the Haar ensemble we proposed for chaotic systems. In \cite{Balasubramanian:2007qv} entropic suppression of variance was also described for typical, random states in a quantum microcanonical ensemble.   In Sec.~\ref{sec:local-chaotic} we are using similar reasoning to argue that  typical energy eigenstates of a chaotic theory will be random combinations of a fixed ``simple'' basis, and so variances will be suppressed via averaging.

\subsection*{General integrable systems}

In this paper we have considered an interacting integrable deformation which is a quadratic function of the local operators $J_3^{(i)}$ in the diagonalized free theory. Consequently, the structure of the operator dynamics as governed by $[H, \:\cdot\,]$ is simplified, which allowed us to obtain analytic results in perturbation theory in Sec.~\ref{sec:perturbation}. General integrable systems can look much more complicated. For instance, in App.~\ref{sec:ising-numerical} we study the ECH matrix for the (integrable) transverse-field Ising model, which appears to be a nontrivially interacting lattice spin model. This model is equivalent to a free fermion model after performing a non-local Jordan-Wigner transformation taking the bosonic spins to fermions. It is natural to choose the $k$-local operators in the theory to be the bosonic spin operators supported only on contiguous size-$k$ regions of the bosonic spin lattice for the purposes of computing complexity. However, the Jordan-Wigner transformation will not respect this split into local and nonlocal operators. Consequently, it is plausible that more general integrable theories behave similar to chaotic systems with respect to complexity, owing to the fact that local operators in the theory are scrambled into the nonlocal sector when the theory is diagonalized.

In full generality, integrable systems in a finite-dimensional phase space can be written in action-angle variables. Much like the dynamics of the harmonic oscillator (a canonical example of an integrable system) consists of ``rotation'' in phase space, the dynamics of these more general systems also consists of periodic motion in phase space, albeit with a possibly action-dependent frequency.\footnote{More precisely, the ``actions" are first integrals of motion. For the one-dimensional harmonic oscillator, the action is proportional to the energy. For the harmonic oscillator, the constant-energy slices foliate phase space by a set of concentric circles, and the dynamics is rigid rotation around these circles. More generally the phase space of an integrable system need only be foliated by topological tori, with the dynamics corresponding to a periodic motion around each torus whose frequency depends on the corresponding action. In fact, the KAM theorem \cite{Kolmogorov, ArnoldKAM} guarantees that most of these tori are preserved given small deformations of the Hamiltonian, so we expect results that hold for integrable systems may also hold for perturbatively chaotic systems.} So there is no guarantee that integrable systems appear free in any basis. Infinite-dimensional integrable systems like the Korteweg - de Vries (KdV) system \cite{Drinfeld:1984qv} or the solitonic Sine-Gordon system \cite{Coleman:1974bu} exemplify this fact. Similarly, a broad class of highly interacting quantum integrable systems consists of the lattice spin models that are exactly solvable using the Bethe ansatz \cite{Bethe}. Nevertheless, the Hamiltonian in all these systems is built out of the commuting charge operators, and so we expect our analysis of Sec.~\ref{sec:integ_loops} to be generalizable to such systems. It would be interesting to further explore whether it is possible to make precise analytic statements about complexity in these highly structured models whose notion of locality in the original variables does not align with locality in the variables that simplify the dynamics.

\subsection*{A version of complexity restricted to local circuit modifications}

In AdS/CFT, tensor networks have proven useful in gaining intuition about properties of the bulk semiclassical theory \cite{Swingle:2009bg, Pastawski:2015qua, Hayden:2016cfa, Milsted:2018san, Milsted:2018yur, Bao:2018pvs, Caputa:2020fbc}.
Roughly speaking, the tensor network lives on a tessellation of a bulk Cauchy slice.
There is an approximate notion of quantum complexity for tensor networks which corresponds simply to counting the number of tensors in a region of the bulk spacetime, and this supports the suggestion that a quantity like bulk wormhole volume should be dual to quantum complexity in a two-sided black hole \cite{Brown:2015bva}.
However, once the complexity saturates at its maximum value (polynomial in the entropy for integrable systems and exponential for chaotic systems), the tensor network which grew to foliate the wormhole interior is no longer expected to be the minimal network, just as the quantum circuit which builds the state will become a non-minimal circuit.
In the geometric language, the linear geodesic will encounter conjugate points or geodesic loops.

As physics is at least approximately local, for a geometric wormhole interior it would be surprising if there could be large correlated fluctuations of geometry which act in concert to decrease the tensor network size.
Such large fluctuations with global changes to tensor network structure would correspond to geodesic loops in the complexity geometry which have little or no relation to the original linear geodesic.
This motivates a notion of ``local complexity", where only local updates to the tensor network (equivalently, the quantum circuit) which decrease the length are allowed.\footnote{This notion was inspired by discussions during the It from Qubit annual meeting in December 2019 and the IAS It from Qubit workshop in December 2020.  A similar notion is discussed as ``pseudo-complexity" in appendix B of \cite{Bouland:2019pvu}.} Of course, if many sequential local updates are made, we can still achieve a large decrease in the size of the tensor network.

In the geometric complexity language, local complexity is computed by starting with the linear geodesic $L$ and flowing downward in the space of paths, where the downward directions all correspond to conjugate points along $L$.
These downward flows will explore the space of paths at least in the neighborhood of $L$, and will find the geodesic of smallest length which is continuously connected by upward flows in path space to $L$.
To our knowledge, local complexity has not been explored, and may be an interesting alternative to the usual complexity-theoretic definition.
We will not explore it in any great detail here, but we will make the following point: local complexity behaves more or less analogously to circuit complexity in chaotic theories like holographic CFTs.
There are conjugate points along $L$ which sit approximately at $t \sim e^S$, which will terminate the linear growth at the expected timescale just as geodesic loops would \cite{Balasubramanian:2019wgd}.
Furthermore, the density of conjugate points along the linear geodesic is roughly constant after an initial growth, as can be seen from a simple calculation in the bi-invariant geometry.
So, we expect multiple exchanges of dominance between many geodesics induced by all of these conjugate points, which should generate the plateau.

Of course, after we encounter the first conjugate point, the remainder along the linear geodesic are not relevant for complexity growth since there is a new geodesic which computes the complexity.
So, in order for this argument to hold, we need a sort of universality among geodesics under the flow from $\mu = 0$ to $\mu \sim e^S$ which ensures that, even as the location of the geodesic changes in path space, the conjugate points which were present at $\mu = 0$ are shifted in roughly the same way as occurs for the linear geodesic.
That is to say, all geodesics at $\mu \sim e^S$ have a constant density of conjugate points around exponential length, just as we expect for the linear geodesic.
We have not proven this, but it seems likely from general intuitions about the complexity growth of chaotic Hamiltonians \cite{Balasubramanian:2019wgd}.

One difference in these notions is that geometric complexity is always bounded by the diameter of the manifold, and local complexity may slightly violate this bound.
However, it is unknown whether anything physical would be associated with such a slight modification of complexity's behavior. If local complexity is really the notion to consider in holography, it will have implications for the complexity of the AdS/CFT dictionary, following the arguments of \cite{Bouland:2019pvu}. 
This is because the calculation of local complexity is essentially a local optimization problem in path space, which should be polynomially computable in general, unlike circuit complexity which would involve searching the entire path space for potential geodesic loops.\footnote{One might try to search for global obstructions by allowing upward flows from $L$, in addition to downward flows in directions given by the Jacobi fields corresponding to conjugate points.  However, there are an infinite number of upward directions in the space of paths with the energy functional as a Morse function, so it is not efficient (and indeed, not clear that it is even possible) to explore path space in this way.}

\subsection*{Acknowledgments}

We thank Steve Shenker for helpful discussion and for motivating us to study the free SYK model, and Pedro Bernardinelli for use of a personal server on which some numerics were performed.
VB, MD, and CL are supported in part by the Department of Energy through grant DE-SC0013528.
VB is supported in part by the Simons Foundation through the It From Qubit Collaboration (Grant
No. 38559) and by the Department of Energy through grant QuantISED DE-SC0020360.
VB also thanks the Aspen Center for Physics, which is supported by National Science Foundation grant PHY-1607611, for hospitality while this work was in
progress.
AK is supported by the Simons Foundation through the It from Qubit Collaboration.
MD is supported by the National Science Foundation Graduate Research Fellowship under Grant No. DGE-1845298.

\appendix

\section{Two-point functions and the super-operator}\label{sec:twoptfunctions}

In this appendix we will write an interesting expression relating the super-operator to the matrix of thermal two-point correlation functions at infinite temperature, 
\beq
R_{IJ}(s) = e^{-S}\mathrm{Tr}\,(T_J e^{isH}T_I e^{-isH}),
\eeq
where $T_I = \{T_{\alpha}, \widetilde{T}_{\dot\alpha}\}$ denotes all the generators. The direct connection between the two comes from the fact that we can rewrite the Heisenberg operators $T_I (s)$ in terms of the $R_{IJ}$ via
\beq
e^{isH}T_I e^{-isH} =  R_{IJ}(s)T_J.
\eeq
with summation on $J$ implied here and below.

Starting with the general expression for the super-operator,
\begin{equation}
\begin{split}
    \textbf{Y}_{\mu}(\delta V(0)) = & \int_0^1 ds e^{iHts} \biggl[ \delta V_L(0) + (1+\mu) \sum_{\dot{\alpha}} \frac{\exp \left( \frac{-i \mu t \lambda_{\dot{\alpha}} s}{1+\mu}  \right)-1}{\lambda_{\dot{\alpha}}}  \delta \tilde{V}^{\dot{\alpha}}(0) [H,\tilde{T}_{\dot{\alpha}}]_L \\
    & + \sum_{\dot{\alpha}} \exp \left( \frac{-i \mu t \lambda_{\dot{\alpha}} s}{1+\mu}  \right) \delta \tilde{V}^{\dot{\alpha}}(0) \tilde{T}_{\dot{\alpha}} \biggr] e^{-iHts} ,
\end{split}
\end{equation}
we first use the trick of Sec.~\ref{sec:syk-numerical} to remove the local projection in the second term via $[H,\tilde{T}_{\dot{\alpha}}]_L = [H,\tilde{T}_{\dot{\alpha}}] - \lambda_{\dot{\alpha}} \tilde{T}_{\dot{\alpha}}$. For the resulting commutator term, we use the fact that (using the notation $\text{ad}_H \mathcal{O}= [H, \mathcal{O}]$)
\begin{align*}
    \frac{d}{dx} e^{\text{ad}_{xH}} \mathcal{O} = e^{\text{ad}_{xH}}  \text{ad}_H\mathcal{O}.
\end{align*}
This identity can be derived simply by comparing the Taylor series of both sides. Recalling the Campbell-Baker-Hausdorff identity $e^{xH} \mathcal{O} e^{-xH} = e^{\text{ad}_{xH}} \mathcal{O}$, this new identity is useful because it converts a commutator into a derivative that can be integrated by parts.

Using these substitutions and evaluating the integration by parts, after a short computation one can write the super-operator as
\begin{equation}
    \textbf{Y}_{\mu}(\delta V(0)) =  -\mu   \sum_{\dot{\alpha}} \phi(-M_{\dot{\alpha}} t) \delta \tilde{V}^{\dot{\alpha}} (0) e^{iHt} \tilde{T}_{\dot{\alpha}} e^{-iHt} + \int_0^1 ds e^{iHts} \biggl[ \delta V_L(0) + (1+\mu)\delta V_{NL} (0) \biggr] e^{-iHts} ,
\end{equation}
where we have defined $M_{\dot{\alpha}} = \frac{\mu \lambda_{\dot{\alpha}}}{1+\mu}$ and $\phi(x) = (\exp (ix) - 1)/(ix)$ as in Sec.~\ref{sec:syk-numerical}. The integral term is nearly the average of the Heisenberg operator $\delta V(t)$ over the time interval $0$ to $t$, but with the nonlocal components given extra weight $(1+\mu)$. This expression can also be written in terms of the two-point functions. Rescaling $s \to s/t$ and defining the time-averaged two-point function
\beq
\overline{R}_{IJ} (t) = \frac{1}{t}\int_0^t ds\,R_{IJ}(s),
\eeq
the resulting expression for the super-operator is simply
\begin{equation}
    \textbf{Y}_{\mu}(\delta V(0)) =  \left[-\mu   \phi(-M_{\dot{\alpha}} t) \delta \tilde{V}^{\dot{\alpha}}_{NL} (0) R_{\dot{\alpha} J} (t)  +  \delta V^{\alpha}_{L} (0) \overline{R}_{\alpha J} (t) + (1+\mu)\delta \tilde{V}^{\dot{\alpha}}_{NL} (0) \overline{R}_{\dot{\alpha} J} (t) \right] T_J  ,
\end{equation}
where summation on $J, \alpha$, and $\dot{\alpha}$ has been left implied.

The last two terms only involve the averaged $R$ matrix, and so have a smooth limit for large $t$. The first term however involves the exact $R$ matrix, which oscillates wildly at late times. It would be interesting for future work to explore further connections between the super-operator and thermal two-point functions.

\section{Symmetries of the ECH matrix in the SYK model}\label{sec:syk-ech-symmetries}

In this appendix, we analyze the symmetry properties of the SYK model to explain some of the structure of the ECH matrix found in Sec.~\ref{sec:ECH}.
We reproduce the SYK Hamiltonian:
\begin{equation}
    H= i^{\frac q 2}\sum_{1 \leq i_1<\dots i_q\leq N} J_{i_1,\dots,i_q} \psi^{i_1} \cdots \psi^{i_q} ,
\end{equation}
where $J_{i_1,\dots,i_q}\sim\mathcal{N}(0,\sigma^2)$, $\sigma^2= \frac{2^{q-1} (q-1)! \mathcal{J}^{2}}{q N^{q-1}}$ and $\{ \psi^i, \psi^j \}=2\delta^{ij}$. 

Starting from the Pauli matrices, we construct $N=2k$ Hermitian Majorana fermions by writing
\begin{equation}
\begin{split}
    \psi^1 &=1\otimes 1\otimes \cdots \otimes \sigma_1 ,\\
    \psi^2 &=1\otimes 1\otimes \cdots \otimes \sigma_2 ,\\
    \psi^3 &= 1 \otimes \cdots \otimes \sigma_1\otimes \sigma_3 ,\\
    \psi^4 &=  1 \otimes \cdots \otimes \sigma_2\otimes \sigma_3 ,\\
     \vdots & \qquad\qquad\quad \vdots\\
    \psi^{2k-1} &= \sigma_1 \otimes \sigma_3 \otimes \cdots \otimes \sigma_3 ,\\
    \psi^{2k} &=  \sigma_2 \otimes \sigma_3 \otimes \cdots \otimes \sigma_3 .
\label{eq:basis-su(n)}
\end{split}
\end{equation}
where the tensor products are taken over $N/2$ slots. 
Similar to CPT symmetries, the Majorana fermions either commute or anticommute (depending on dimension of the Hilbert space) with the following special operators defined by
\begin{align}
    \text{fermion number}: F & = i^{N/2} \prod_{j=1}^N \psi^j,\\
    \text{time reversal}: \mathcal{T} &= i^{N(N-1)/2} \prod_{j=1}^{N/2} \psi^{2j},\\
    \text{charge conjugation}: \mathcal{C} & = i^{N(N-1)/2}\prod_{j=1}^{N/2} \psi^{2j-1} .
\end{align}

There are certain $N$-dependent additional relations between these operators which are important,
\begin{align}
    [\mathcal{T}, F]=[\mathcal{C},F]&=0, \quad N=0,4 \mod 8,\\
    \{\mathcal{T}, F\}=\{\mathcal{C},F\}&=0, \quad N=2,6\mod 8 .
\end{align}
These symmetry relations are also observed in classifying topological insulators \cite{Ryu_2010} and can be traced back to the Bott periodicity in homotopy groups of classical groups.

We now explain the curious fact noted in Fig.~\ref{fig: ECH matrix q=4 syk} that the off-diagonal matrix elements $R_{mn}$ of the ECH matrix are exactly symmetric between the even and odd sectors.
There is no a priori reason that operator matrix elements in different superselection sectors should be exactly equal.
As we will see, it follows from the form of the ECH matrix and the enhanced symmetries above.
For the $q=4$ model, the Hamiltonian commutes with $F$.
Write $\ket{n,\pm}$ for the simultaneous eigenstates of $H$ and $F$.
These satisfy $H\ket{n,\pm} = E_n\ket{n,\pm}$ and $F\ket{n,\pm} = \pm\ket{n,\pm}$.
Let $N=2,6$ mod 8 (in particular, this includes $N=14$).
Then the operator $\mathcal{T}$ anticommutes with $F$.
In fact, up to a phase, its inverse $\mathcal{T}^{-1}$ is actually just its adjoint $\mathcal{T}^\dagger$, since it is a phase times a product of Majorana fermions which all square to 1.
This means we must have
\begin{equation} 
\mathcal{T}^\dagger T_i \mathcal{T} = e^{-i\phi} T_i ,
\end{equation}
where we have made the phase explicit and also written $T_i$ for any traceless Hermitian generator of $\mathfrak{su}(2^{N/2})$.
A basis of such generators is given by the $2^N-1$ appropriately Hermiticized products (excluding the identity) of the Majorana fermions \eqref{eq:basis-su(n)}.

The crucial point now is that the time reversal operator exchanges the $F$ superselection sectors due to the anticommutation relation:
\begin{equation} 
F \mathcal{T} \ket{n,\pm} = -\mathcal{T} F \ket{n,\pm} = \mp \mathcal{T} \ket{n,\pm} ,
\end{equation}
so, because $[H,F]=0$, we must have
\begin{equation}
    \mathcal{T}\ket{n,\pm} = \ket{\pi_n,\mp} .
\end{equation}
It may be the case that $\mathcal{T}$ permutes the energy levels by $\pi$; however, this does not change the conclusion because these permutations will cancel up to an overall phase in the expression $\mathcal{T}^{\dagger} T_i \mathcal{T}$.
Now all that's left is to analyze the matrix element magnitudes:
\begin{equation}
\begin{split}
    |\bra{m,+} T_i \ket{n,+}|^2 & = |\bra{m,+} e^{i\phi} \mathcal{T}^\dagger T_i \mathcal{T} \ket{n,+}|^2 \\
    & = |\bra{m,-} e^{i\phi} T_i \ket{n,-}|^2 \\
    & = |e^{i\phi}|^2 \bra{m,-}  T_i \ket{n,-}|^2 \\
    & = |\bra{m,-}  T_i \ket{n,-}|^2 .
\end{split}
\end{equation}
Therefore, the plus and minus charge sectors have identical ECH matrix elements.
This explains why the two diagonal blocks in Fig.~\ref{fig: ECH matrix q=4 syk} are exactly equal, rather than only approximately equal. By Bott periodicity, this argument should also apply to $N=10$, and we numerically verified that for $N=10,14$ the diagonal blocks are the same (up to machine precision) while for $N=12,16$ the diagonal blocks have similar but different values.

\section{Mixed-field Ising model}\label{sec:ising-numerical}
One can also check the suppression of the matrix elements in the ECH criterion in the mixed-field Ising model. This model consists of a spin-1/2 chain with nearest neighbor interactions and external fields in the parallel and transverse directions. The 2-local Hamiltonian is:
\begin{equation}
    H=\sum_{j=1}^N \sigma^z_j \sigma^z_{j+1}+h\sigma^z_j+g\sigma^x_j.
\label{isingHam}
\end{equation}

\begin{figure}[t]
\centering
\subfigure[] {\includegraphics[width=0.45\textwidth]{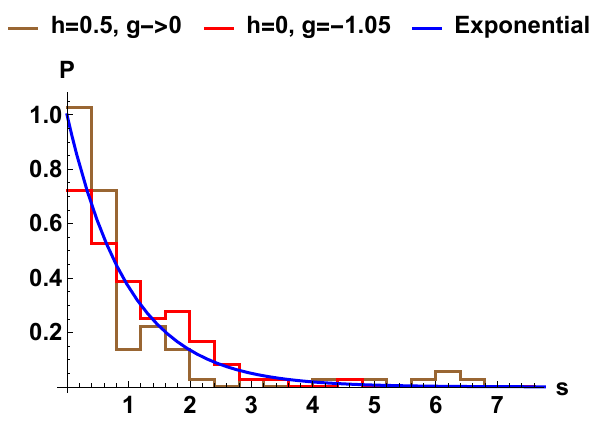} 
}
\subfigure[]{\includegraphics[width=0.45\textwidth]{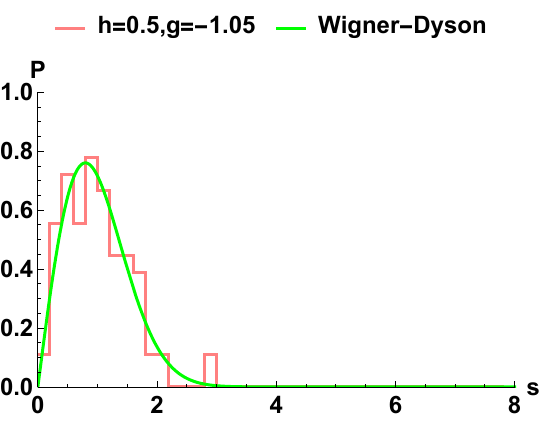} 
}  
\caption{Level spacings for the mixed field Ising model with $N=10$ at three different points in parameter space. (a) compares two parameter choices, one of which corresponds to a free model and other to an interacting integrable model, to the exponential distribution, while (b) shows a choice of parameters in the chaotic regime compared to the Wigner-Dyson distribution.
}
\label{fig:levelspacing plots}
\end{figure}

This model can demonstrate either chaotic or integrable behavior depending on the regime of the parameter space $(h,g)$. This can be observed from the level spacing statistics after exact diagonalization of the model with different choices of $(h,g)$ \cite{ED} as shown in Fig.~\ref{fig:levelspacing plots}. We focus on three  cases, two of which are integrable and one of which is chaotic. The first choice takes $h = 0$ but $g \neq 0$. This is the transverse-field Ising model, which is an interacting integrable model. Although it can be mapped to free fermions through the Jordan-Wigner transformation, the physical interpretation of the model is usually through interacting hard-core bosons, so we refer to it as the ``interacting integrable" choice. The second model takes $g = 0$ and $h \neq 0$, which can be reduced to the Ising model in the absence of external fields, so this model is effectively free. The third choice takes $h \neq g \neq 0$ and is generally chaotic. We choose $(h,g)=(0.5,-1.05)$ as a particular chaotic point in the space of  couplings, $(h,g)=(0,-1.05)$ as an example of interacting integrable couplings, and $(h,g)=(0.5,0)$ as the non-interacting integrable choice. For the integrable choices of couplings, the level spacings are exponentially distributed. By contrast, in the chaotic case   the level spacing distribution is roughly Wigner-Dyson.  When $(h,g)$ takes intermediate values between any of these combinations, the level spacing distributions interpolate between the Wigner-Dyson and exponential distributions. This model also has nontrivial momentum sectors which can lead to zero modes if left unfixed; here we generally consider momentum eigenstates in the $k=1$ momentum sector.

We represent the spin chain operator algebra $\mathfrak{su}(2^N)$ with tensor products of Pauli operators $\sigma_i^x$, $\sigma_i^y$, and $\sigma_i^z$ where $i=1,\dots ,N$ is the site index. Since the spin-chain with nearest-neighbor interaction still retains a notion of spatial locality, we define simple operators to be operators which are supported on at most $k$ sites (that is, they act as the identity outside a $k$-site contiguous region). For example, if we consider easy operators to be at most 2-local, $\sigma^z_3 \sigma^z_4$ is an easy operator, but $\sigma^z_3 \sigma^z_6$ is not. 

We expect that the off-diagonal matrix elements of the ECH matrix $R_{mn}$ will be suppressed by \eqref{eq:ratio}, which for the Ising model is
\begin{equation}
   s_{MF}(N,k)=\frac{N }{4^N-1} \sum_{j=1}^k 3^j .
\label{eq:ratio-mf}
\end{equation}

\begin{figure}[t]
\centering
\includegraphics[scale=0.5]{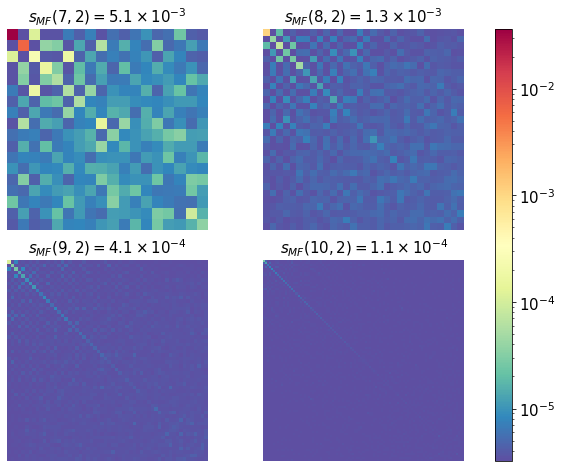}
\caption{The ECH matrices of the mixed-field Ising model of length $N=7,8,9,10$ at the maximally chaotic point $h=0.5,g=-1.05$, with $k=2$-local operators taken as simple.}
\label{fig:k2chaoticIsing}
\end{figure}

Below, we compute the ECH matrices for spin chains of length $N=7,8,9,10$ with $k=2$-local operators taken as simple. As in the case of the SYK model, we expect that in general the off-diagonal entries of the ECH matrix are suppressed according to \eqref{eq:ratio-mf}, and that the variance in the chaotic regime is $O(1)$. 
We numerically verify these properties in Fig.~\ref{fig:k2chaoticIsing} taking $k=2$-local operators to be simple to demonstrate the behavior. And this behavior persists independent of the choice of degree of locality $k$, as long as it is larger than the degree of interaction of the Hamiltonian.

\begin{figure}[t]
\centering
\subfigure[] {\includegraphics[width=0.45\textwidth]{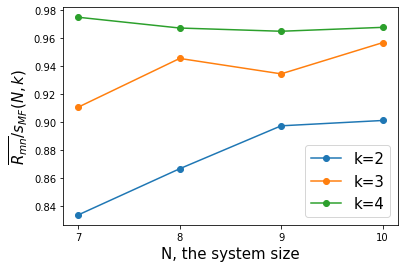} }
\subfigure[]{\includegraphics[width=0.45\textwidth]{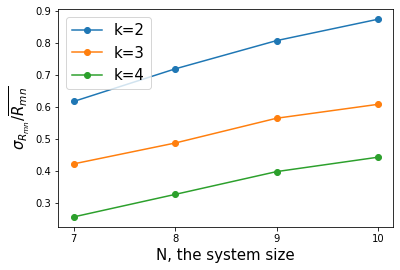} }
\caption{Comparison of the average and standard deviation of the $R_{mn}$ matrix in the maximally chaotic regime with varying sizes.}
\label{fig:chaoticIsing_varying_k}
\end{figure}

\begin{figure}[t]
\centering
\includegraphics[scale=0.5]{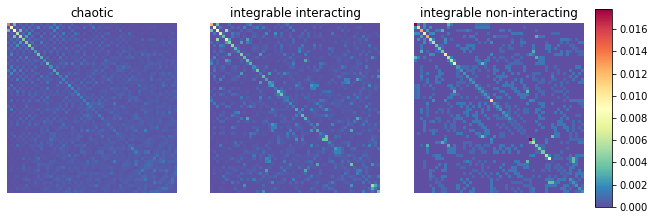}
\caption{ECH matrices of the mixed-field Ising models of length $N=9$ at the free, integrable interacting, and maximally chaotic points, with $k=2$-local operators taken as simple.
}
\label{fig: N9_k2_Rmn_compare}
\end{figure}

In Fig.~\ref{fig: N9_k2_Rmn_compare} we compare the ECH matrix elements between the different choices of couplings at fixed $N=9$. Because the choice of the specific momentum sector has eliminated all the potential symmetries in the system, the ECH matrix entries does not show the same symmetry-sector structure as the $q=4$ SYK model. The greater degree of suppression of the off-diagonal matrix elements can be easily seen in the heat map.
Another difference between the regimes comes from comparing the free theory to the two interacting theories. The distribution of the overlaps in the free theory is again a discrete distribution with a sizeable mass at 0. On the other hand, the overlap-distribution at the chaotic point and the interacting integrable point look more like a density function, analogous to what happens in the $q=3,4$ SYK models.

\begin{figure}[t]
\centering
\includegraphics[
width=0.45\textwidth]{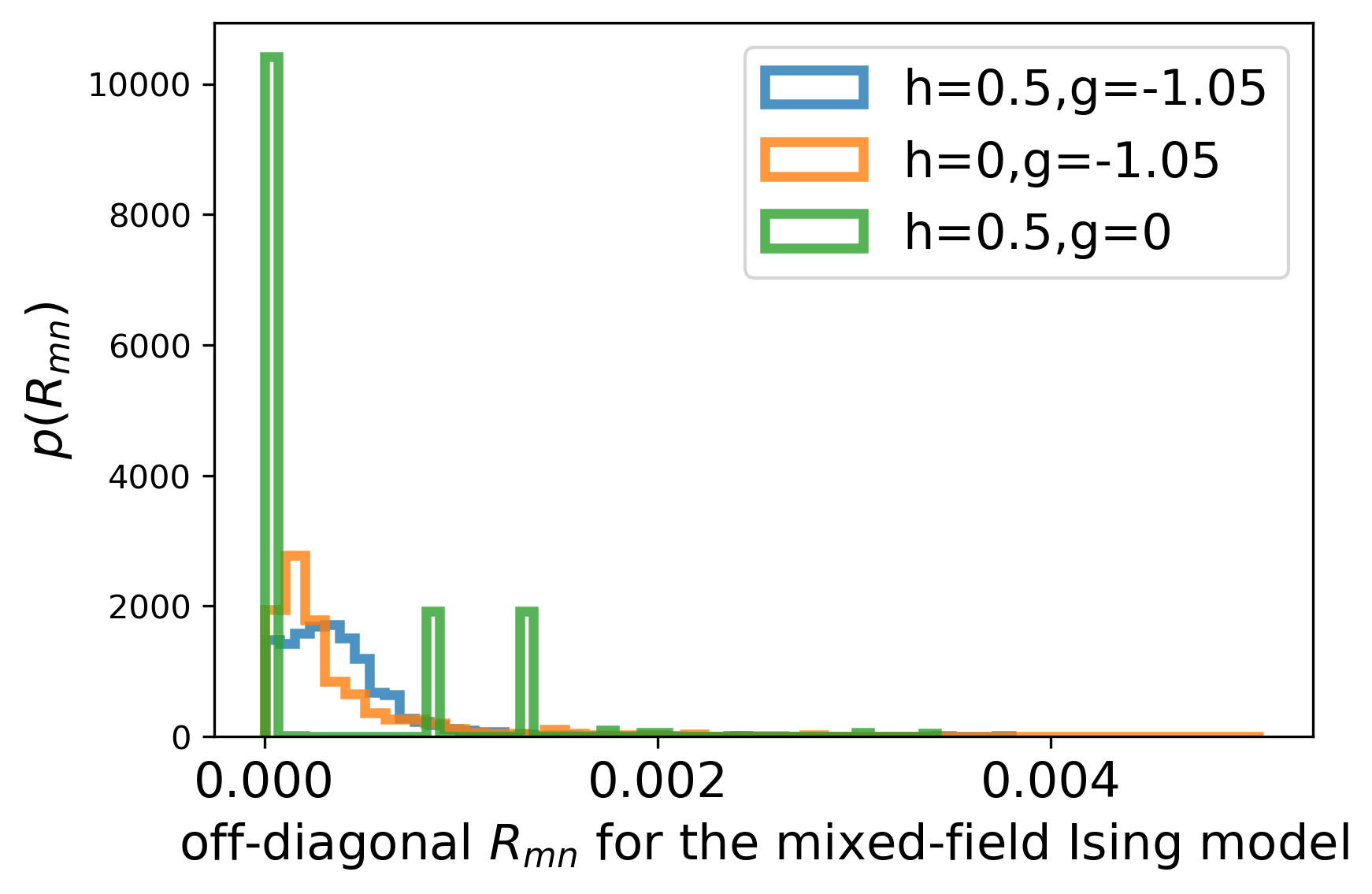}
\caption{Distributions of ECH matrix elements for the mixed-field Ising model in the integrable, integrable-interacting and chaotic regimes, calculated using spin chain of length $N=9$, with $k=2$-local operators taken as simple.}
\label{fig: ECH dist Ising}
\end{figure}

\begin{figure}[t]
\centering
\subfigure[] {\includegraphics[width=0.45\textwidth]{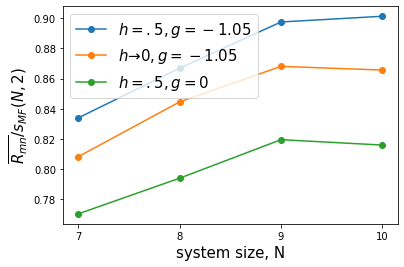} \label{fig: average Rmn compare} }
\subfigure[]{\includegraphics[width=0.45\textwidth]{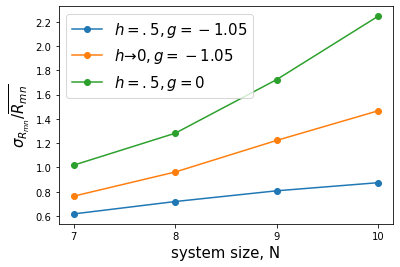} \label{fig: standard deviation compare} }  
\caption{Comparison of the average and standard deviation of the $R_{mn}$ matrix in the chaotic and integrable regimes showing (a) the scaling of the average off-diagonal element divided by the proportion of simple operators ($N=9$, $k=2$), and (b) the scaling of the standard deviation with system size ($k=2$).}
\label{fig:ratio-stdev-scaling}
\end{figure}

We have also investigated the behavior of the distributions of off-diagonal entries of $R_{mn}$ as $N$ is increased in the various cases (Fig.~\ref{fig: ECH dist Ising}). Because of the large number of vanishing overlaps,   the average value of an off-diagonal entry of $R_{mn}$ in the integrable case is smaller than that of the chaotic case (Fig.~\ref{fig: average Rmn compare}). However, the variance in the chaotic case remains significantly smaller when $N\to \infty$; it appears to grow very slowly and approximately linearly while the variance in the free theory grows approximately quadratically (Fig.~\ref{fig: standard deviation compare}). 

\bibliographystyle{JHEP}
\bibliography{main}

\end{document}